\documentclass[aps,preprint,amsmath,floatfix,nofootinbib]{revtex4}

\usepackage{graphicx}
\begin{document}

\title{ Diffuse $\gamma$-rays and $\bar{p}$ flux from dark matter
annihilation --- a model for consistent results with EGRET and cosmic ray data}

\author{Xiao-Jun Bi$^{1,2}$}
\email{bixj@mail.ihep.ac.cn}
\author{Juan Zhang$^1$}
\author{Qiang Yuan$^1$}
\affiliation{$^1$ Key laboratory of particle astrophysics, IHEP, 
Chinese Academy of Sciences, Beijing 100049, P. R. China \\
$^2$ Center for High Energy Physics,
Peking University, Beijing 100871, P. R. China}

\begin{abstract}

In this work we develop a new propagation model for the Galactic cosmic rays
based on the GALPROP code, including contributions from dark matter
annihilation. Its predictions of the Galactic diffuse $\gamma$ ray spectra
are compatible with the EGRET data in all sky regions. It also gives consistent
results about the diffuse $\gamma$ ray longitude and latitude distributions.
The B/C, $^{10}$Be/$^9$Be, proton,
electron and antiproton spectra are in agreement with cosmic 
ray measurements as well.
In this model we have taken a universal proton spectrum throughout the Galaxy
without introducing large fluctuation, 
considering the proton energy loss is negligible.
The dark matter annihilation signals are ``boosted'' after taking the 
contributions from subhalos into account.
Another interesting feature of this model is that
it gives better description of the diffuse $\gamma$ rays when taking the
source distribution compatible with supernova remnants data, which is
different from previous studies.
\end{abstract}

\maketitle

\section{introduction}

Cosmic ray (CR) propagation is a complex process involving diffusion by magnetic
field, energy losses and spallation by interactions with the interstellar medium
(ISM). Diffuse Galactic $\gamma$-rays are produced via the decay of neutral pion 
and kaon, which are generated by high energy cosmic nuclei interacting 
with interstellar gas, and via energetic 
electron inverse Compton (IC) scattering and bremsstrahlung. 
The $\gamma$ rays are not deflected by the magnetic field and the
ISM is transparent to $\gamma$-rays below a few TeV \cite{jlzhang}. 
Therefore, the observation
of the diffuse $\gamma$-ray spectra and distribution is a valuable 
diagnosis of the self-consistency of propagation models, 
the distribution of CR sources and the
ISM. 
The Galactic diffuse $\gamma$ rays has been measured by EGRET \cite{egret,hunter}
and exhibits an excess above $\sim$ 1 GeV compared to prediction \cite{hunter}.
The theoretical calculations are based on a conventional CR model, 
whose nuclei and electron spectra in the whole Galaxy are taken to be
the same as those observed locally.

The discrepancy has attracted much attention
\cite{spec,smr00,volk,opt,boer-fit,boer-susy,boer-gas}
since it was first raised. It may either indicate a non-ubiquitous       
proton or electron spectrum, 
or the existence of new exotic sources of diffuse $\gamma$-ray emission.

Many efforts have been made to solve the ``GeV excess'' problem within the 
frame of CR physics, such as adopting different CR spectra 
\cite{spec,smr00,opt}, or assuming more important contribution 
to diffuse $\gamma$-rays from CR sources \cite{volk}.
A brief review of these efforts is given in \cite{opt}.
In that paper
an ``optimized'' propagation model has been built by directly fitting the 
observed diffuse $\gamma$-ray spectrum.  
This ``optimized''  model  introduces interstellar 
electron and proton intensities that are different from the local ones and reproduces
all the CR observational data at the same time. 
Up to now, it seems to be the best model to explain the EGRET diffuse
$\gamma$-ray data based on CR physics.
However, this ``optimized''  model is fine tuned  
by adjusting the electron and proton injection spectra,  
while keeping the injection spectra of heavier nuclei  
unchanged, as in the conventional model, so that the B/C ratio is not  
upset. Furthermore a large scale proton spectrum different from the locally
measured one might not be reasonable,   
since the proton diffusion time scale is much smaller than its energy 
loss time scale, 
which tends to result in a large scale universal proton spectrum
within the Galaxy apart from some specific sources. 
Unlike protons, the electron spectrum may have large spatial 
fluctuation due to their 
fast energy losses from IC, bremsstrahlung, 
ionization and the stochastic sources \cite{sm01b}.

Another interesting solution, given by de Boer  
et al. \cite{boer-fit,boer-susy,boer-gas}, 
is that the ``GeV excess'' is attributed to dark matter (DM)
annihilation from the Galactic halo, where the
DM candidate is the neutralino from the supersymmetry (SUSY).
By fitting both the background spectrum shape from cosmic nuclei collisions
and the signal spectrum shape from dark matter annihilation (DMA) 
they found the EGRET data could be well explained \cite{boer-fit}.
This suggestion is very interesting and impressive, due to the fact that 
in 180 independent sky regions\,\footnote{The division at low energies 
may be problematic as the EGRET point spread function is about $6^\circ$ 
and non-Gaussian at low energy.},  all the discrepancies between
data and the standard theoretical prediction can be
well explained by a single spectrum from DMA with $m_\chi = 50 \sim 70$ GeV.
Furthermore, by fitting the spatial distribution of the diffuse $\gamma$-ray
emission they reconstructed the DM profile,
with two rings supplemented to the smooth halo.
The ring structure seems also necessary to explain the damping 
in the Milky Way rotation curve \cite{boer-fit} 
and the gas flaring \cite{boer-gas}. 
However, the DMA solution to the ``GeV excess'' also meets a great challenge
because of its prediction of the antiproton flux.
In de Boer's model, this flux is more than 
one order of magnitude greater than data \cite{bergs}.
The overproduction of antiprotons comes from two factors: 
a universal ``boost factor''  $\sim 100$ of the diffuse $\gamma$-rays
boosts the local antiproton flux by the same amount;  
the two rings introduced to account for the diffuse $\gamma$-ray flux 
enhance the antiproton flux greatly 
since they are near the solar system and are strong antiproton sources.
In their work, de Boer et al. did not try to
develop a propagation model. 
Instead they focused on \textit{reconstruction} of the DM profile by
fitting the EGRET data.
They need a ``boost factor'' to enhance the contribution from DMA.
The background contribution from pion decay is arbitrarily normalized 
in order to fit data best.


In the present work we try to build a propagation model to explain
the EGRET diffuse $\gamma$-ray data based on both Strong's and 
de Boer's models while overcoming their difficulties.
In our model the diffuse $\gamma$-ray comes from both CRs and DMA directly. 
On one hand we do not introduce a  different 
interstellar proton spectrum from the local one;  
on the other our model gives consistent $\bar{p}$ flux 
even when including contribution from DMA. 
Furthermore we do not need the large ``boost factor'' to DMA or 
renormalization factor to CR contribution.
Actually, the $\gamma$-ray flux from DMA is boosted by taking the subhalos into account. 
The diffuse $\gamma$-ray spectra at different sky regions and its 
profiles as functions of Galactic longitude and latitude are well consistent
with EGRET data. 
In a previous paper \cite{0611783},  we have briefly introduced our model.
Full details are given in the present paper.

The paper is organized as follows.
We describe the calculation of the DMA contribution in Section II. 
In Section III, we focus on the conventional CR model. 
As underlined, it explains the EGRET data, but produces too large $\bar{p}$ flux. 
In Section IV, we present our new propagation model and its  
predictions for the diffuse 
$\gamma$-ray spectra and profiles, and the $\bar{p}$ flux. 
Finally we summarize and conclude in Section V.

\section{dark matter annihilation} 

Here, we calculate the diffuse  $\gamma$-ray flux from DMA.
The frame of minimal supersymmetric extension 
of the standard model (MSSM) is retained, where
we assume that DM consists of the lightest neutralinos.
A pair of neutralinos in the Galactic halo 
can annihilate into leptons, quarks and gauge bosons.  
Their decay products include a $\gamma$-ray continuum and 
thus contribute to the $\gamma$-rays diffuse emission
produced by CR interaction with the ISM.

\subsection{Particle factor}

The flux of $\gamma$ rays from the 
neutralino annihilation is given by
\begin{equation}
\label{flux}
\Phi(E)=\frac{\langle\sigma v\rangle}{2m_{\chi}^2}\frac{dN}{dE}
\int{dV \frac{\rho^2}{4\pi d^2}} = \frac{1}{4\pi}
\frac{\langle\sigma v\rangle}{2m_{\chi}^2} \frac{dN}{dE} \times \frac{1}{d^2}
\int \rho^2(r) r^2 dr d\Omega \ ,
\end{equation}
where $\langle\sigma v\rangle$ is the averaged neutralino annihilation 
cross section times
relative velocity, $\frac{dN}{dE}$ is the differential flux
in a single annihilation,
$m_{\chi}$ is the mass of neutralino,
$d$ is the distance between the detector and the $\gamma$-ray source,
and $\rho(r)$ is the spherically-averaged DM distribution,
determined by numerical simulation or by observations.
The flux in Eq.~(\ref{flux}) is determined by two
independent factors, the first one only depending on the
DM particle nature (mass, strength of interaction and so on), 
the second one depending on the DM distribution only. 
The first factor is denoted as ``particle factor'' and 
the second as ``astrophysics factor''.

\begin{figure}
\includegraphics[scale=1]{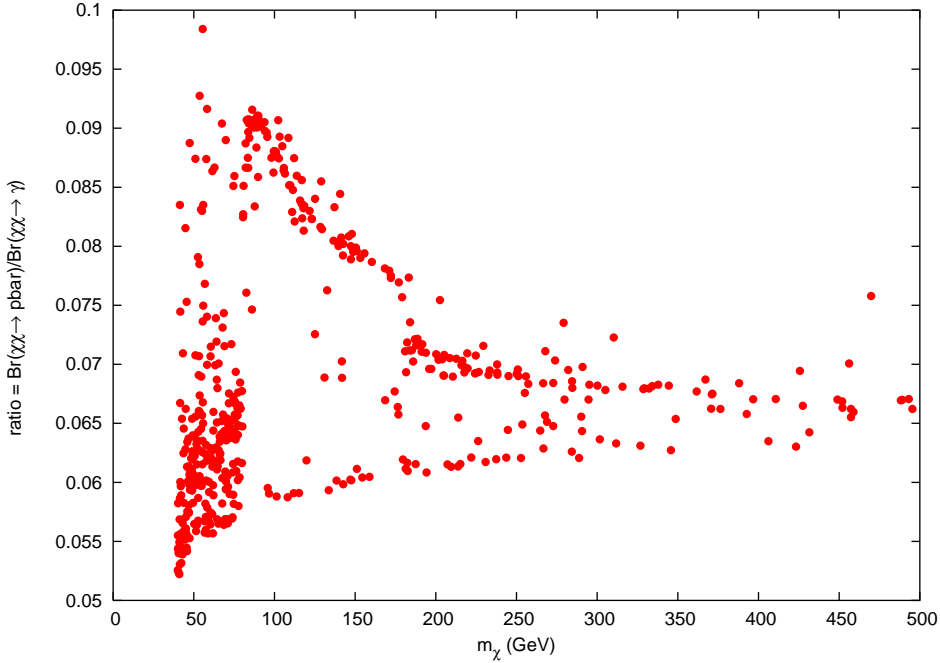}
\caption{\label{ratio}
$\frac{Br(\chi\chi\to {\bar p})}{Br(\chi\chi\to \gamma)}$
as a function of neutralino mass. Thresholds for ${\bar p}$ and
$\gamma$-ray kinetic energies are taken as $0.5$ GeV.
}
\end{figure}

The ``particle factor'' is calculated by doing a random scan
in the SUSY parameter space and choosing models which could satisfy all the collider and cosmological constraints.
However, there are more than one hundred free SUSY breaking parameters,
even for the R-parity conservative MSSM.
A general practice in phenomenological studies is to assume
some simple relations between the parameters.
Following the assumptions in DarkSUSY \cite{darksusy} we take 
seven free parameters during the calculation of DM 
production and annihilation, i.e., the higgsino mass 
parameter $\mu$, the wino mass parameter $M_2$,
the mass of the CP-odd Higgs boson $m_A$, the ratio of the Higgs
Vacuum expectation values $\tan\beta$, the scalar fermion mass parameter
$m_{\tilde{f}}$, the trilinear soft breaking parameter $A_t$
and $A_b$.  All the sfermions are taken with a common soft-breaking mass
parameter $m_{\tilde{f}}$; all trilinear
parameters are zero except those of the third family; the bino and wino
have the
mass relation, $M_1=5/3\tan^2\theta_W M_2$, coming from the unification
of the gaugino mass at the grand unification scale.

A random scan in the 7-dimensional parameter space of MSSM 
is performed using the package DarkSUSY \cite{darksusy}. 
We assume these parameters to range as follows:
$50$ GeV $< |\mu|,\ M_2,\ M_A,\ m_{\tilde{f}} < 10$ TeV,
$1.1 < \tan\beta < 61$, $-3m_{\tilde{q}} < A_t, \ A_b < 3m_{\tilde{q}}$,
$\text{sign}(\mu)=\pm 1$.
We find the $\gamma$-ray spectrum
with $m_\chi=40 - 50$ GeV (after being added to the background  
$\gamma$-ray spectrum) can fit the EGRET data very well, which
is consistent with the result of de Boer et al. \cite{boer-fit}.
However, the branching ratio to $\gamma$ rays varies by a few order
of magnitude for different models, and only models with large branching
ratio into $\gamma$-rays can account for the ``GeV excess''. 
We also calculate the branching ratio 
into antiprotons since the $\bar{p}$ flux is a sensitive
test of the DM model.

In Fig.~\ref{ratio} we give the ratio of neutralino
annihilation into ${\bar p}$ and $\gamma$-rays
as a function of neutralino mass. We find $\frac{Br(\chi\chi\to {\bar p})}
{Br(\chi\chi\to\gamma)} \approx \frac{1}{20} - \frac{1}{10}$
above the threshold energy $E_{th} = 0.5$ GeV for $m_\chi=40 - 500$ GeV.
For light neutralinos, the variation of the ratio in the
parameter space is at most as large as a factor of $2$. 
For heavy neutralinos the variation is very small, 
and the reason is that ${\bar p}$ and
$\gamma$-rays are coming from the same final states,
thus they are closely related. 
Below, we choose  $m_\chi =48.8$ GeV 
to explain the EGRET data which predicts $\Omega h^2 = 0.09$
and $\frac{Br(\chi\chi\to {\bar p})}
{Br(\chi\chi\to\gamma)} \approx 0.055$ for the threshold energy
$E_{th} = 0.5$ GeV. Actually for $m_\chi$ between $40 \sim 50$ GeV
the annihilated $\gamma$-ray spectra are almost identical.
A few models also show similar branching ratio as in our chosen model
(e.g.,  for $m_\chi =41.13, 41.21$ GeV). 

\subsection{Astrophysics factor}

The astrophysics factor determining the annihilation fluxes is defined as
$\Phi^{astro}=\int_{V}\frac{\rho^2}{d^2} dV$ with $d$ the
distance to the source of $\gamma$-ray emission, $\rho$ the
density profile of DM, and $V$ the volume in which annihilation
taking place. In the de Boer model, the authors adopt a cored Galactic halo
model with two DM rings, boosting the $\gamma$-ray flux
by a factor of the order $\sim 100$. In our work we take a cuspy
NFW \cite{nfw97} (or Moore \cite{moore}) profile. Especially we take
the contribution from subhalos into account to enhance the
annihilation signals.

High resolution simulations of cosmological structure evolution
reveal that in the cold DM scenario
the structures form hierarchically and a large number of substructures
survive in the galactic halos. 
A fraction of about 10\% of the total halo mass may have survived tidal
disruption and appear as distinct and self-bound substructures 
inside the virialized host halos.
The existence of substructures will enhance the annihilation rate greatly
by enhancing the astrophysics factor in Eq.~(\ref{flux}).

The mass function and spatial function of subhalos are given by 
N-body simulations. A simple analytical fit to simulations 
allows to write the probability of a subhalo with mass $m$ appearing at position 
$r$ as \cite{diemand}
\begin{equation}
\label{dis}
n(m,r)=n_0 \left( \frac{m}{M_{vir}} \right)^{-1.9} \left(
1+\left( \frac{r}{r_H} \right)^2 \right)^{-1}\ , 
\end{equation} 
where $n_0$ is the normalization factor so that about $10\%$ of the halo 
mass is enclosed in subhalos, 
$M_{vir}\approx 1.0\times 10^{12} M_{\odot} $ is the Galactic halo mass, 
$r_H=0.14r_{\text{vir}} \approx 29$ kpc is the core radius for the 
distribution of subhalos. 
The result given above agrees well with that of another recent 
simulation by Gao et al. \cite{gao}. 

The DM density profile within each subhalo is taken as 
the NFW \cite{nfw97}, Moore \cite{moore} and a cuspier form \cite{zhao} as 
$\rho = \frac{\rho_s}{(r/r_s)^\gamma ( 
1+r/r_s)^{3-\gamma}}$ with $\gamma = 1.7$. The last form 
is favored by the simulation conducted by Reed et al. \cite{reed}, 
which gives that $\gamma = 1.4 - 0.08\log(M/M_*)$,  
for the subhalo mass of $0.01{M_*}\sim 1000{M_*}$ with a large scattering, 
increasing as the subhalo mass decreases. 
Small halos with large $\gamma = 1.5 - 2$  
are also found by Diemand et al. \cite{minihalo}. 
We take $\gamma = 1.7$ for the whole range 
of subhalo masses as a simple approximation. 
 
We calculate the concentration parameter $c_v$ by adopting the semi-analytic  
model of Bullock et al. \cite{bullock}, which describes it as a function
of the viral mass and redshift. We adopt the mean $c_v-m_{sub}$ relation
at redshift zero.
The scale radius is then determined by $r^{nfw}_s=r_v/c_v$,
$r^{moore}_s=r^{nfw}_s/0.63$ and $r^\gamma_s=r^{nfw}_s/(2-\gamma)$ 
for the three density profiles respectively. 
Another factor determining the $\gamma$-ray flux is the core radius, 
$r_{\text{core}}$, 
within which the DM density should be kept constant due to the balance 
between the annihilation rate and the rate of DM particles  
falling into this region \cite{berezinsky}. 
The core radius $r_{\text{core}}$ is approximately 
$10^{-8} \sim 10^{-7}$ kpc for $\gamma = 1.7$ and about
$10^{-9} \sim 10^{-8}$ kpc for the Moore profile. 

Along a direction $(\theta,\phi)$, 
the subhalos contribute to the ``astrophysical factor'' 
$\Phi^{astro}=\int_{l.o.s} \Phi^{sub}dN_{sub}(\theta,\phi) $, 
where $\Phi^{sub} = \int_{V_{sub}}\frac{\rho^2}{d^2} 4\pi r^2 dr$ is the astrophysical 
factor for a single subhalo, and $N_{sub}$ is the number density of subhalos. 
When we calculate the integration along the line-of-sight 
starting from the Sun, we get the 
Jacobian determinant as $\frac{r'}{r}$, with 
$r'$ the distance from the subhalo to the Sun. 
The minimal subhalos can be as light as $10^{-6} M_{\odot}$ 
as shown by the recent simulation conducted by 
Diemand et al. \cite{minihalo}, while the maximal mass 
of substructures is taken to be $0.01 M_v$ \cite{bi}. 
The tidal effects are taken into account based on the ``tidal approximation'' 
\cite{bi}, where subhalos are disrupted near the Galactic center (GC). 
The total signal flux comes from the annihilation in the 
subhalos and in the smooth component. 
 
\begin{figure} 
\includegraphics[scale=1]{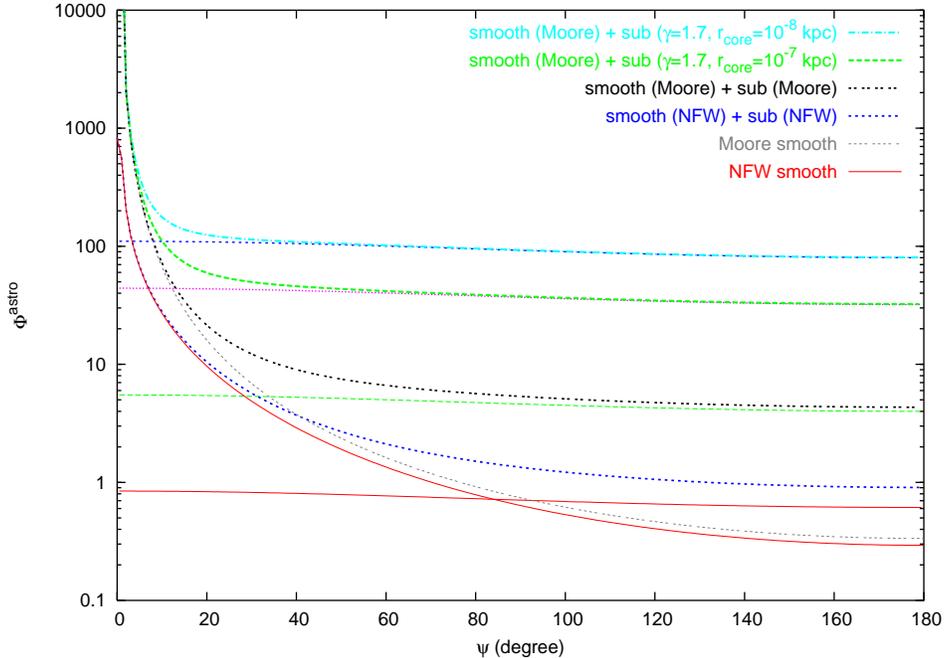}  
\caption{\label{density}
The astrophysical factor $\Phi^{astro}$ (in unit of $(GeV/cm^3)^2\ kpc\ Sr^{-1}$) 
 from different directions.
The almost horizontal lines correspond to the contributions from
subhalos only.
}
\end{figure}

In Fig.~\ref{density},  we show the factor $\Phi^{astro}$ from the
smooth component, the subhalos and the total contribution as a 
function of the direction to the GC.
The $\Phi^{astro}$ from subhalos is almost isotropic in all directions  
as the Sun is near the GC compared with the viral radius $r_{vir}$ of the Galactic halo.
The largest enhancement for $\gamma = 1.7$  subhalos is observed
at large angles and can reach 2 orders of magnitude. This enhancement
depends on the value of $r_{\text{core}}$, while
for the Moore profile the enhancement is about one order of magnitude,
and for for the NFW profile only about $ 3$ times larger.
The $\Phi^{astro}$ for Moore and NFW profiles is not sensitive to $r_{\text{core}}$ \cite{bi}.
We also notice that near the GC there is no enhancement. This is
actually a very important difference from the model given by de Boer
where the ``boost factor'' is universal \cite{boer-fit}.
Given the factor $\Phi^{astro}$ and the SUSY model we can predict the
$\gamma$-ray flux from neutralino annihilation.

In the next section we give the background diffuse $\gamma$-rays
from CR interactions with the ISM.

\section{The conventional model}

The optimized model of Strong et al. reproduces the diffuse $\gamma$ rays
assuming interstellar proton and electron spectra different from those locally 
measured. However, the required fluctuation of the proton spectrum may be not realistic.
The works of de Boer et al. have strongly indicated that DMA may 
account for the diffuse $\gamma$ ray excess \cite{boer-fit}. 
Therefore our first attempt is to build a propagation model 
including contribution from DMA based on
the conventional CR model assuming universal CR spectra.

\begin{figure}
\includegraphics[scale=0.45]{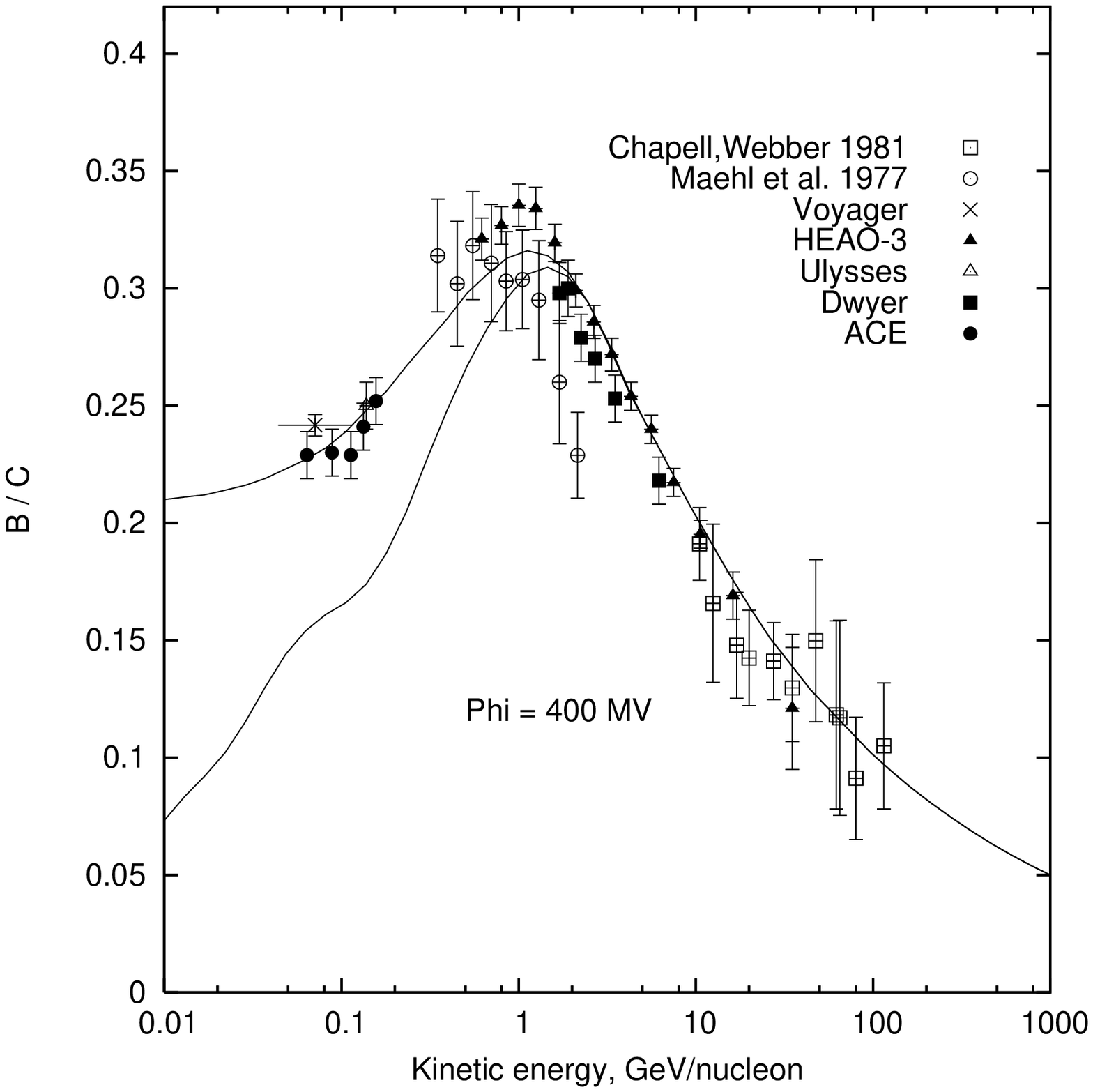}
\includegraphics[scale=0.45]{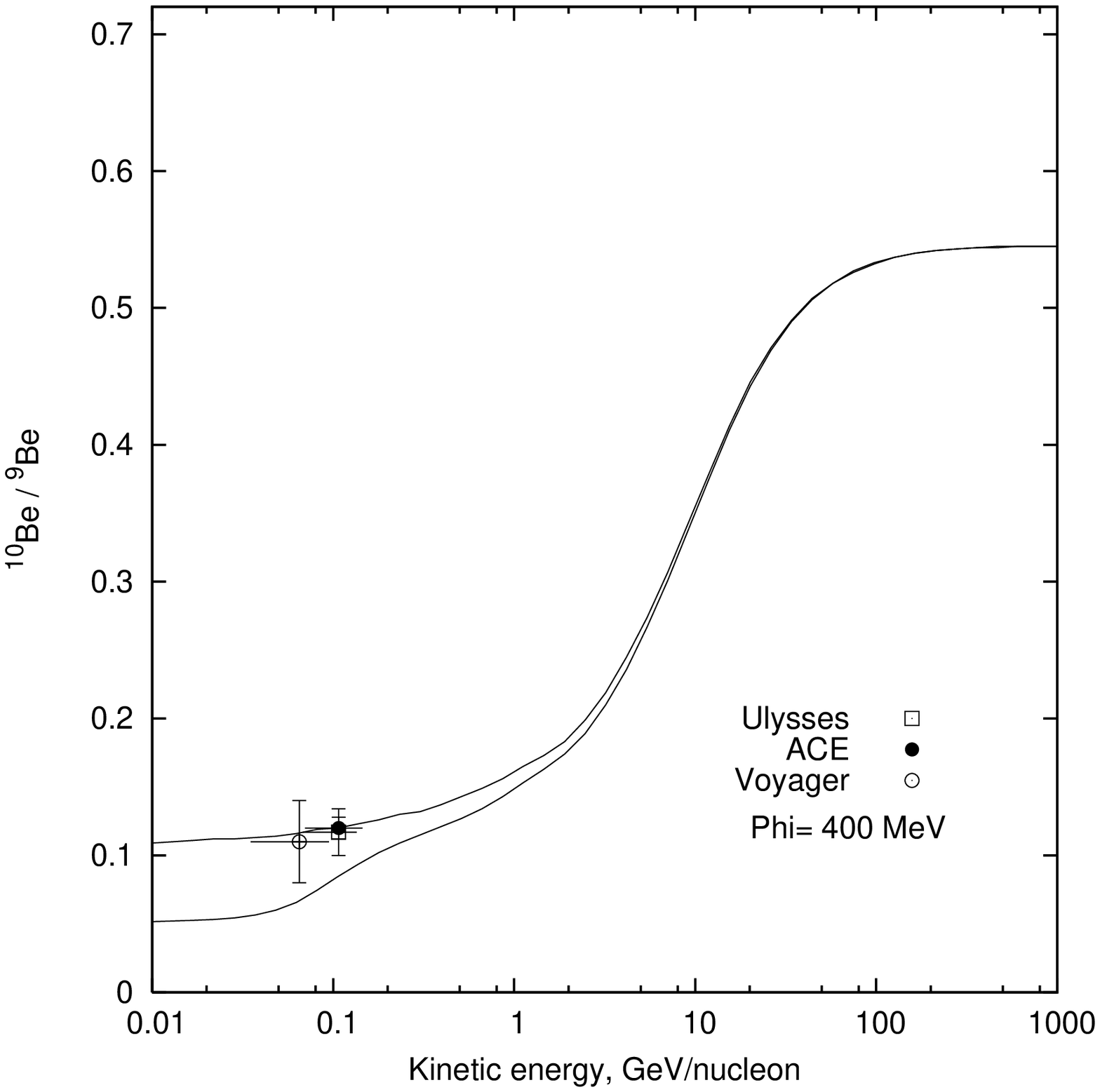}
\includegraphics[scale=0.45]{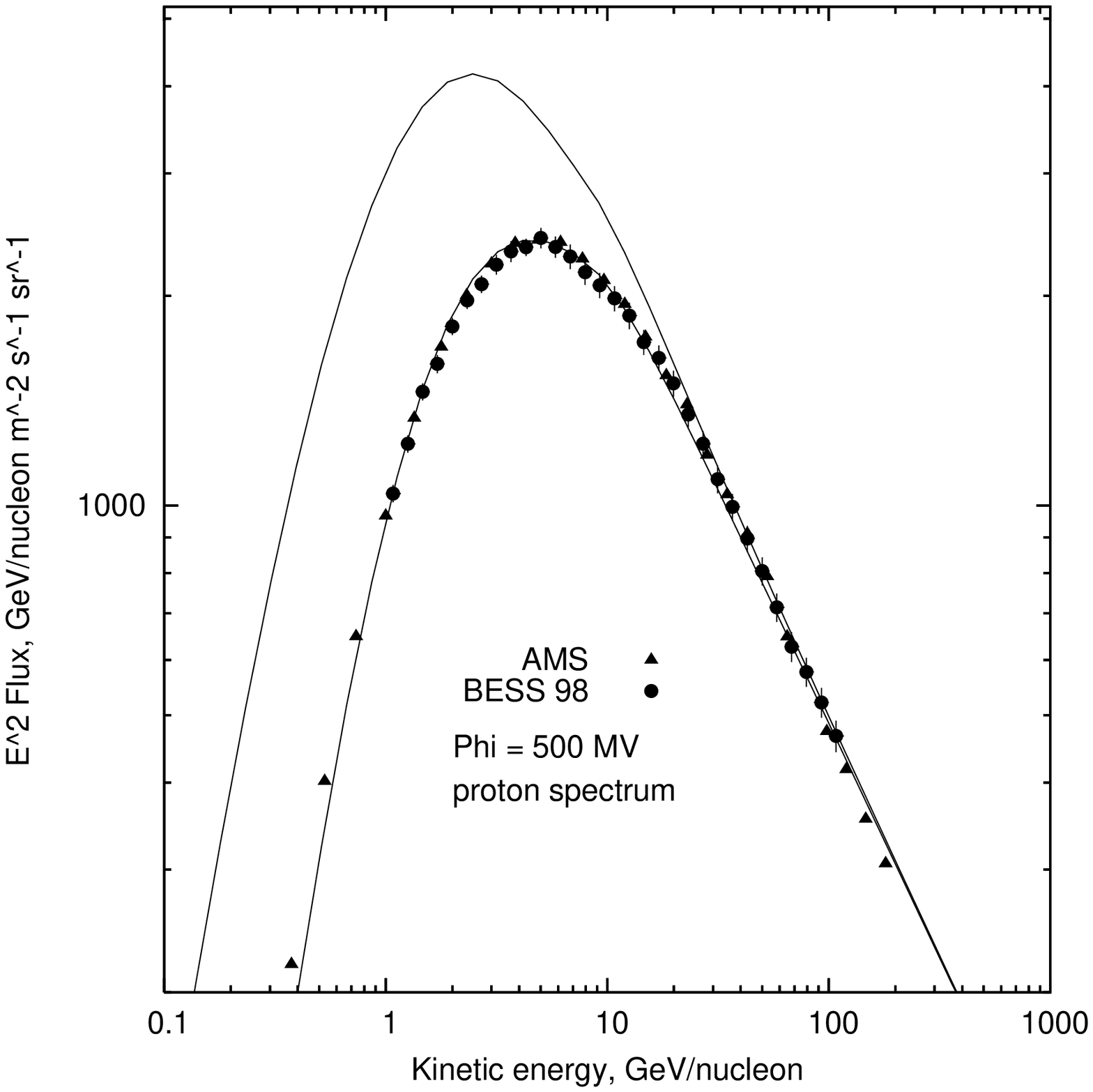}
\includegraphics[scale=0.45]{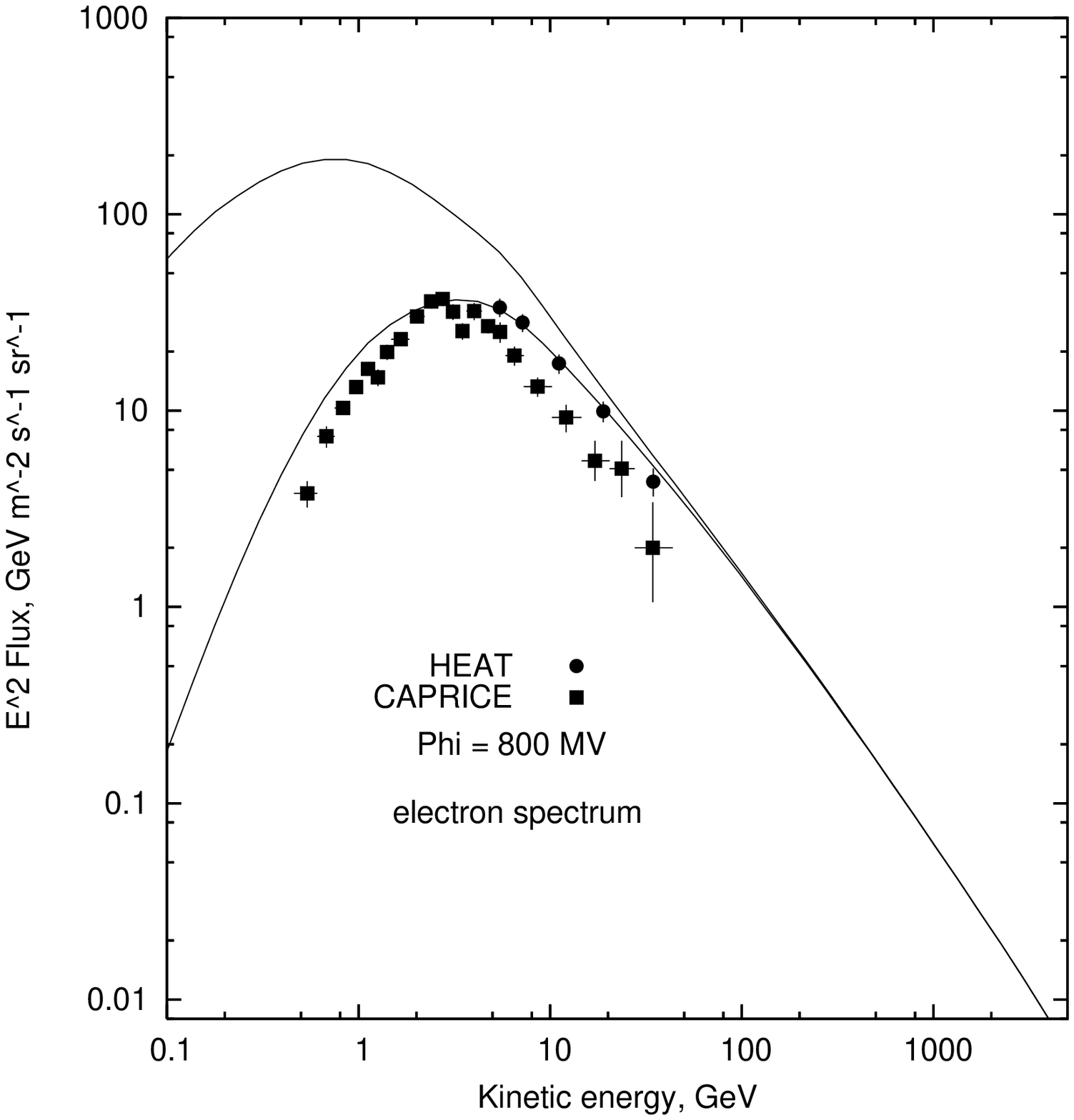}
\caption{\label{conv-crs}
Cosmic ray results in our conventional model.
Lower and upper curves for B/C ratio correspond to
solar modulated and local interstellar (LIS) values respectively.
For $^{10}$Be/$^{9}$Be ratio, the flat curve near 
0.1 GeV/nucleon corresponds to the modulated flux and the other one is LIS.
On the contrary, the lower curves for proton and electron spectra are the modulated
ones and the upper are LIS.
B/C data are from ACE\cite{ace}, Ulysses\cite{ulysses},
Voyager\cite{voy}, HEAO-3\cite{heao} and others \cite{bcdata};
$^{10}$Be/$^{9}$Be data from Ulysses\cite{connell}, ACE\cite{aceBe} and Voyager\cite{voyBe};
Proton data from BESS98\cite{bess98} and AMS\cite{amsp};
electron data from CAPRICE\cite{cap} and HEAT\cite{barw}.
}
\end{figure}

We adopt the GALPROP package \cite{galprop,smr00} 
to calculate the propagation of CRs
and production of Galactic background diffuse $\gamma$ rays.
It was shown that an explanation of the EGRET data in \textit{all} sky 
directions in the conventional model
is not an easy task, even including the contribution from DMA|considering
that DMA provides only a single extra spectrum from neutralino annihilation.
To give the best fit to the EGRET data, de Boer et al. \cite{boer-fit} 
have to introduce arbitrary renormalization factors in different directions
for the $\gamma$ ray spectra given in the conventional model \cite{opt}.
Since we are trying here to build a propagation model, we will not 
introduce any renormalization factors, but rather explain the $\gamma$ ray data
by adjusting the propagation parameters 
after adding the contribution from DMA.

After a lot of tests, we came to the following propagation parameters.
The scale height of the propagation halo $z_h$ takes the same value 4\,kpc
as that taken by Strong et al. in their conventional and optimized models \cite{opt}.
The nuclei injection spectra share the same power law form in rigidity,
and nuclei up to $Z=28$ and all relevant isotopes are included.
The CR injection spectra are given in Table \ref{inj_para}.
We adopt the diffusive reacceleration propagation model.
The spatial diffusion coefficient is given as a function of rigidity 
in the form 
\begin{equation}
\label{diff}
D(\rho) = \beta D_0(\rho/\rho_0)^\delta\, ,
\end{equation}
where $\beta = v/c$, $D_0=5.4\times\,10^{28}$\,cm$^2$\,s$^{-1}$, 
$\rho_0=4$\,GV, and $\delta=0.34$. 
The Alfv\'en speed to describe the reacceleration process is $v_A=34$\, km\,s$^{-1}$.
These propagation parameters well describe the observed B/C ratio, 
the $^{10}$Be/$^{9}$Be ratio, and the local measured proton and electron spectra, 
as shown in Fig.~\ref{conv-crs}.

\begin{table}
\caption{\label{inj_para} Cosmic ray injection spectrum parameters.}
\begin{ruledtabular}
\begin{tabular}{lccc}
& Injection index &  Break rigidity & Normalization at E\\
& below/above break rigidity & GV & m$^{-2}$ sr$^{-1}$ s$^{-1}$ GeV$^{-1}$ at GV\\
\hline
Nuclei Spectra\footnotemark[1] & 1.82 / 2.36 & 9 & $4.9\times10^{-2}$ at 100\\
Electron Spectrum\footnotemark[1] & 1.60 / 2.54 & 4 & $4.86\times10^{-3}$ at 34.5\\
\hline
Nuclei Spectra\footnotemark[2] & 1.86 / 2.36 & 10 & $5\times10^{-2}$ at 100\\
Electron Spectrum\footnotemark[2] & 1.50 / 2.54 & 6  & $1\times10^{-2}$ at 34.5\\
\end{tabular}
\end{ruledtabular}
\footnotetext[1]{Spectra used in our conventional model;}
\footnotetext[2]{Spectra used in our new model.}
\end{table}

\begin{figure}
\resizebox{16cm}{8cm}{\includegraphics{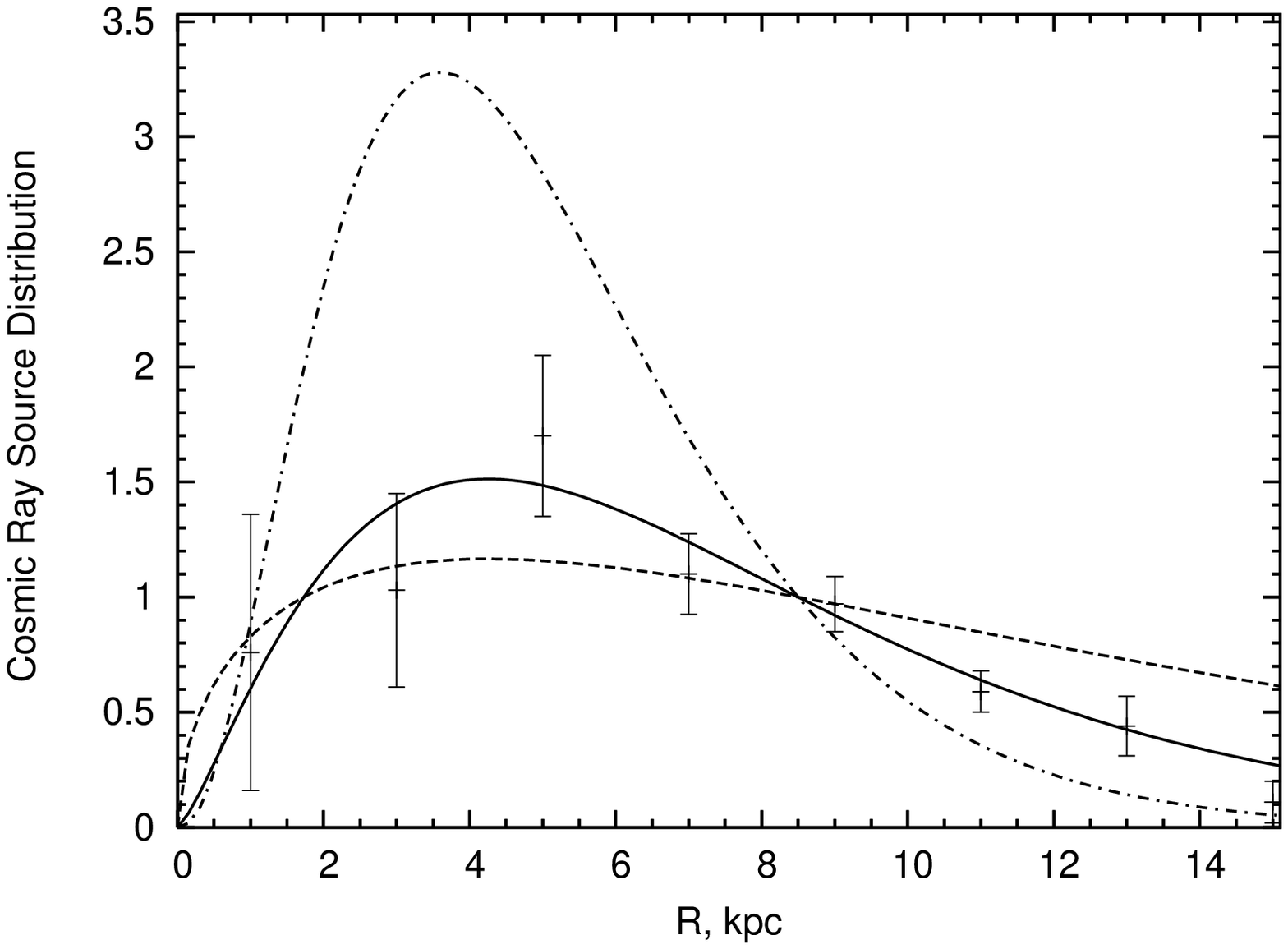}}
\caption{\label{s-d}
Cosmic ray source distribution along the Galactocentric radius R. 
The source distribution is in arbitrary units.
Dash-dotted line: based on pulsars\cite{pulsar};
dashed line: used by Strong et al.\cite{opt}
and in our conventional model\cite{0611783};
solid line: adopted in our new model. 
Vertical bars are SNR data points\cite{case-bha}.
These distributions are normalized at R=8.5\,kpc.
}
\end{figure}

A major uncertainty in calculating the diffuse Galactic $\gamma$-ray
emission is the distribution of molecular hydrogen, since the derivation of
H$_2$ density from the CO data is problematic \cite{sm04}.
The CO-to-H$_{2}$ conversion factor $X_{\text{CO}}$ from COBE/DIRBE
studies given by Sodroski et al. \cite{sodroski}
is about $2-5$ times greater than the value given by Boselli et al.
\cite{boselli} in different Galactocentric radius. The later is based on the
measurement of Galactic metallicity gradient and the
inverse dependence of $X_{\text{CO}}$ on metallicity \cite{sm04}.
The value of $X_{\text{CO}}$ is then normalized to
the $\gamma$-ray data \cite{sm04}.
Strong et al. have derived the  $X_{\text{CO}}$ by fitting the EGRET diffuse
$\gamma$-ray data directly \cite{sm96}.
A constant $X_{\text{CO}} = (1.9\pm 0.2) \times 10^{20}$ cm$^{-2}/$(K km s$^{-1})$ for
$E_\gamma = 0.1 - 10$ GeV was given \cite{sm96}. 
However, observations of particular
local clouds yield lower values $X_{\text{CO}} = 0.9-1.65
\times 10^{20}$ cm$^{-2}/$(K km s$^{-1})$ \cite{boselli}.
Since the fit by Strong et al. to the EGRET data 
in \cite{sm96} assumes 
only the background contributions, we expect that it gives a larger $X_{\text{CO}}$
than is the case with the new DMA component.
We find that a smaller $X_{\text{CO}}\sim 1.0 \times 10^{20}$ molecules
cm$^{-2}/$(K km s$^{-1})$ can give a much better description of the EGRET data
below 1 GeV. After taking DMA into account, the global fitting is greatly
improved. We assume for $X_{\text{CO}}$ a constant value independent of the
radius $R$. 
As shown in Ref.~\cite{sm04}, the simple form of  $X_{\text{CO}}$ is compensated 
by an appropriate form of the CR sources.
We have taken the radial distribution of CR sources in the form
of $(\frac{R}{R_{\odot}})^{\alpha}\exp(-\frac{\beta(R-R_{\odot})}{R_{\odot}})$,  
with $\alpha=0.5$, $\beta=1.0$, $R_{\odot}=8.5$ kpc,  
and limiting the sources within $R_{max}=15$ kpc,
which are adjusted to best describe the diffuse $\gamma$-ray spectrum.
The source distribution is shown in Fig.~\ref{s-d} (dashed line).

\begin{figure}
\includegraphics[scale=0.35]{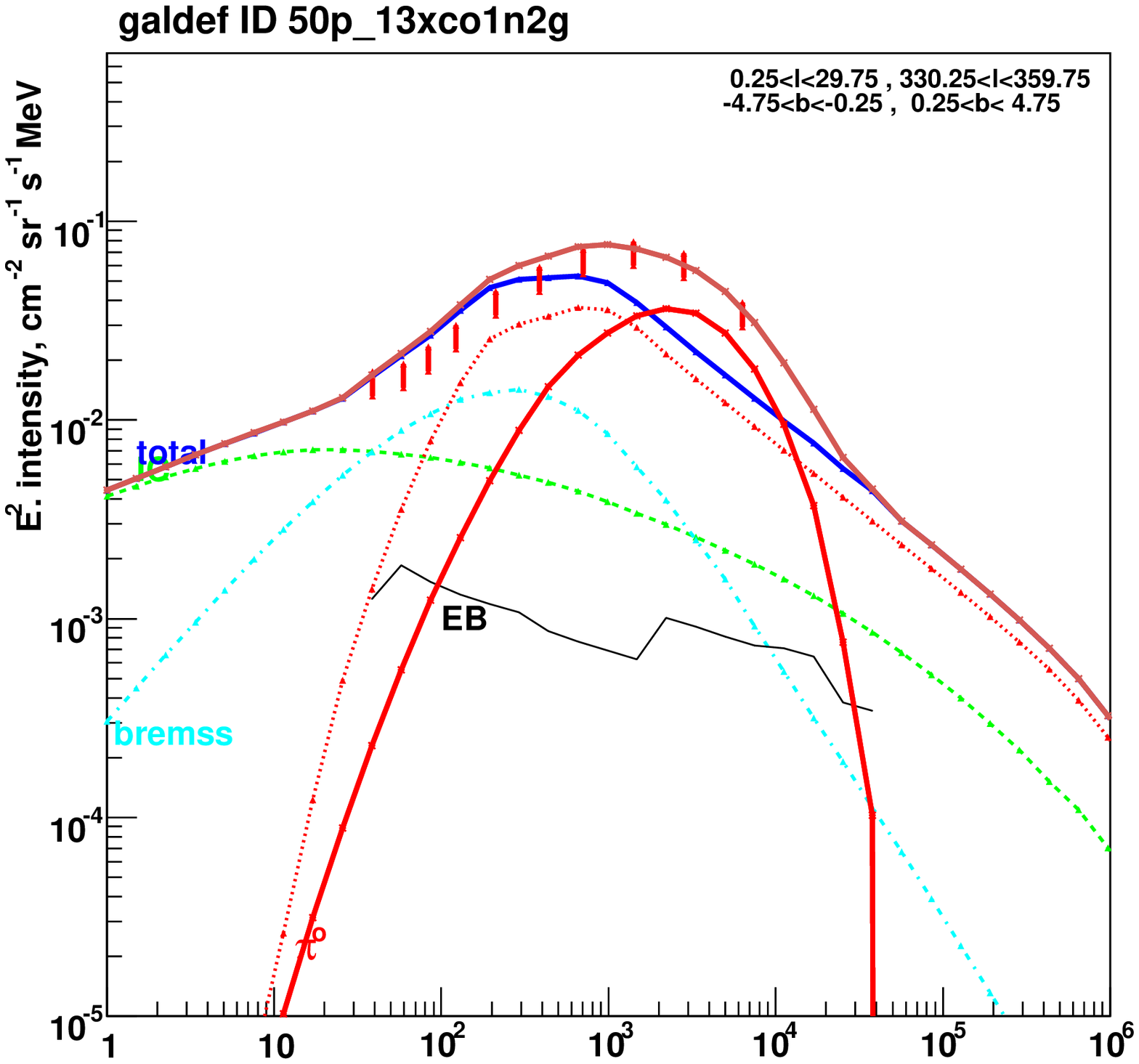}
\includegraphics[scale=0.35]{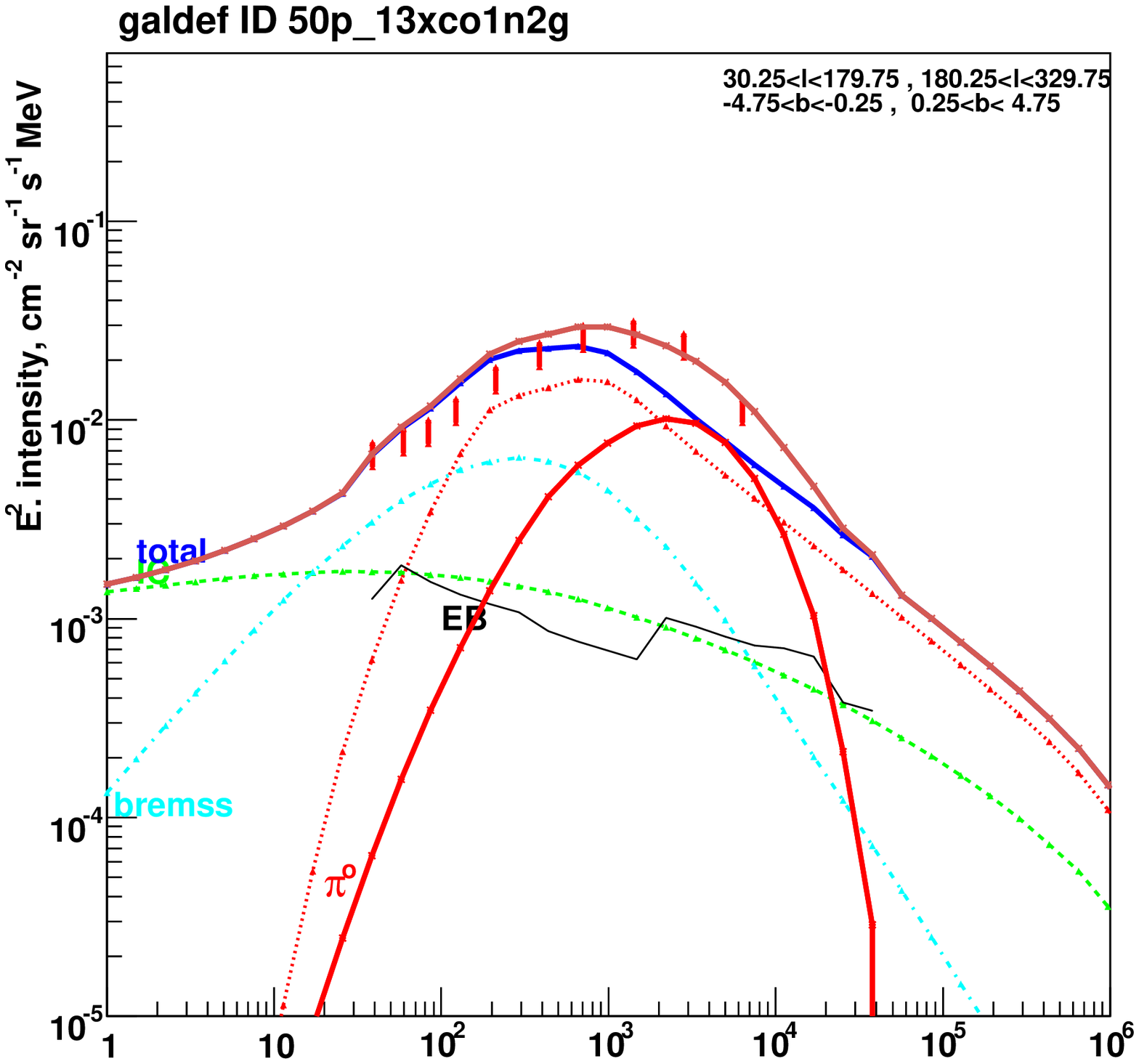}
\includegraphics[scale=0.35]{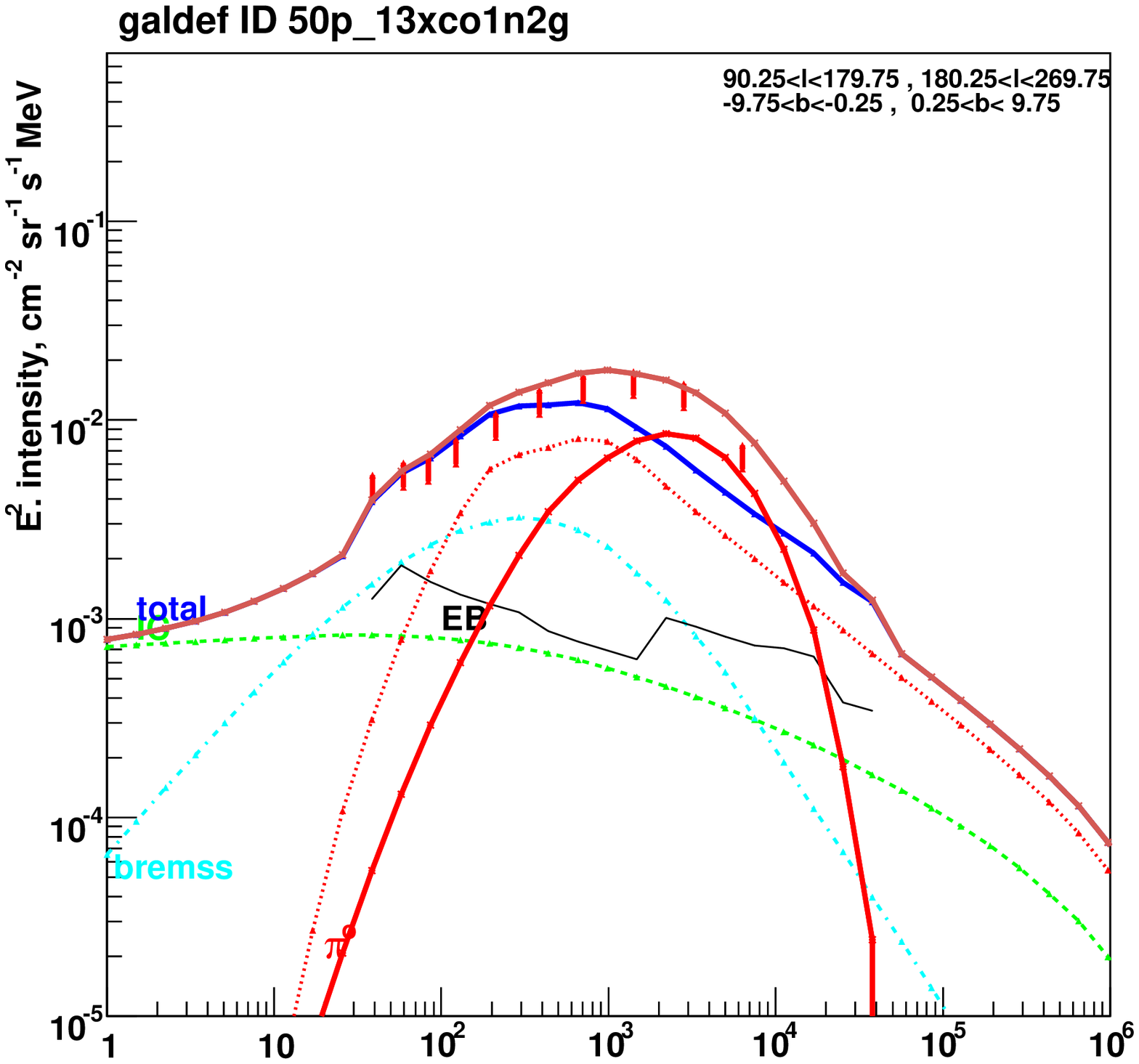}
\includegraphics[scale=0.35]{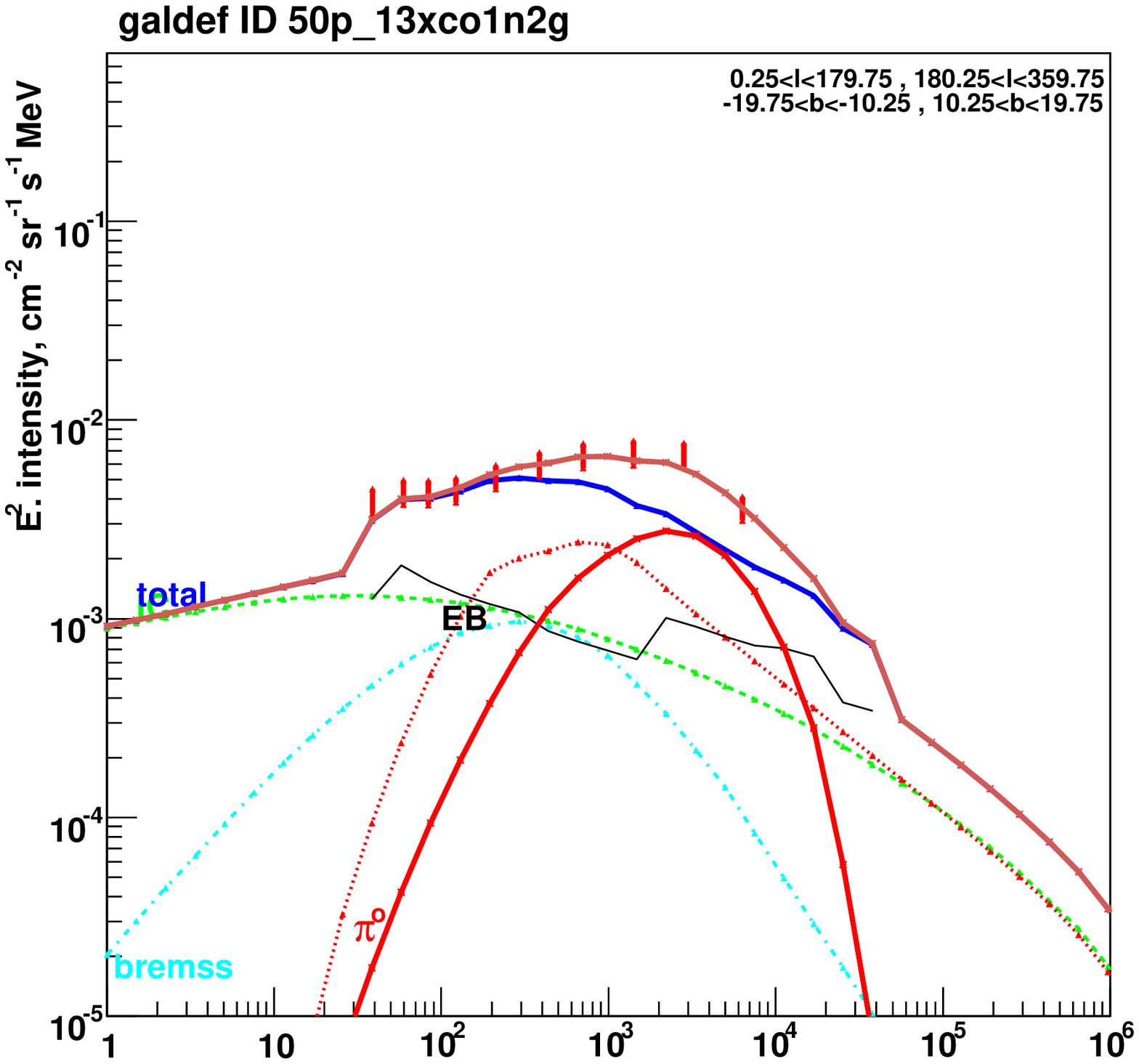}
\includegraphics[scale=0.35]{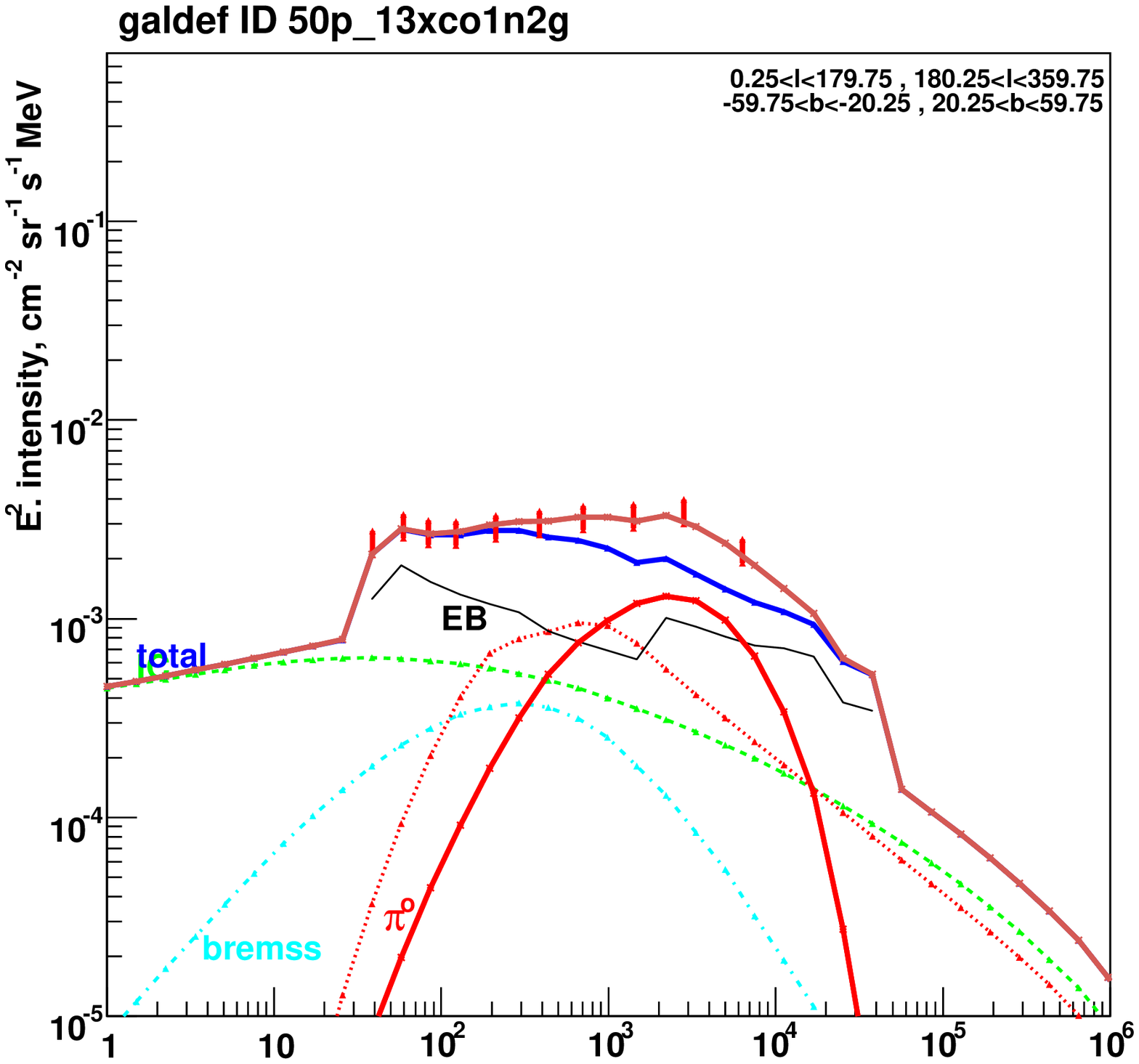}
\includegraphics[scale=0.35]{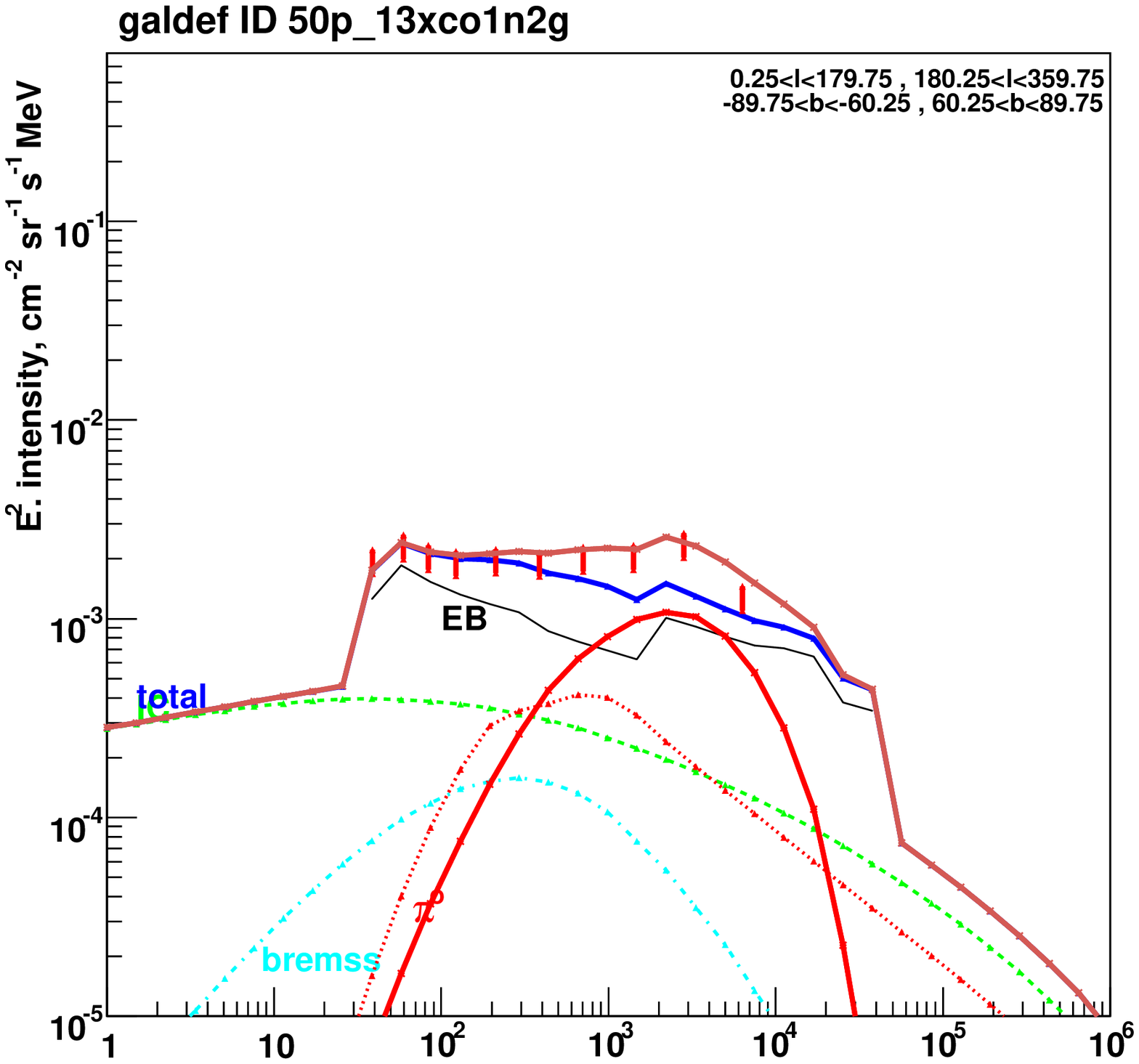}
\caption{\label{result}
Spectra of diffuse $\gamma$-rays for different sky regions
(top row, regions A, B, middle C, D, bottom E, F).
The model components are $\pi^0$ decay, inverse Compton,
bremsstrahlung, EGRB and DMA (dark red curve).
}
\end{figure}

Following Strong et al. \cite{opt}, we divide the whole sky into
six regions: region A corresponds to the ``inner
radian'', region B is the Galactic plane excluding the inner radian,
region C is the ``outer Galaxy'', regions D and E cover higher
latitudes at all longitudes, and region F describes the ``Galactic poles''.
The calculated diffuse $\gamma$ spectra in the six different sky regions
are given in Fig.~\ref{result}. 
The diffuse $\gamma$ ray background includes contributions from
$\pi^0$ decays produced by nuclei collisions, IC scattering off 
the interstellar radiation field (ISRF), bremsstrahlung by electrons,
and the isotropic extragalactic $\gamma$ ray background (EGRB).
For the regions A, B, C and D, i.e. the Galactic plane and the 
intermediate latitude regions, the $\pi^0$ decay contribution
is dominant. At high latitude regions E and F the EGRB becomes
more important. To account for the GeV excess the peak of the DMA $\gamma$ ray 
spectrum has to have similar magnitude as the background.
After adding the diffuse $\gamma$ ray emission from DMA to the 
Galactic background, 
we obtain a perfect agreement with EGRET measurements for all
sky directions.

The result in Fig.~\ref{result} is really a success, considering that
we simply add the two kinds of diffuse $\gamma$ ray contributions together,
without introducing any arbitrary normalization for the background $\gamma$ rays
or ``boost factors'' for the DMA contribution.
It should be noted that including the enhancement by subhalos
dose not exclude the ring-like structures proposed by de Boer 
\cite{boer-fit,boer-gas}.
That is natural since taking the subhalos into account
only enhances the signals coming from the smooth component,
but does not mimic the ring-like structure, which can fit
the EGRET data at different directions \cite{boer-fit}.
 
\subsection{Flux of $\bar{p}$ }

\begin{figure}
\includegraphics[scale=0.5]{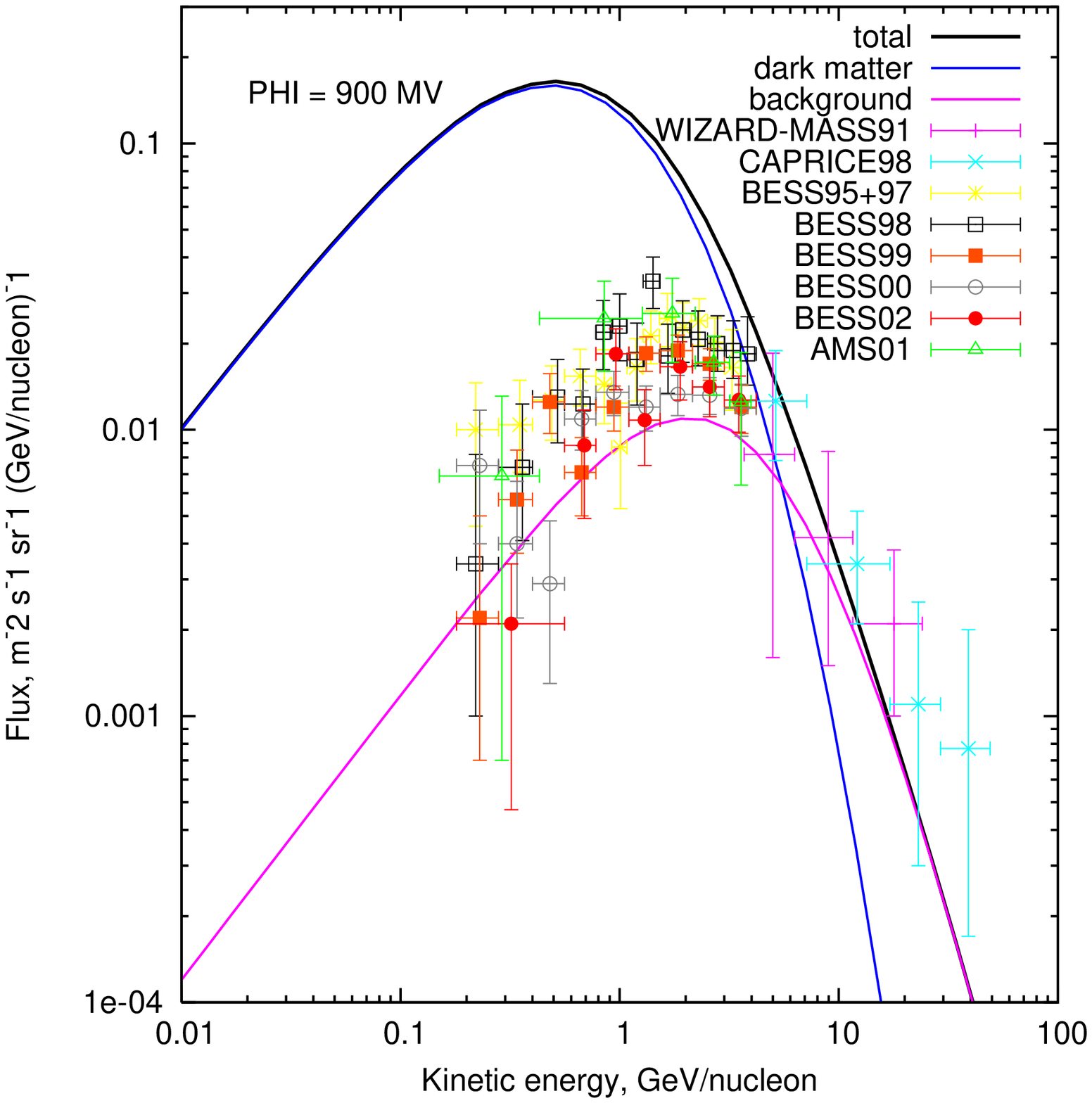}
\caption{\label{pbar}
Solar modulated flux of antiprotons in the conventional model.
The upper/black curve represents the total $\bar{p}$ flux.
For the two lower curves, the one dominating at energies below several GeV
is the DMA contribution and the other dominating at energies above 
several GeV stands for the contribution from CR interactions.
$\bar{p}$ data are from WIZARD-MASS91\cite{wiz}, CAPRICE98\cite{cap98},
BESS95+97\cite{ori00}, BESS98\cite{bs98}, BESS99\cite{bs99-00},
BESS00\cite{bs99-00}, BESS02\cite{bs02}, AMS01\cite{ams01}.
}
\end{figure}

We have seen in Fig.~\ref{ratio} that the
branching ratios to $\gamma$ rays and to $\bar{p}$ are closely related.
Therefore the $\bar{p}$ flux is a sensitive test of the DMA scenario 
to solve the ``GeV excess'' problem. 
In \cite{bergs}, Bergstr\"om et al. have claimed that the de Boer model
has been ruled out due to the overproduction of $\bar{p}$ flux. In their work 
they adopted a simple propagation model similar to that adopted in 
DarkSUSY \cite{darksusy}. We now check the $\bar{p}$ flux in our 
propagation model.

We first calculate the source term of $\bar{p}$ produced by 
neutralino annihilation adopting the same SUSY model used for
the diffuse $\gamma$ ray data,
\begin{equation}
\label{source}
\Phi_{\bar p}(r, E) = \frac{\langle \sigma v \rangle \phi(E)}{2m_\chi^2}
\langle \rho(r)^2 \rangle\, ,
\end{equation}
where $\phi(E)$ is the differential $\bar{p}$ flux at energy $E$ due
to a single annihilation, and
\begin{equation}
\langle \rho(r)^2 \rangle =\rho_{\text smooth}^2 + 
\langle \rho_{\text sub}^2 \rangle \, ,
\end{equation}
are contributions from the smooth DM component and subhalos.  
The contribution from subhalos is given by 
\begin{equation}
\langle \rho(r)_{\text sub}^2 \rangle =
\int_{m_{min}}^{m_{max}} n(m,r)\left( \int \rho^2 dV \right) {\text{d}}m \, ,
\end{equation}
where $n(m,r)$ is the number density of subhalos with mass $m$ at radius $r$,
and $ \int \rho^2 dV$ is the astrophysical factor for a single subhalo
with mass $m$. All subhalo parameters are taken identical to those used to 
account for the diffuse $\gamma$ rays.

We then calculate, in the same propagation model, the ${\bar p}$ spectrum at 
Earth by incorporating the source term of Eq.~(\ref{source}) in GALPROP.
Fig.~\ref{pbar} shows the background, the signal and the 
total ${\bar p}$ fluxes. 
The background antiproton flux s lower than the 
data; this has been discussed in \cite{apj565,opt}. This is another
hint at the necessity of an exotic signals besides the ordinary CR secondaries.
The ${\bar p}$ flux in our model is about one order of magnitude 
greater than the measured data at energies lower than $1$ GeV. 
However, we notice that our prediction is a few times lower than
that by Bergstr\"om et al. \cite{bergs}, when they adopted the median
set of propagation parameters, which are similar to the propagation
parameters in our conventional model.
The difference may be due to two reasons:
first we adopt a different propagation model; second|and maybe more important|,
we do not adopt a universal ``boost factor'' for the diffuse $\gamma$ rays
from DMA as de Boer and Bergstr\"om et al. did. The enhancement by
subhalos tends to boost the $\gamma$ ray and $\bar{p}$ fluxes at large
radii, as shown in Fig.~\ref{density}. Therefore subhalos boost $\gamma$ rays
more than  $\bar{p}$ since only these $\bar{p}$ produced within the
Galactic diffusion region contribute to the $\bar{p}$ flux on Earth.

However, these effects are not enough to give a  $\bar{p}$ flux 
consistent with the data. This is related to the presence of two DM rings 
near the solar system, that enhance the antiproton flux greatly
as they are strong $\bar{p}$ sources. We have to resort to a new propagation
model to suppress the $\bar{p}$ further and give a consistent description
of all data.

\section{A new propagation model}

Inheriting the advantages of our conventional model in the last section,
a new propagation model is intended to be built to account for the Galactic diffuse  
$\gamma$ rays, and at the same time the $\bar{p}$ flux. 
Degeneracies exist between the propagation
parameters, mainly between the diffusion coefficient $D(\rho)$ and the height
of the diffusion region $z_h$. Different sets of parameters can all explain
the CR data but lead to very different signals from DMA \cite{maurin}. If we adopt smaller 
$z_h$ we can adjust $D(\rho)$ at the same time to give similar prediction of
the CR data. However adopting smaller $z_h$ leads to smaller $\bar{p}$ flux from DMA since only $\bar{p}$ 
sources within the diffusion
region contributes to the flux on the Earth. The degeneracy leads 
to about one order
of magnitude uncertainties of the $\bar{p}$ flux when adopting different 
sets of propagation parameters in Bergstr\"om et al. \cite{bergs}.
Therefore it is straightforward to consider a smaller height of
the diffusion region to suppress the DMA signal. 

The diffusion halo height is determined by fitting the CR data.
It is found that the average gas density CRs crossed during their 
travel to the Earth is about
0.2 atoms/cm$^3$, which is significantly lower than the average
gas density in disk of about 1 atom/cm$^3$. This is explained by the 
possibility that CRs are confined in a larger diffusion region than the gaseous disk,
spending a longer time outside the gaseous disk. The height of the 
halo represents the volume of this diffusion region.
In \cite{asr01}, Strong and Moskalenko derived  $z_h=3-7$ kpc
for the four types of radioactive nuclei isotopes of $^{54}$Mn/Mn,
$^{36}$Cl/Cl, $^{26}$Al/$^{27}$Al and $^{10}$Be/$^{9}$Be, while the
data of $^{10}$Be/$^{9}$Be favored a smaller halo of $1.5-6$ kpc. 
Maurin et al. found that several
settings of propagation parameters with halo height $z_h$ ranging from 1 kpc
to 15 kpc could well fit the observational B/C ratio data
\cite{maurin01,maurin02}.  
Certainly the diffusion halo height should be at least as large as
the scale height of the ionized component of the interstellar gas,
$\sim700$\,pc \cite{Gaisser}.  
The halo height is also constrained by the diffuse $\gamma$ ray emission
at the intermediate latitude, since too small halo height will lead to 
too low $\gamma$ ray flux at at these latitudes.

Since the gas is not smoothly distributed in the
disk, CRs may travel in low density regions in the disk and may not
diffuse to so large a region as usually considered.
As shown in \cite{chandran}, the gas density
and the strength of magnetic fields are higher than average within 
molecular clouds: the molecular clouds can act as magnetic mirrors to reflect 
and confine the charged CR particles. In this scenario CRs may travel in a low 
gas density region rather than in a region with the average density. 
Nevertheless, the effect of molecular cloud reflection
is hard to quantify.
We assume the gas density that CRs crossed is lowered 
compared with the
average gas density in the disk by a constant factor, which should be
of the order of $1$ since the conventional diffusion model has been
very successful in describing the CR transportation.
We find that a reduction factor $\sim$1.5 and $z_h=1.5$ kpc can 
reproduce all the data very well, which will be shown below.
It should be noted that the reduction of the gas density crossed by CRs
is the key point of this propagation model. 

Taking this smaller halo height, we then adjust the diffusion 
coefficient and the Alfv\'en speed, so that we can reproduce the 
secondary to primary ratios, e.g. the B/C ratio.
In the present model, parameters in Eq.~(\ref{diff}) are taken as
$D_0 = 1.3\times 10^{28}$ cm$^2$ s$^{-1}$, $\rho_0=4$GV and $\delta=0.34$.
The Alfv\'en speed $v_A$ is taken as 19 km s$^{-1}$.
The injection spectra of nuclei and electrons are given in Table \ref{inj_para}.


The CR source distribution adopted in this model is shown in
Fig.~\ref{s-d}, together with 
the source distribution adopted in \cite{sm04}, which 
is the same as the pulsar distribution \cite{pulsar}, and that 
adopted in \cite{smr00,opt}, which is obtained by fitting the EGRET  
data with a constant X$_{\text{CO}}$ value.
The vertical bars are SNR data points from \cite{case-bha}.
The flat source distribution derived by fitting the EGRET diffuse 
$\gamma$ ray data is not in agreement with the distribution of 
SNRs \cite{sm04}. As for the pulsar distribution, it is so peaked that 
it would 
aggravate the problem of reproducing the relatively smooth EGRET
flux profile along the Galactic Plane \cite{Grasso}.
In addition, the pulsar distribution is probably
not sufficiently reliable to trace the SNRs accurately.
In fact, only a proportion of about one in $\sim 100$ known pulsars 
appears to be convincingly associated directly with SNRs \cite{psrsnr}.
It is very interesting that in our new model, the source distribution 
consistent with SNRs data (solid line in Fig.~\ref{s-d})
can give a better description of the diffuse $\gamma$ ray data with
a constant CO-to-H$_2$ conversion factor X$_{\text{CO}} \sim 1.0\times10^{20}$  
molecules cm$^{-2}/$(K km s$^{-1})$ than the other two distribution functions.  
The radial source distribution has the same form as the previous one 
with 
$\alpha=1.35$, \,$\beta=2.7$. 

\begin{figure}
\includegraphics[scale=0.45]{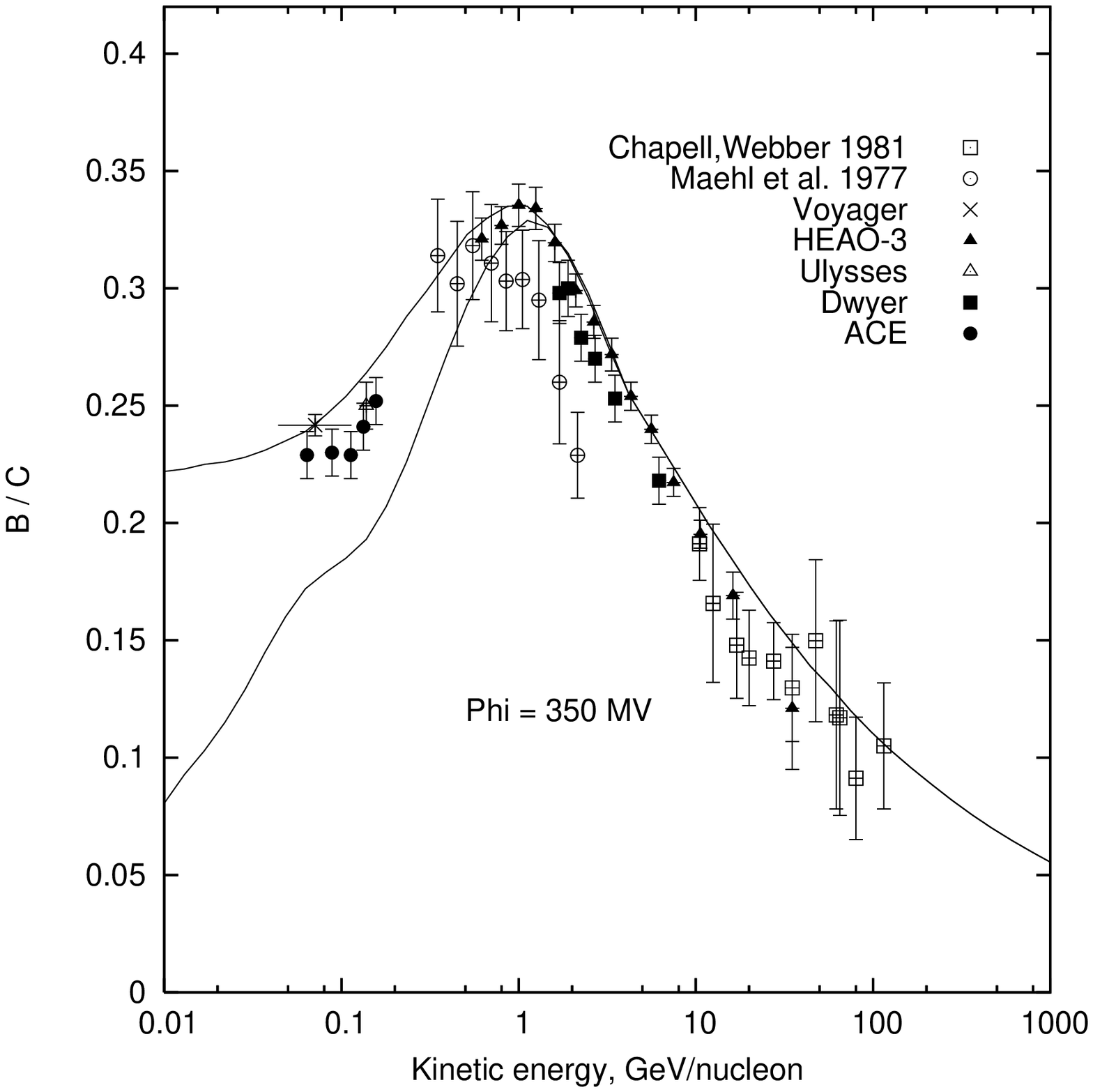}
\includegraphics[scale=0.45]{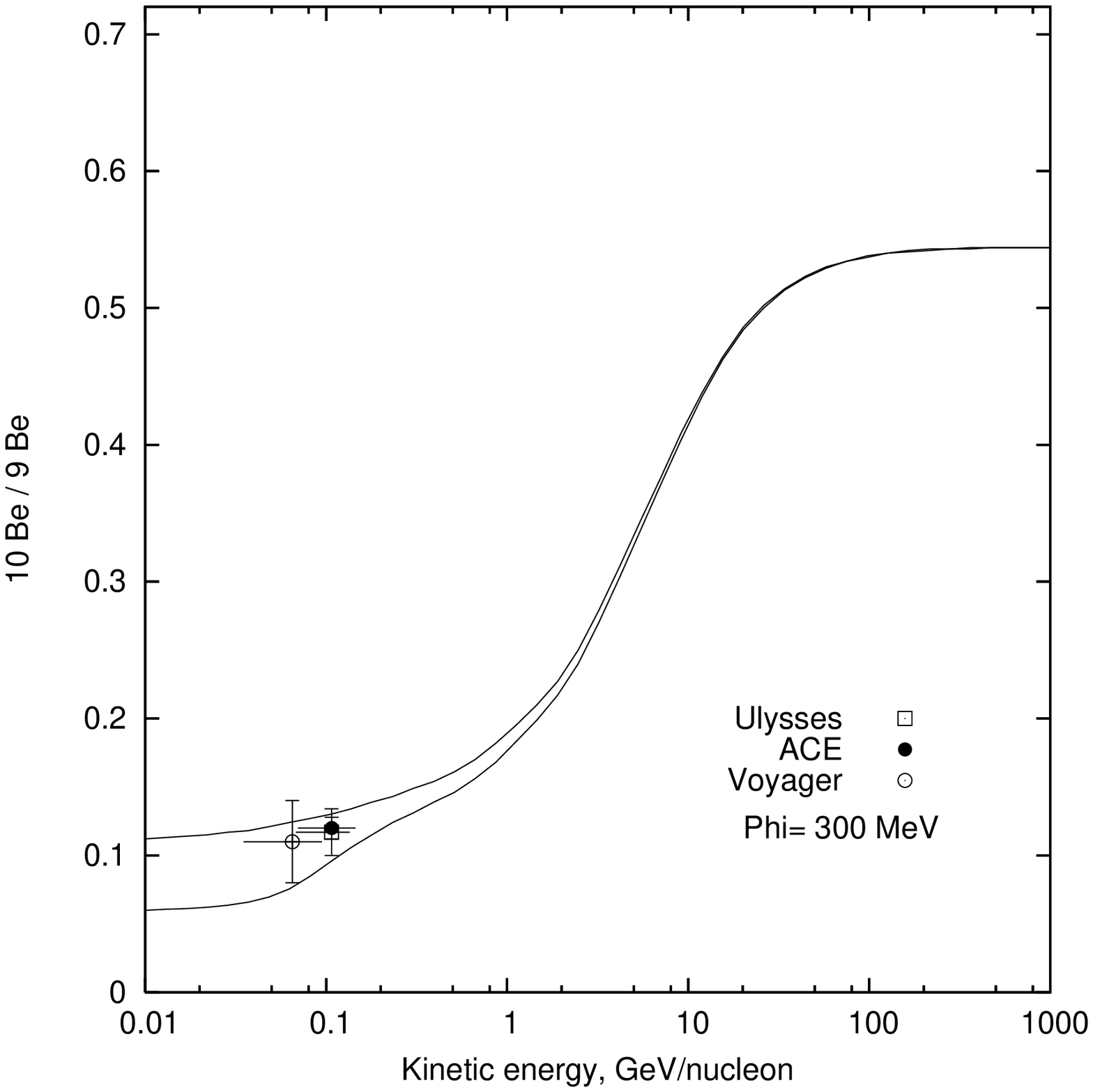}
\includegraphics[scale=0.45]{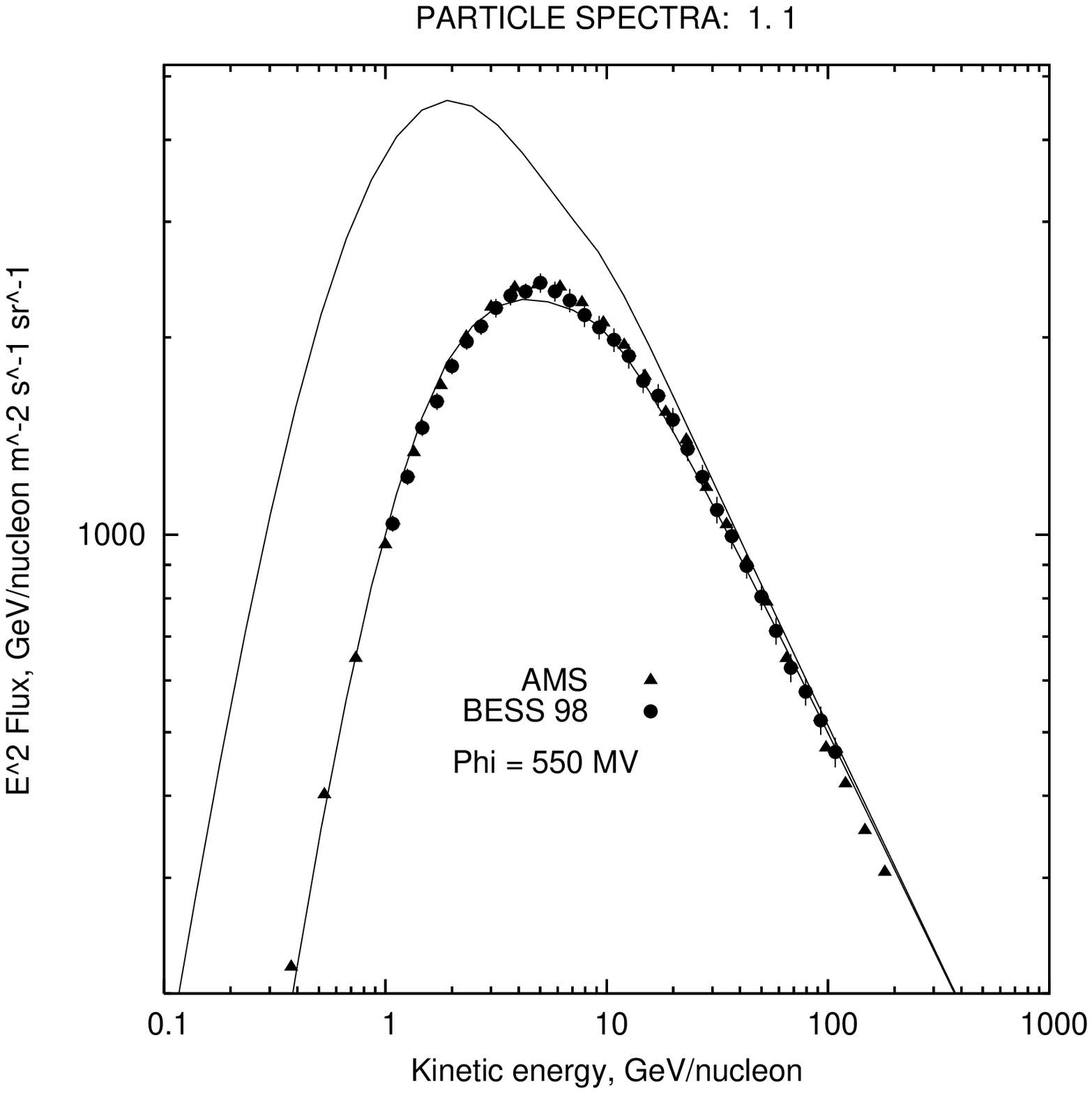}
\includegraphics[scale=0.45]{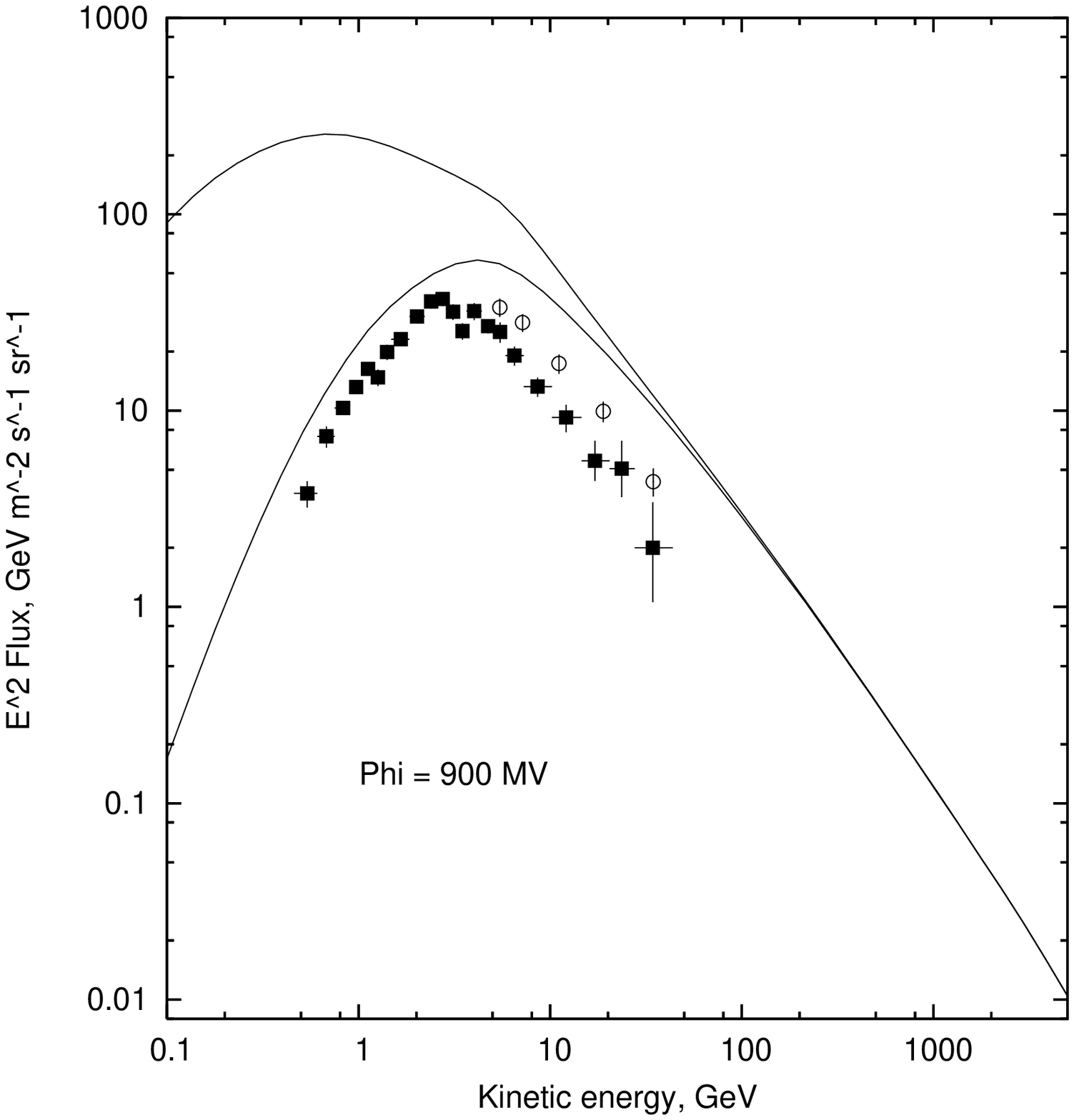}
\caption{\label{cr-sp}
The CR spectra in our new model.
Curves and data are the same as in Fig.~\ref{conv-crs}.
}
\end{figure}

In Fig.~\ref{cr-sp}, we show the results of B/C, $^{10}$Be/$^9$Be,
spectra of protons and electrons as predicted by this new propagation 
model. The stable secondary to primary ratio B/C and the unstable to stable
secondaries $^{10}$Be/$^9$Be show that our model reproduces the 
experimental date very well.
In this model the interstellar proton spectrum is taken to be the 
locally observed spectrum since the energy loss of proton is negligible.
The interstellar electron spectrum has an intensity 
normalization at high energies different from the local one, similar to 
the electron spectrum adopted by Strong et al. in \cite{opt} but with smaller 
fluctuations.
We will show below that this model can also reproduce the diffuse
$\gamma$ ray and $\bar{p}$ data well.

\subsection{Diffuse $\gamma$-rays and $\bar{p}$ flux}

\begin{figure}
\resizebox{14cm}{10cm}{\includegraphics{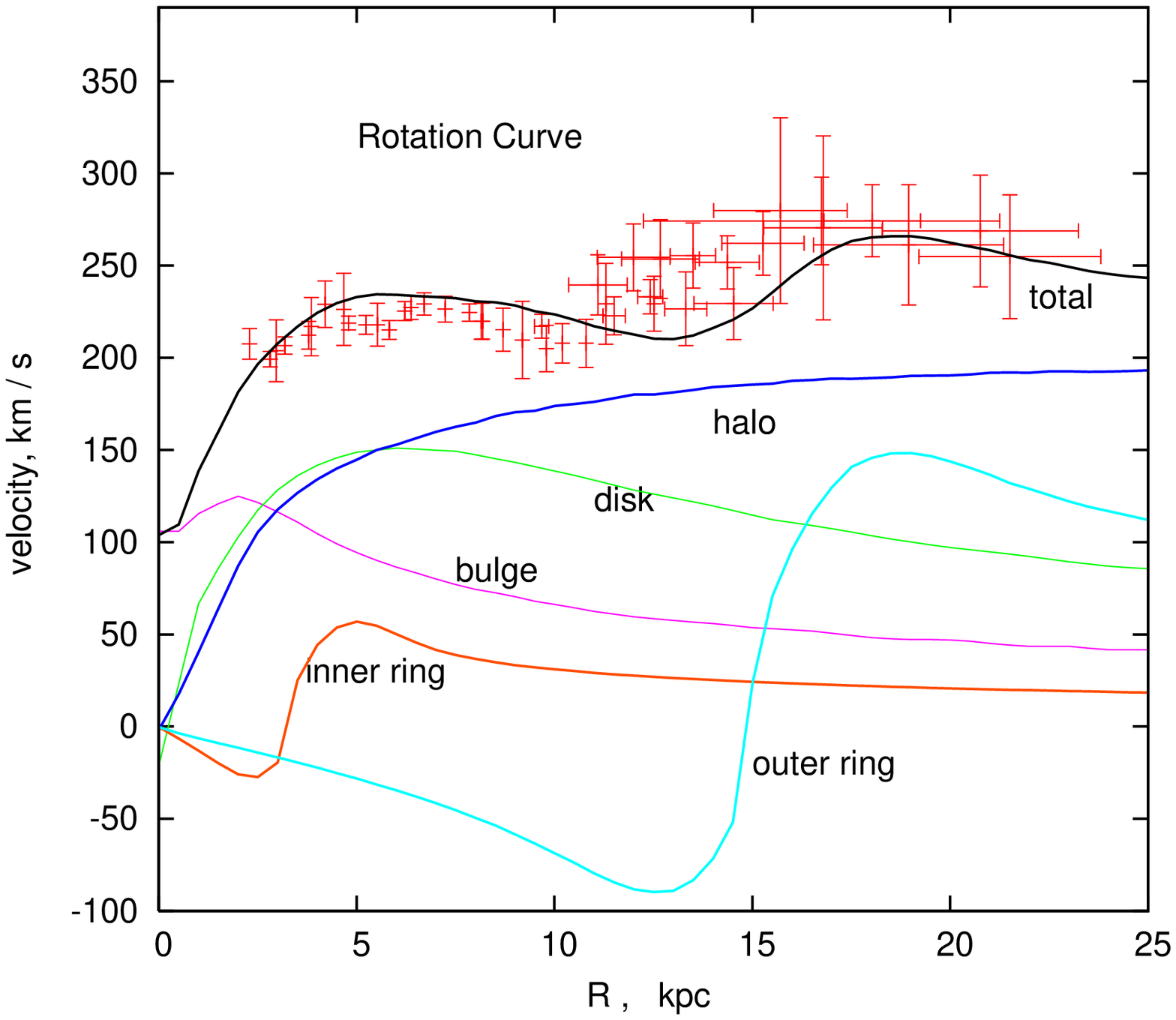}}
\caption{\label{r-c}
The rotation curve in our new propagation model.}
\end{figure}%

In order to give $\gamma$ ray and $\bar{p}$ fluxes consistent 
with experimental data, we also changed the DM ring parameters slightly:
the inner ring is now located at R =\,3.5 kpc and 
the outer ring is moved from R =\,14\,kpc to 16\,kpc.
On one hand, the position of rings are not crucial parameters
to fit the EGRET data: indeed, a slight change will not vary the prediction of
diffuse $\gamma$ rays much. On the other hand, the rings are mainly helpful
to explain the diffuse $\gamma$ rays at intermediate latitudes; there are not enough
to account for the GeV excess elsewhere, such as in regions
D and E (if only the smooth DM component is included). However, our 
consideration of subhalos helps to enhance $\gamma$ ray emissions at large
radii as shown in Fig.~\ref{density}. Therefore, we do not need so large
$\gamma$ ray emissions from the two rings 
to contribute to the intermediate latitudes as done in the de Boer model,
and we can move them slightly far away from the solar system.
It is interesting to note that the analysis of the HI gas flaring by 
Kalberla et al. also favored a DM ring located at a large radius of 
$\sim17.5$ kpc \cite{flaring}.

We have checked that these soft modulation of ring parameters
did not change the rotation curve significantly.
Fig.~\ref{r-c} shows the rotation curve for the present model.
The contributions from the DM smooth halo, the two DM rings, 
and the bulge and disk are included.
It can be seen that the rotation curve is consistent
with data.


\begin{figure}
\includegraphics[scale=0.35]{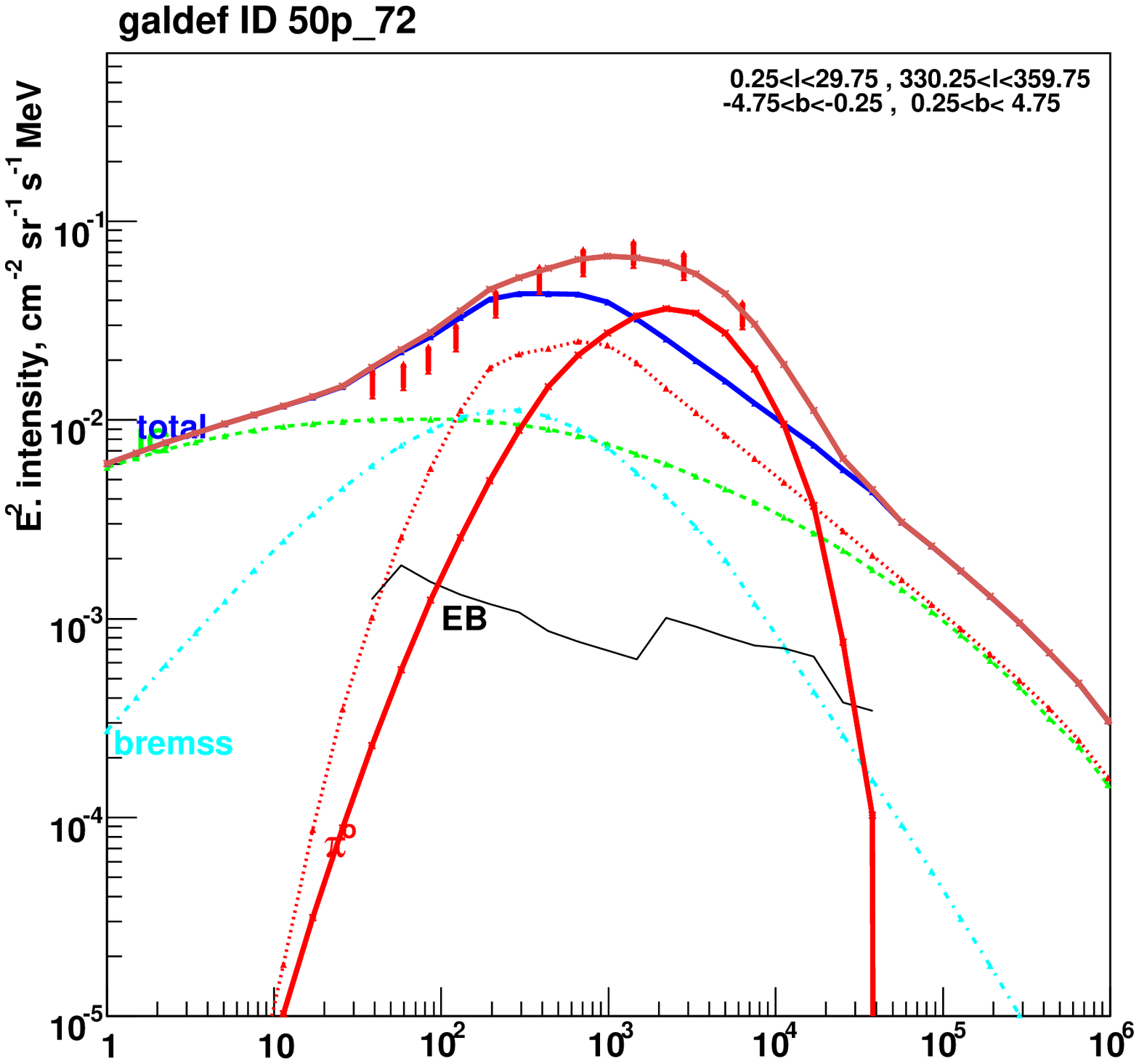}
\includegraphics[scale=0.35]{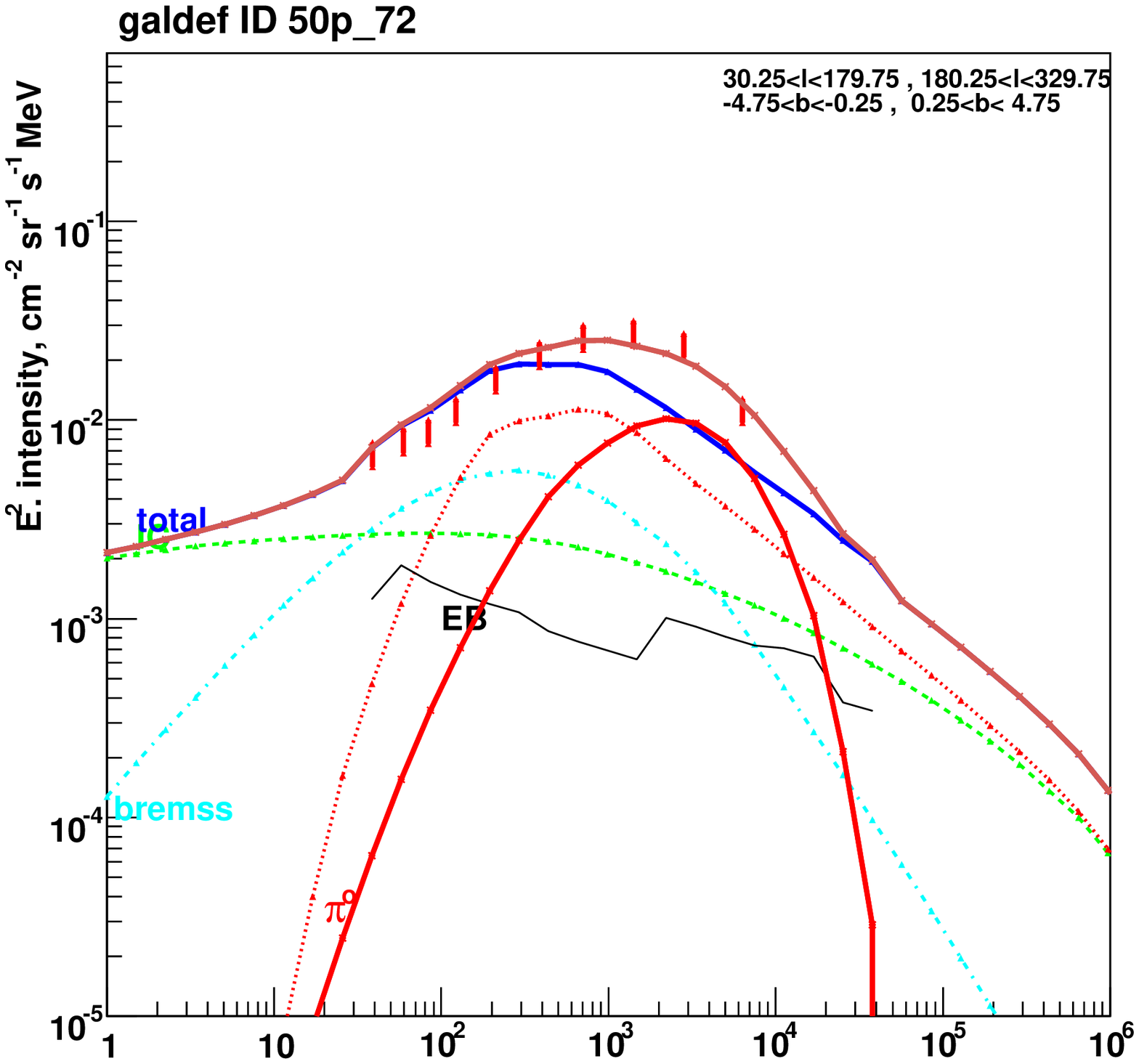}
\includegraphics[scale=0.35]{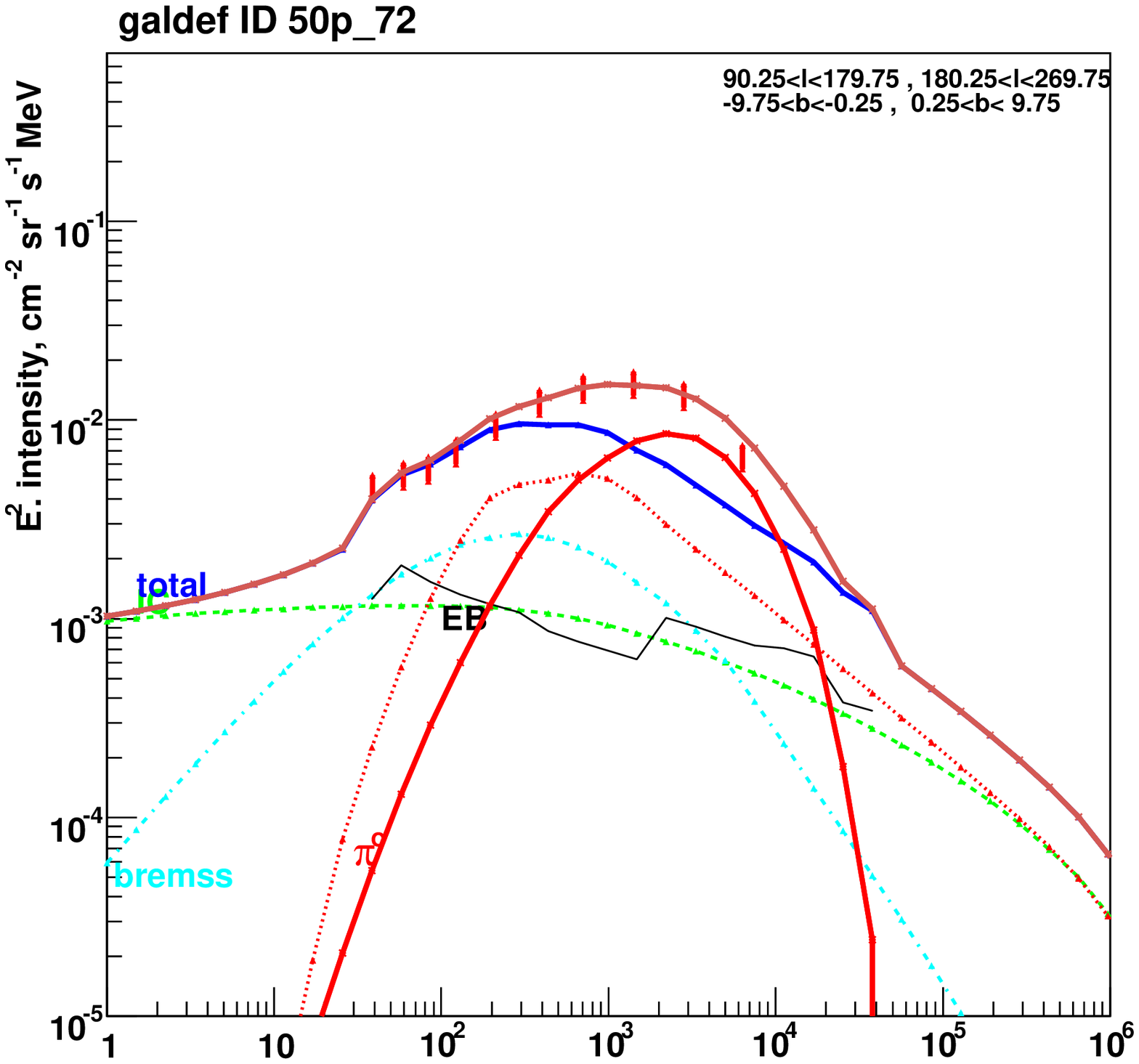}
\includegraphics[scale=0.35]{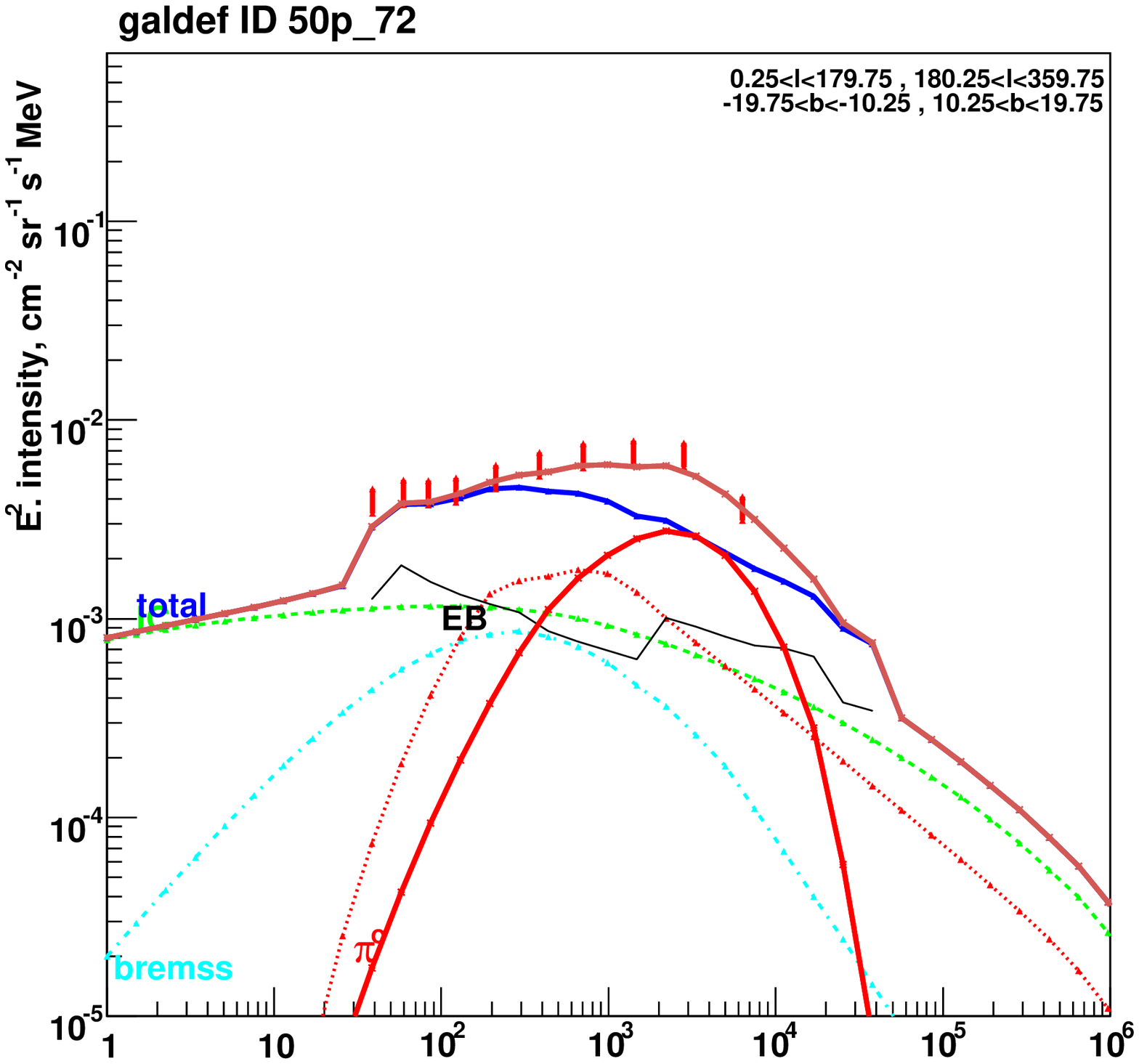}
\includegraphics[scale=0.35]{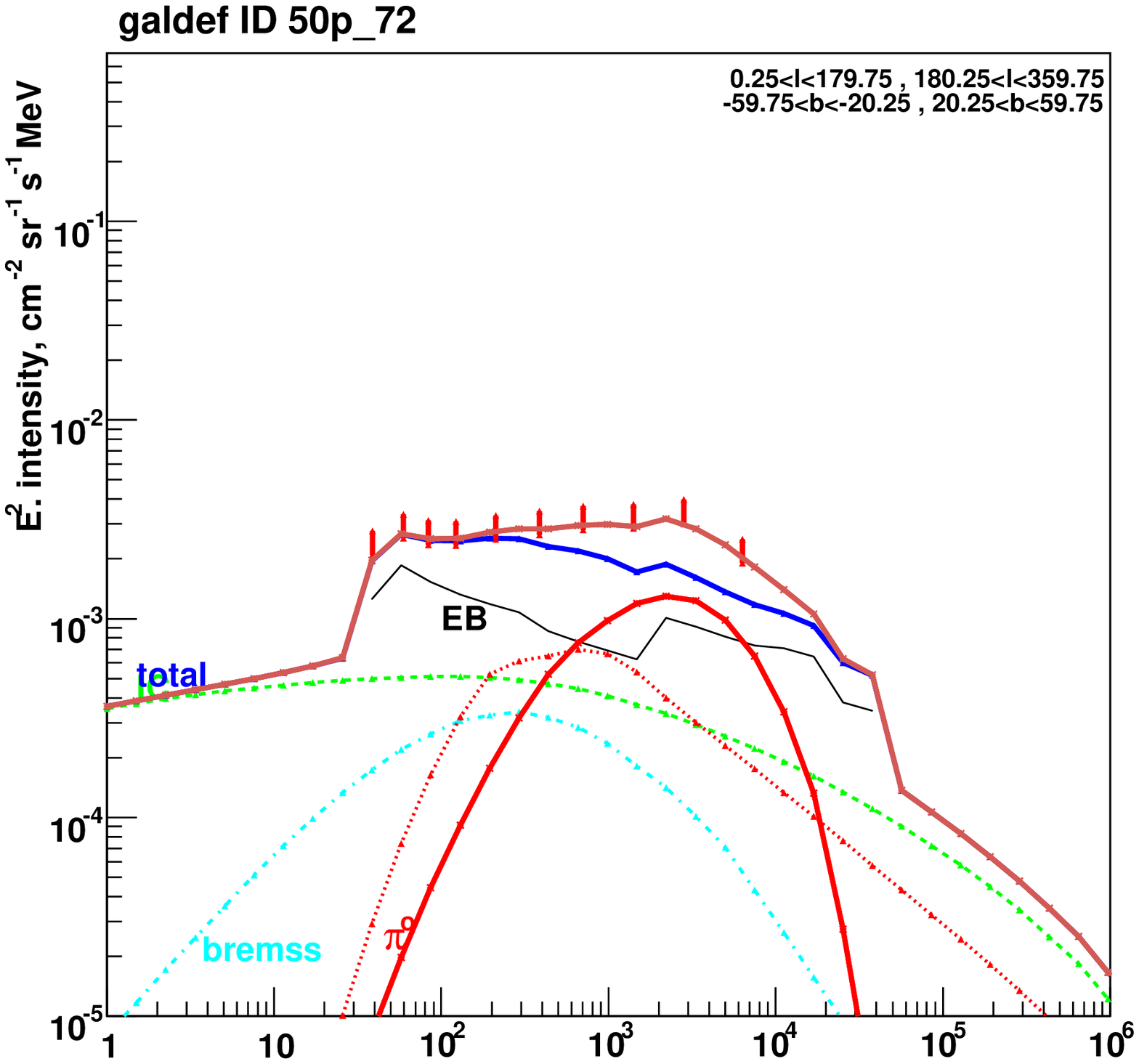}
\includegraphics[scale=0.35]{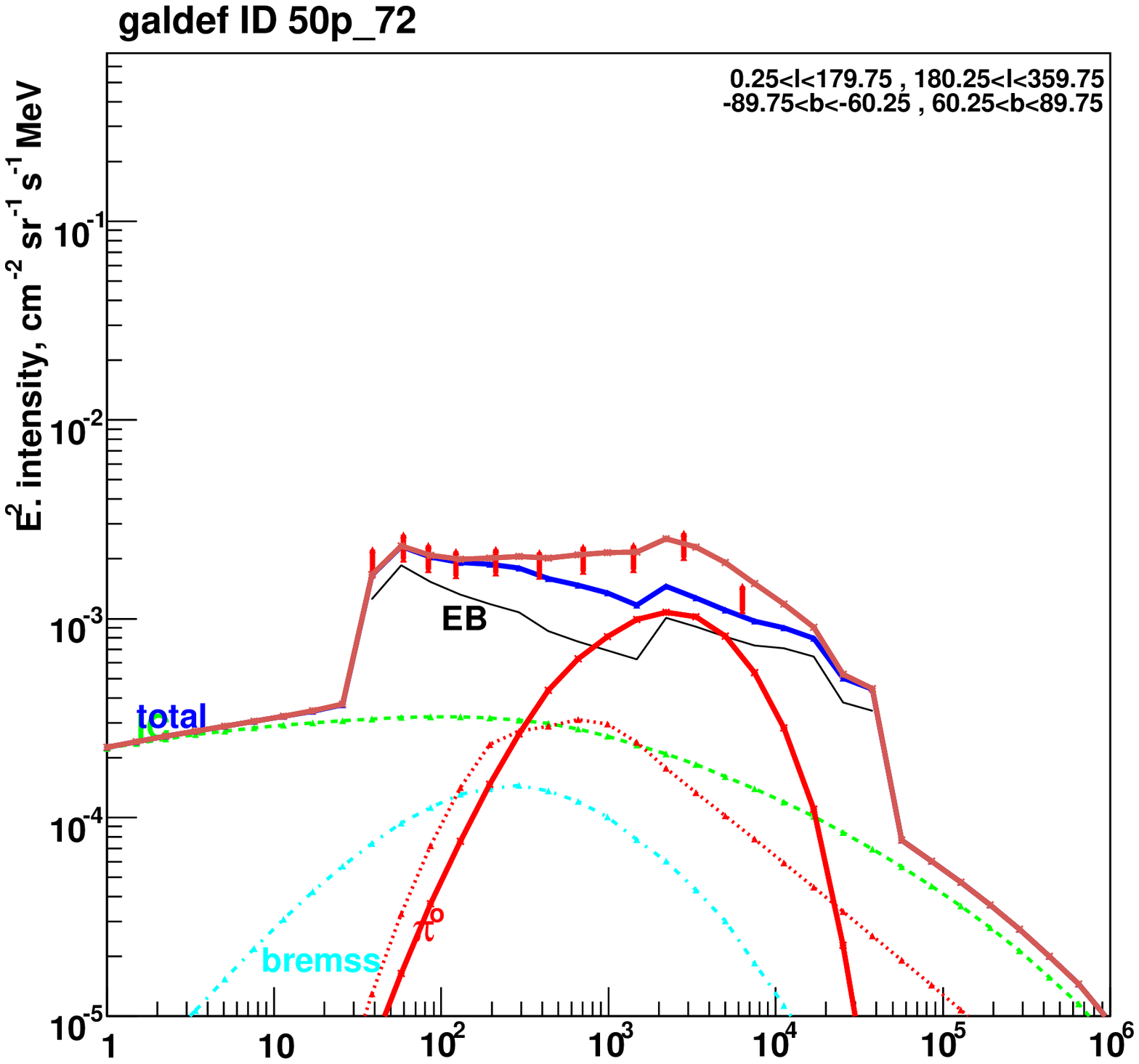}
\caption{\label{gamma}
Diffuse gamma ray spectra in the
six sky regions predicted by the new propagation model
(top row, regions A, B, middle C, D, bottom E, F).
The various contributions are $\pi^0$ decay, inverse Compton,
bremsstrahlung, EGRB and DMA (dark red curve).
}
\end{figure}

Adding the background diffuse $\gamma$ rays 
from CRs and those from DMA {\em directly},
we find the calculated diffuse $\gamma$ rays are well
consistent with observations.
The diffuse $\gamma$ ray spectra in the six sky regions are shown in 
Fig.~\ref{gamma}.
The prediction in the new propagation model is in good
agreement with EGRET data.

\begin{figure}
\includegraphics[scale=0.25]{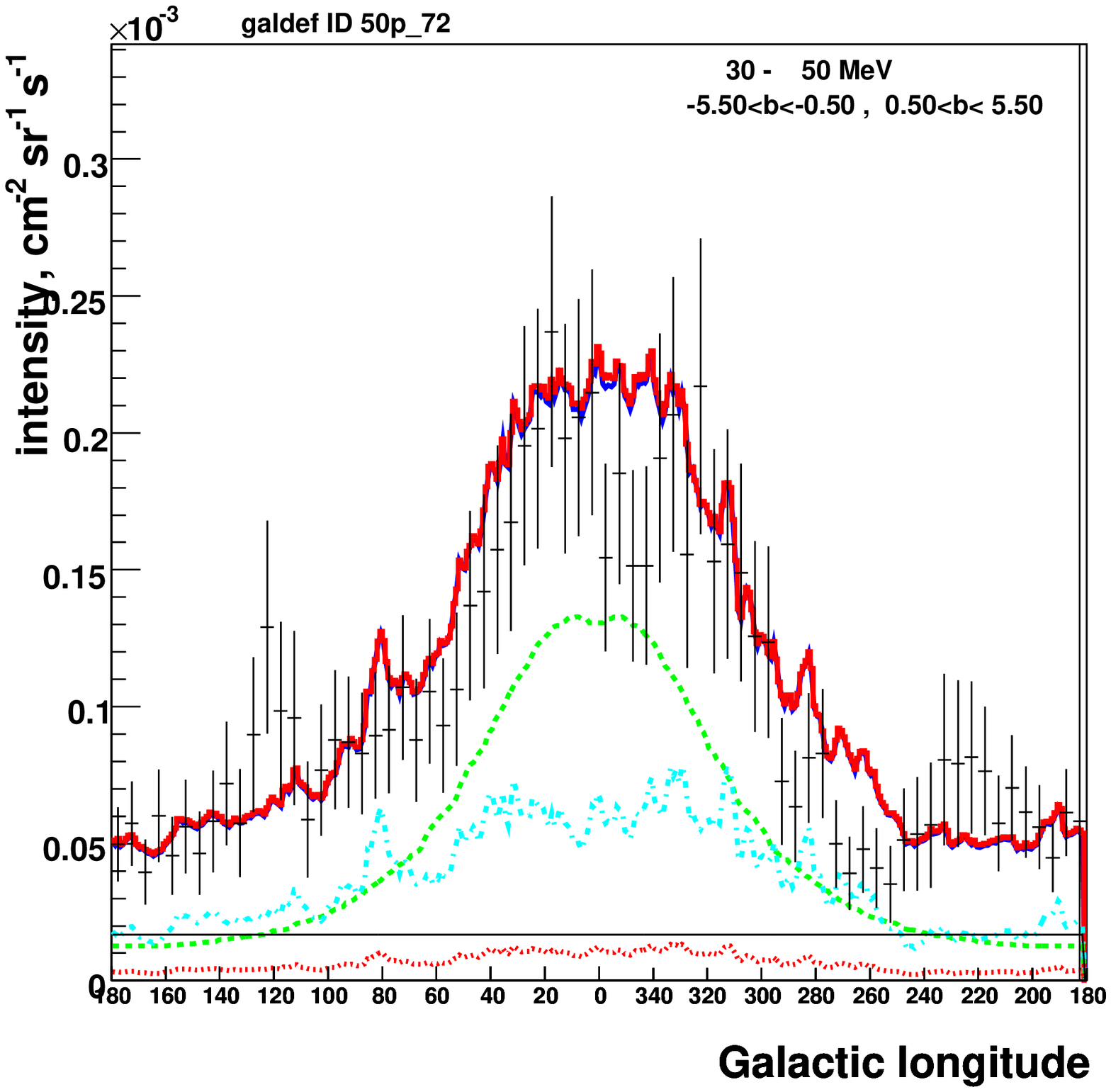}
\includegraphics[scale=0.25]{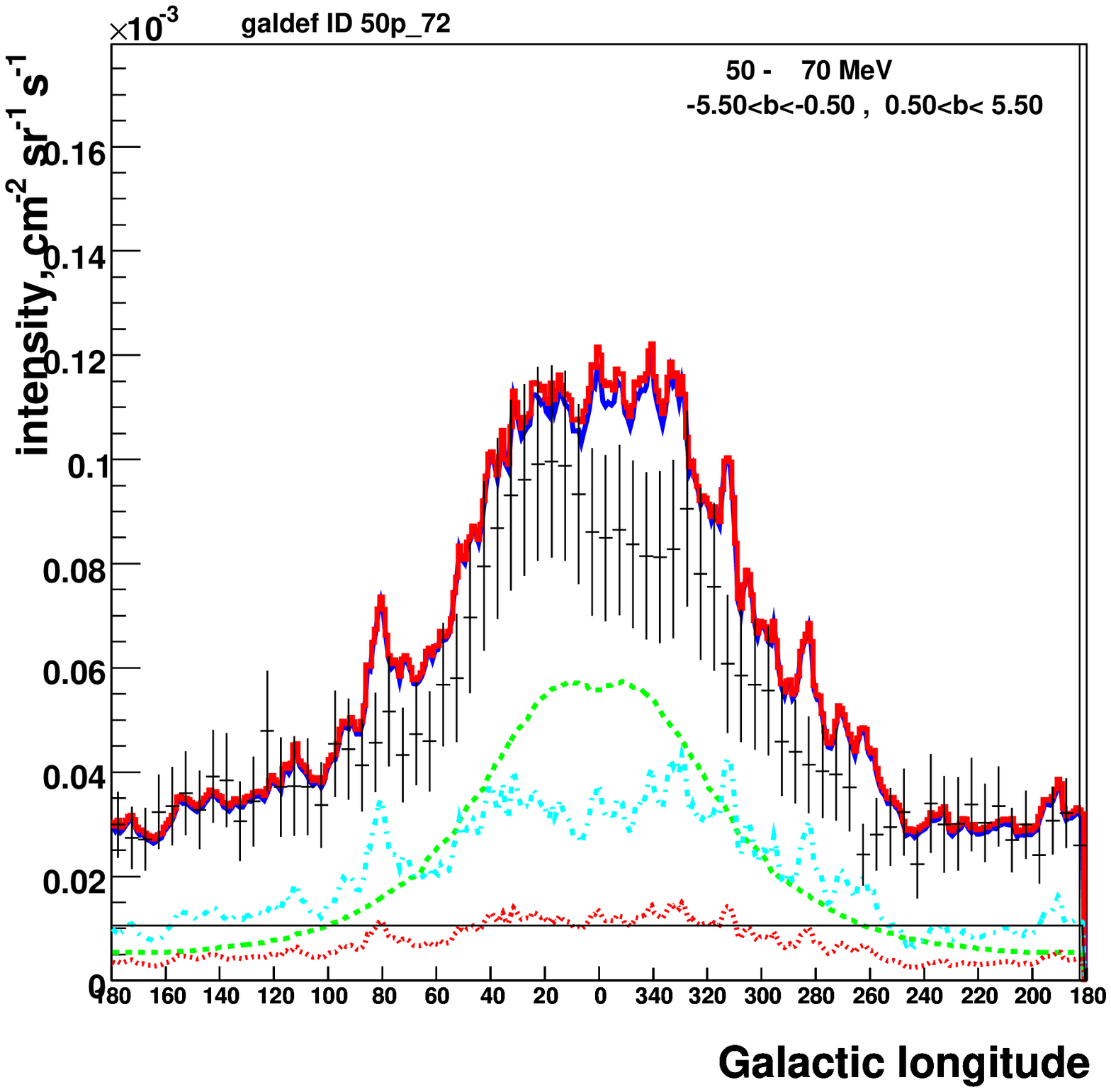}
\includegraphics[scale=0.25]{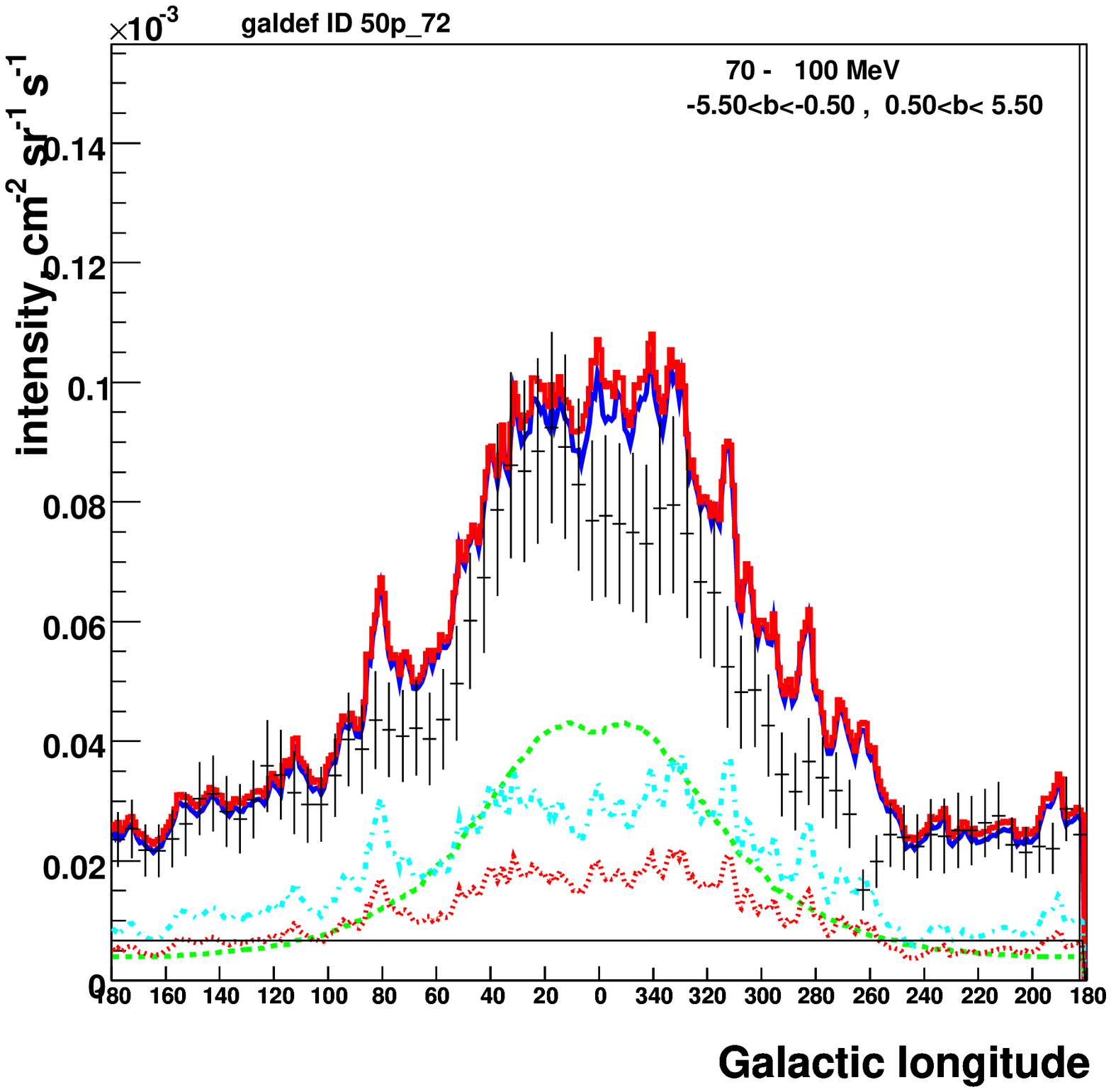}
\includegraphics[scale=0.25]{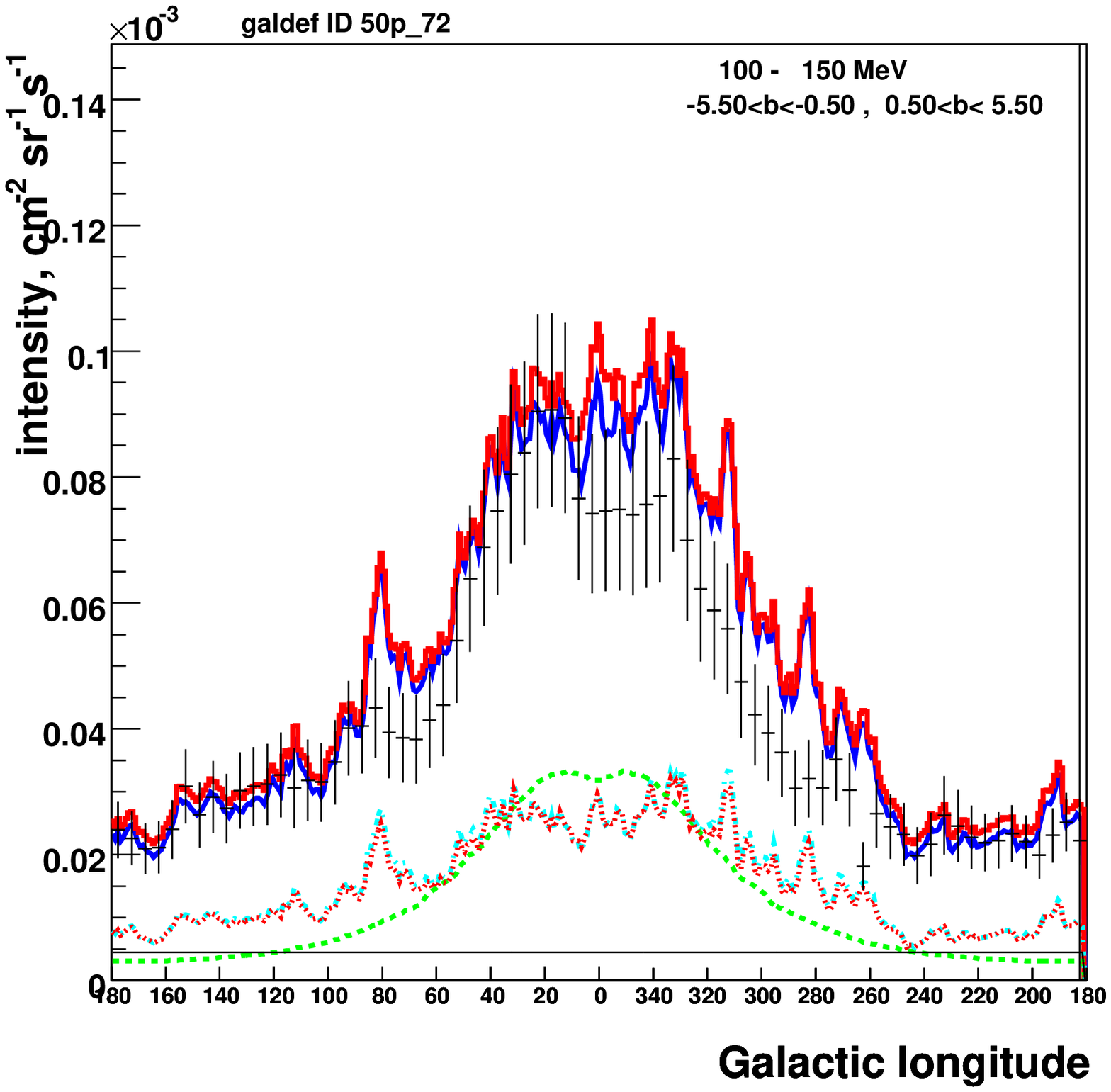}
\includegraphics[scale=0.25]{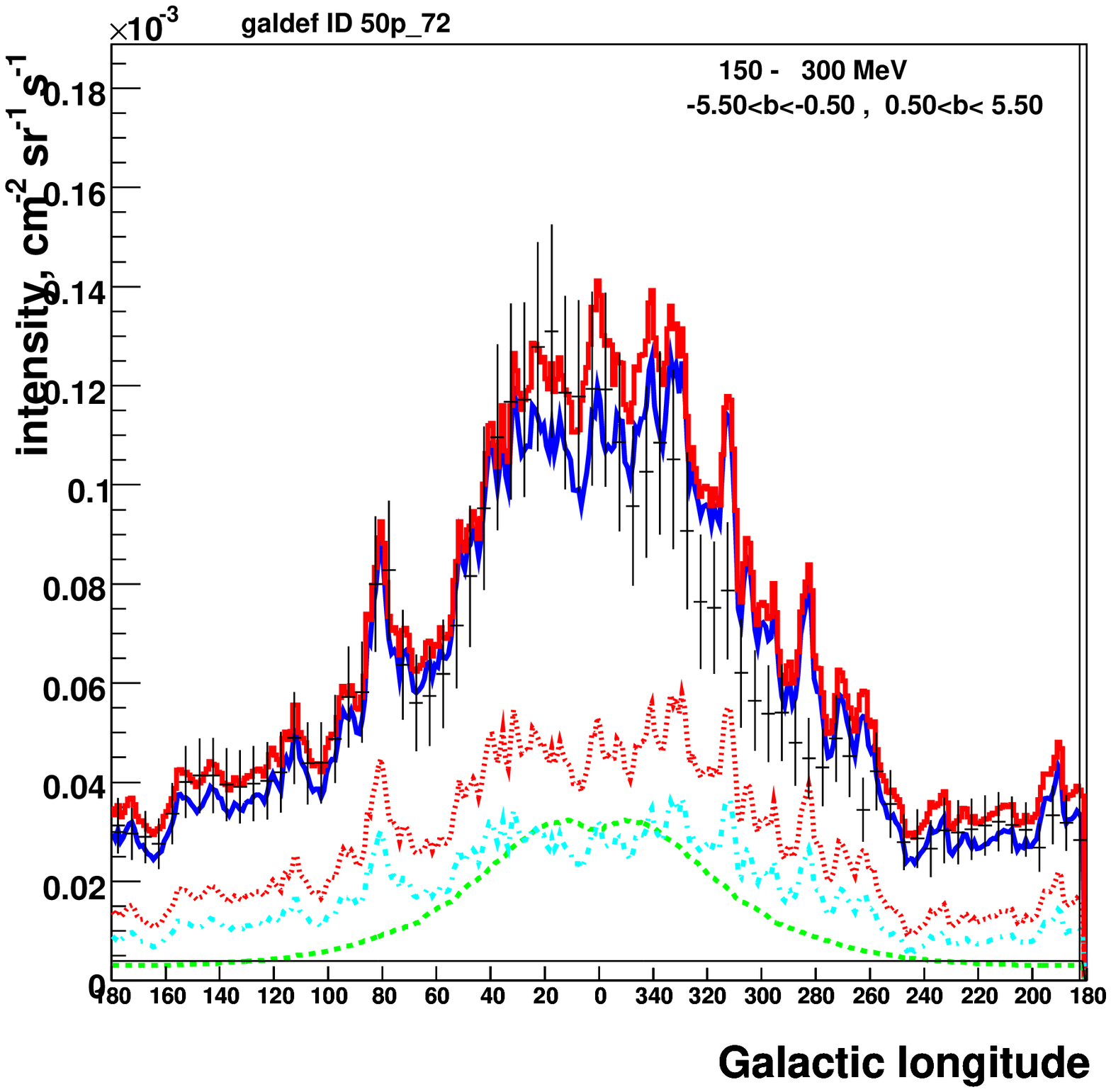}
\includegraphics[scale=0.25]{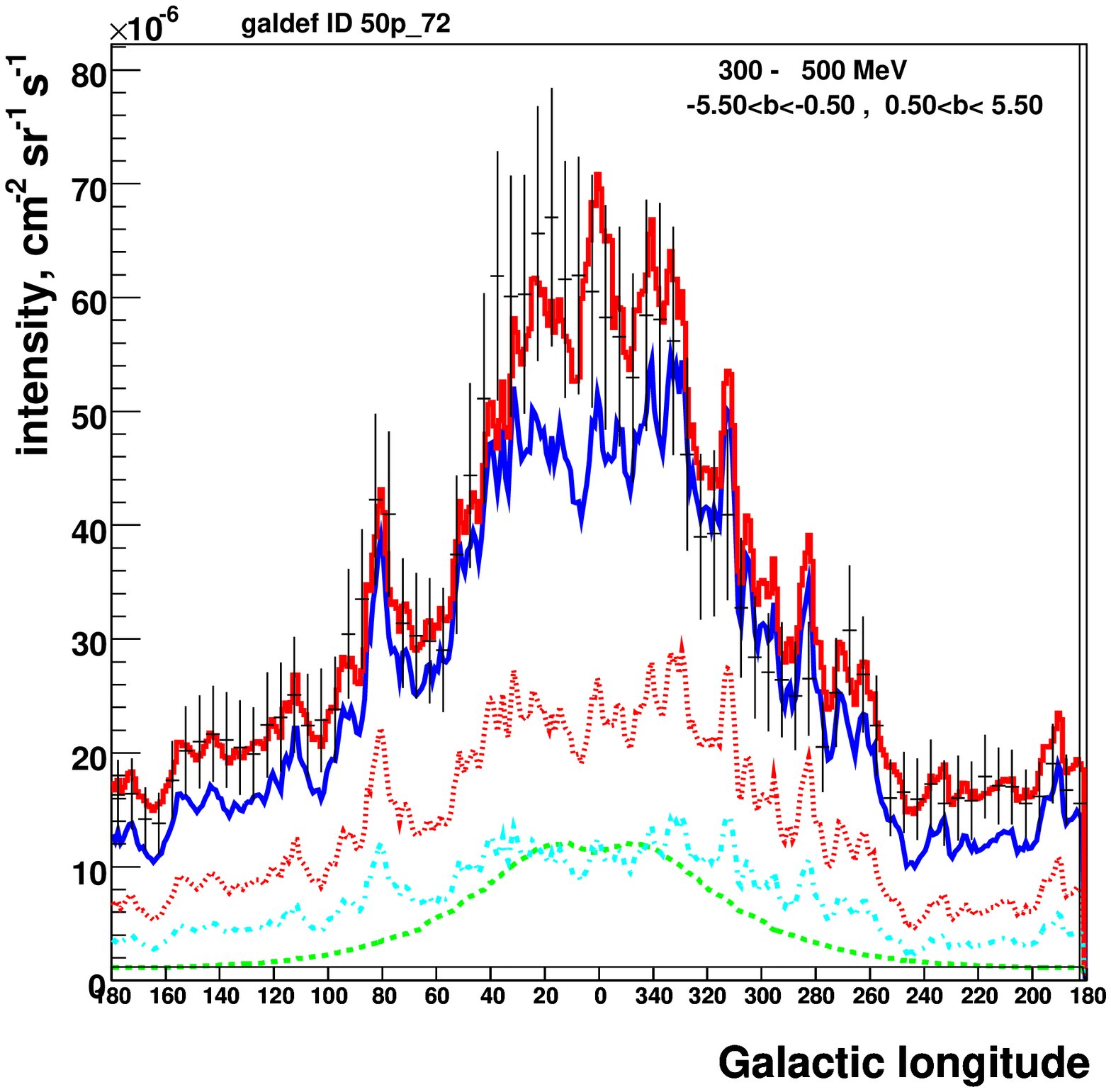}
\includegraphics[scale=0.25]{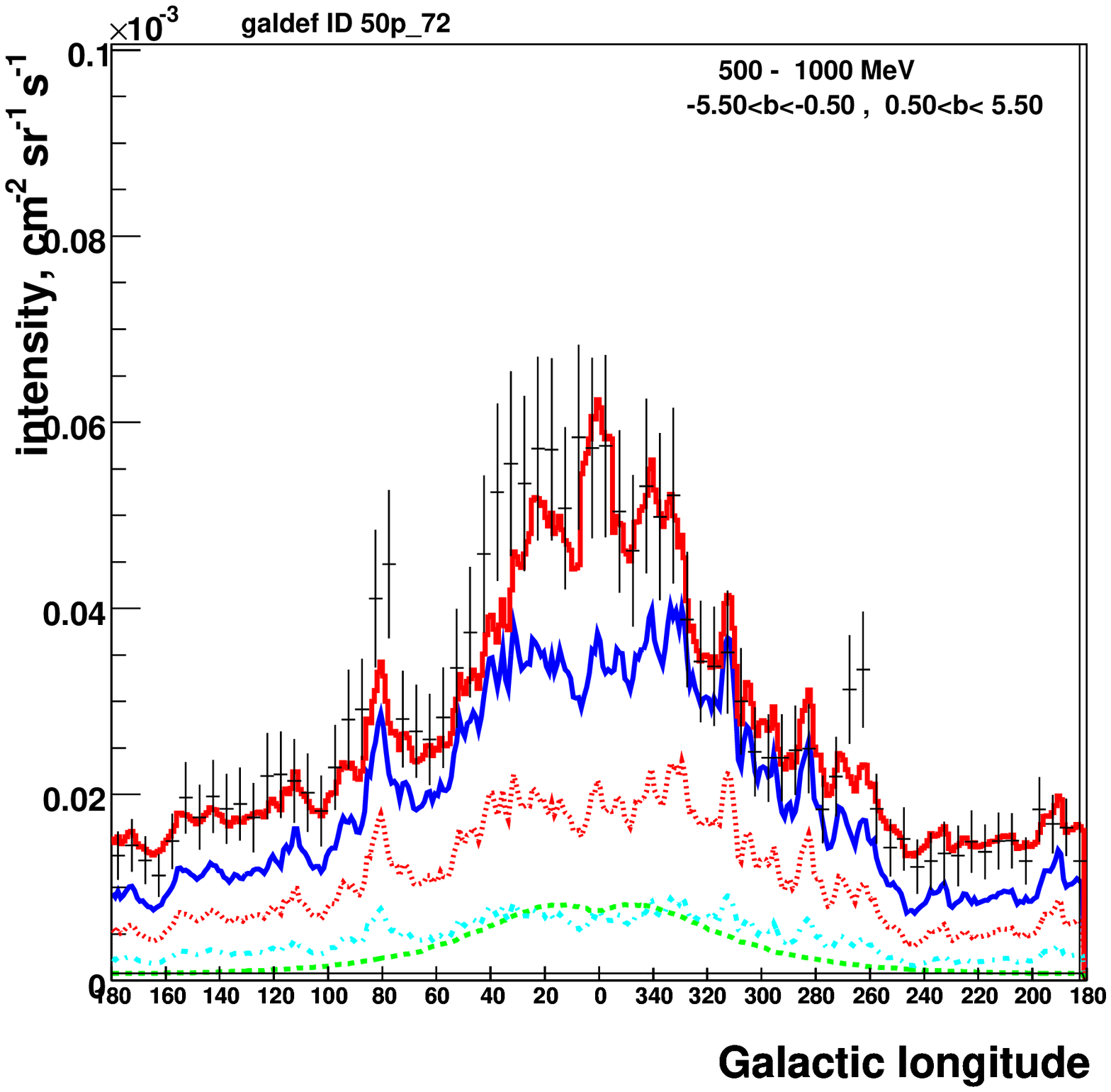}
\includegraphics[scale=0.25]{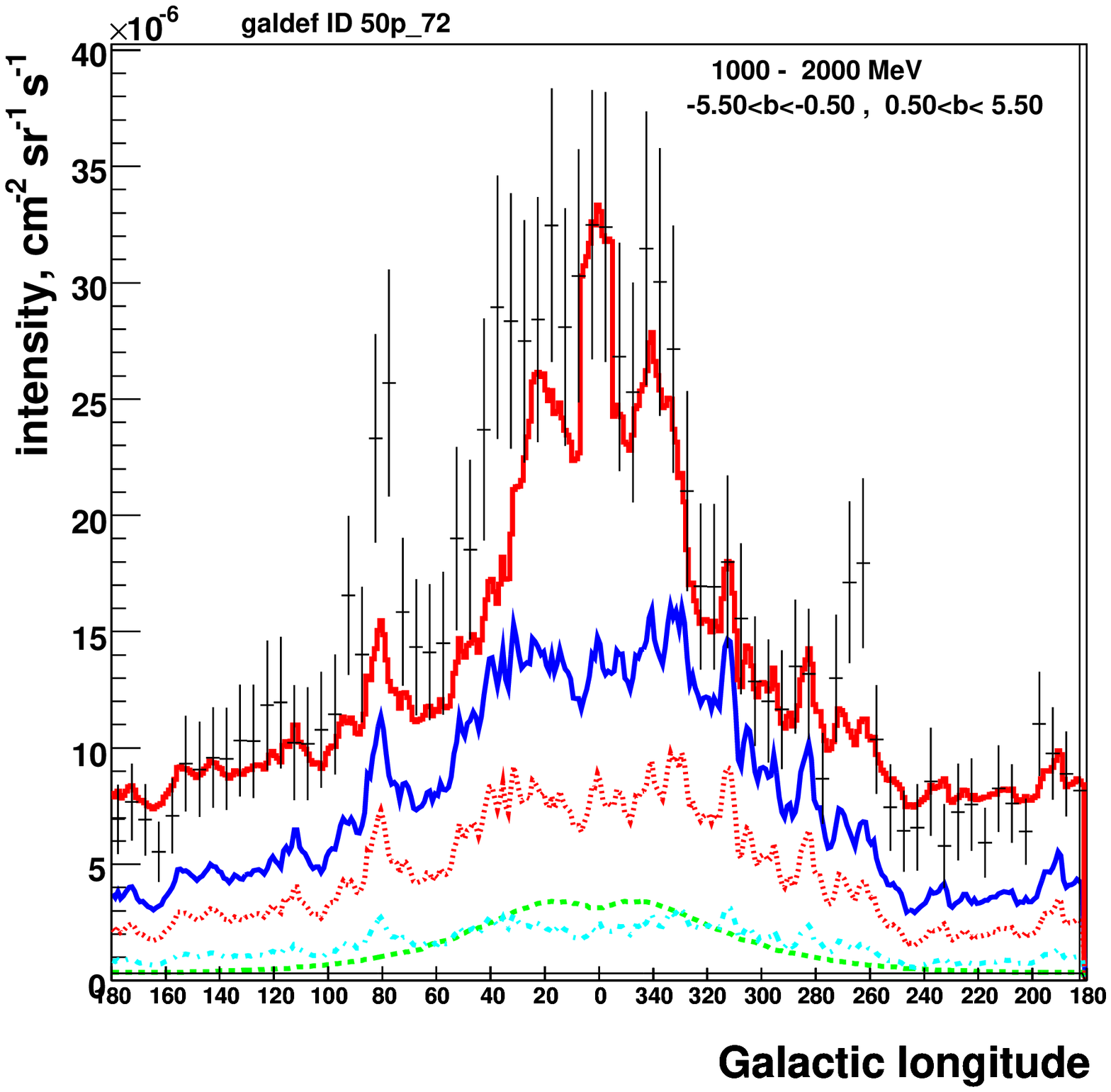}
\includegraphics[scale=0.25]{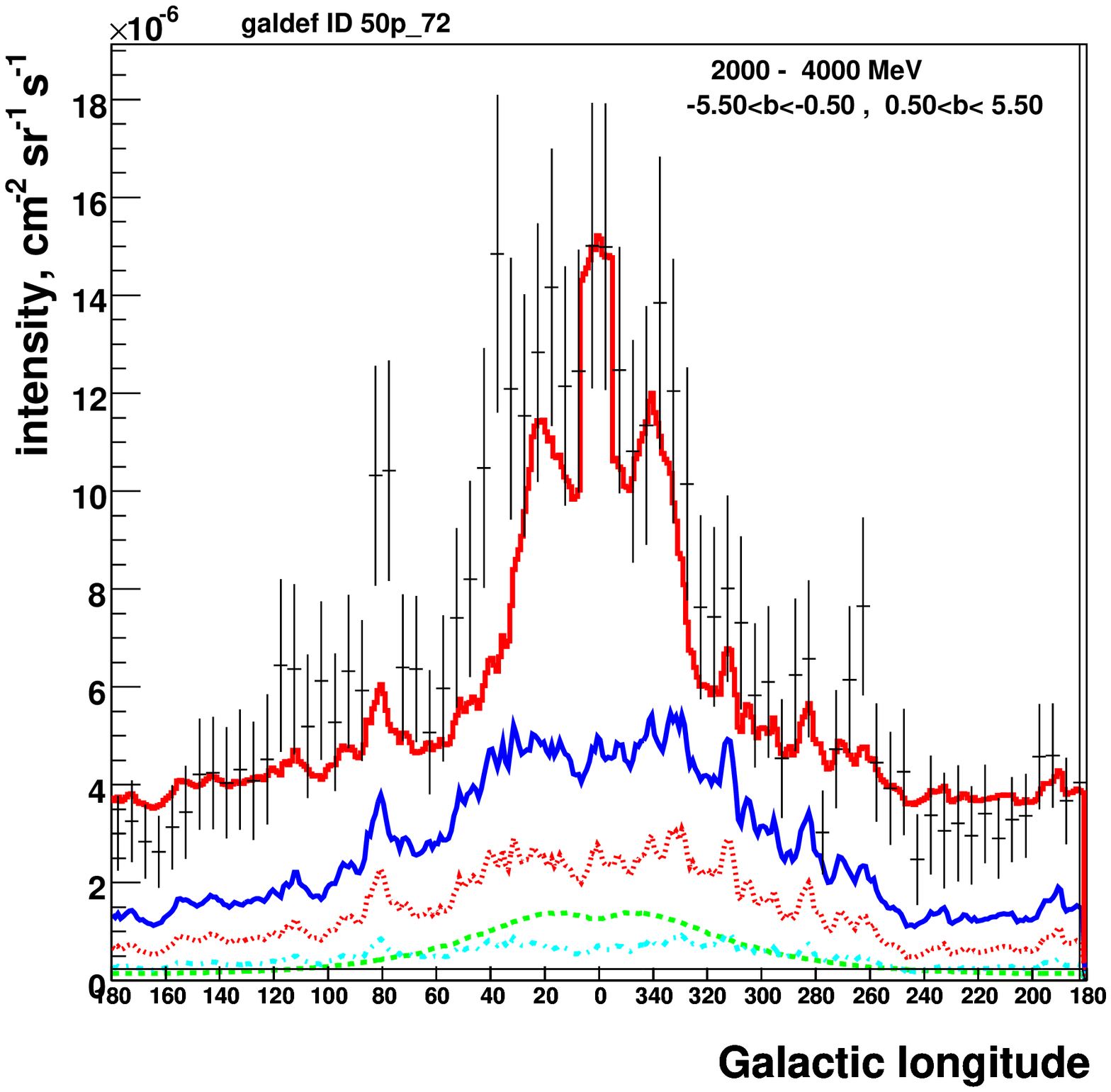}
\includegraphics[scale=0.25]{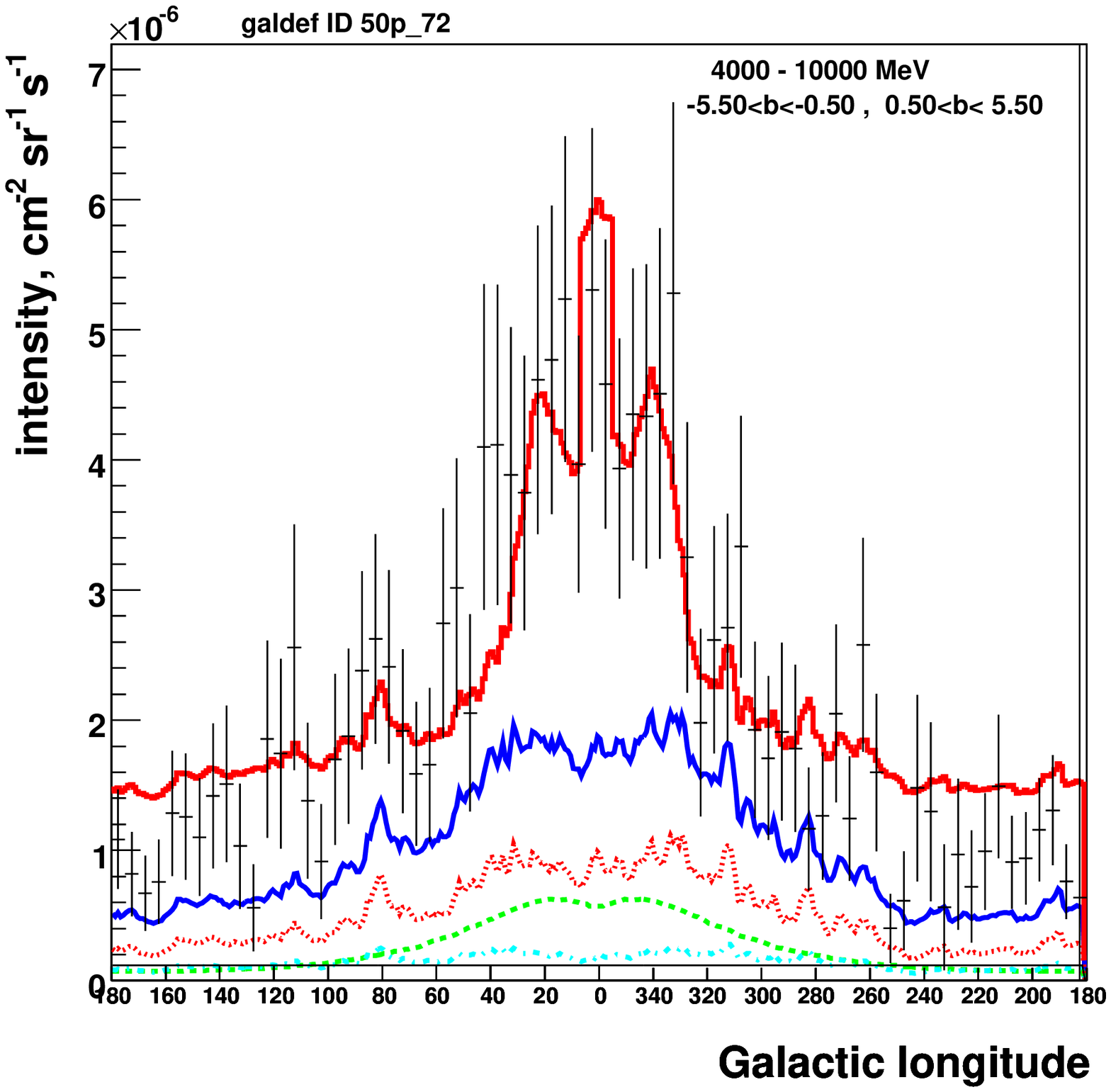}
\caption{\label{l_pf}
Longitude profiles at low latitudes ($|b|<5.5$) in the new propagation model,
compared with EGRET data in 10 energy ranges from 30 MeV to 10GeV.
}
\end{figure}

\begin{figure}
\includegraphics[scale=0.25]{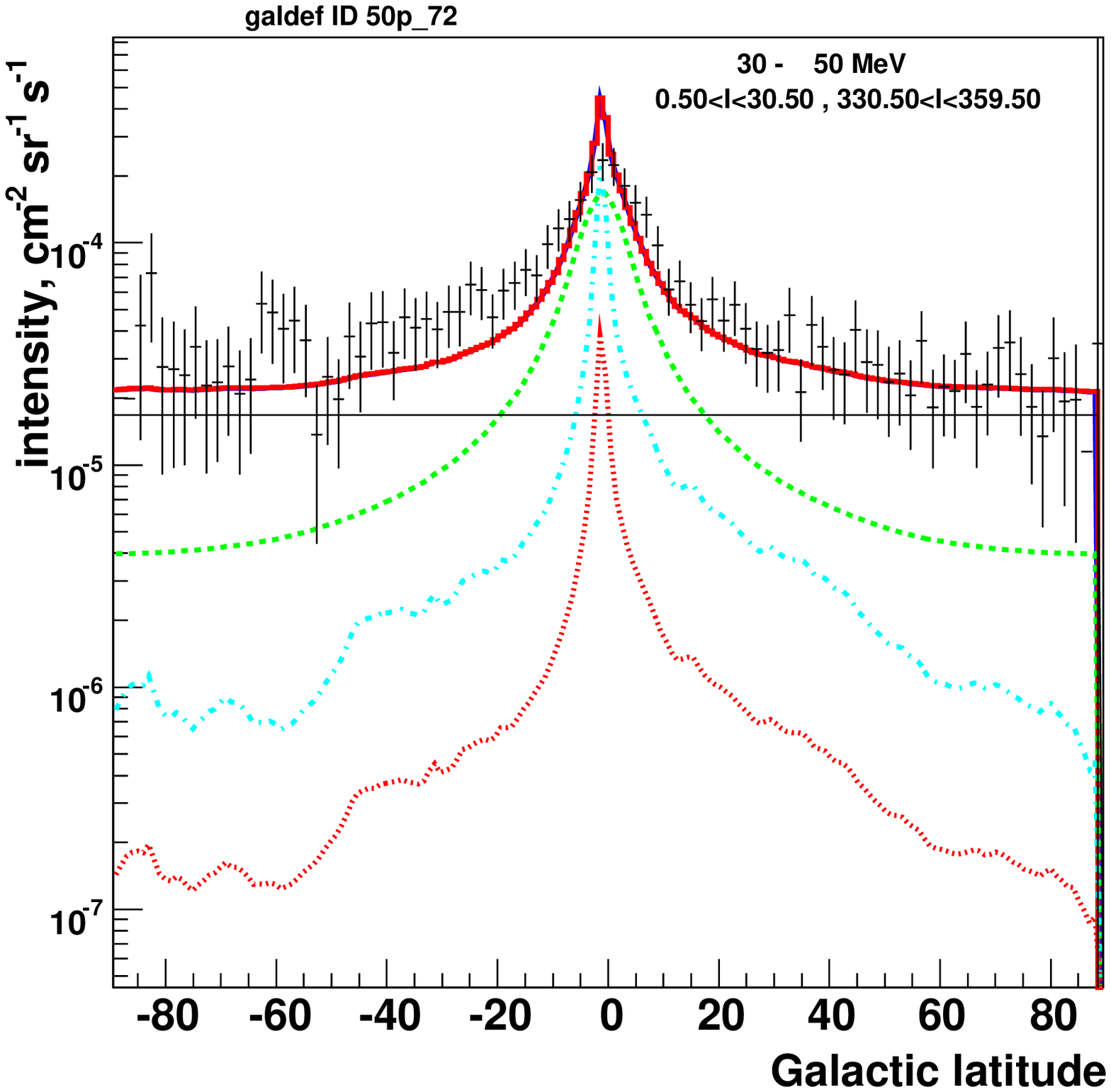}
\includegraphics[scale=0.25]{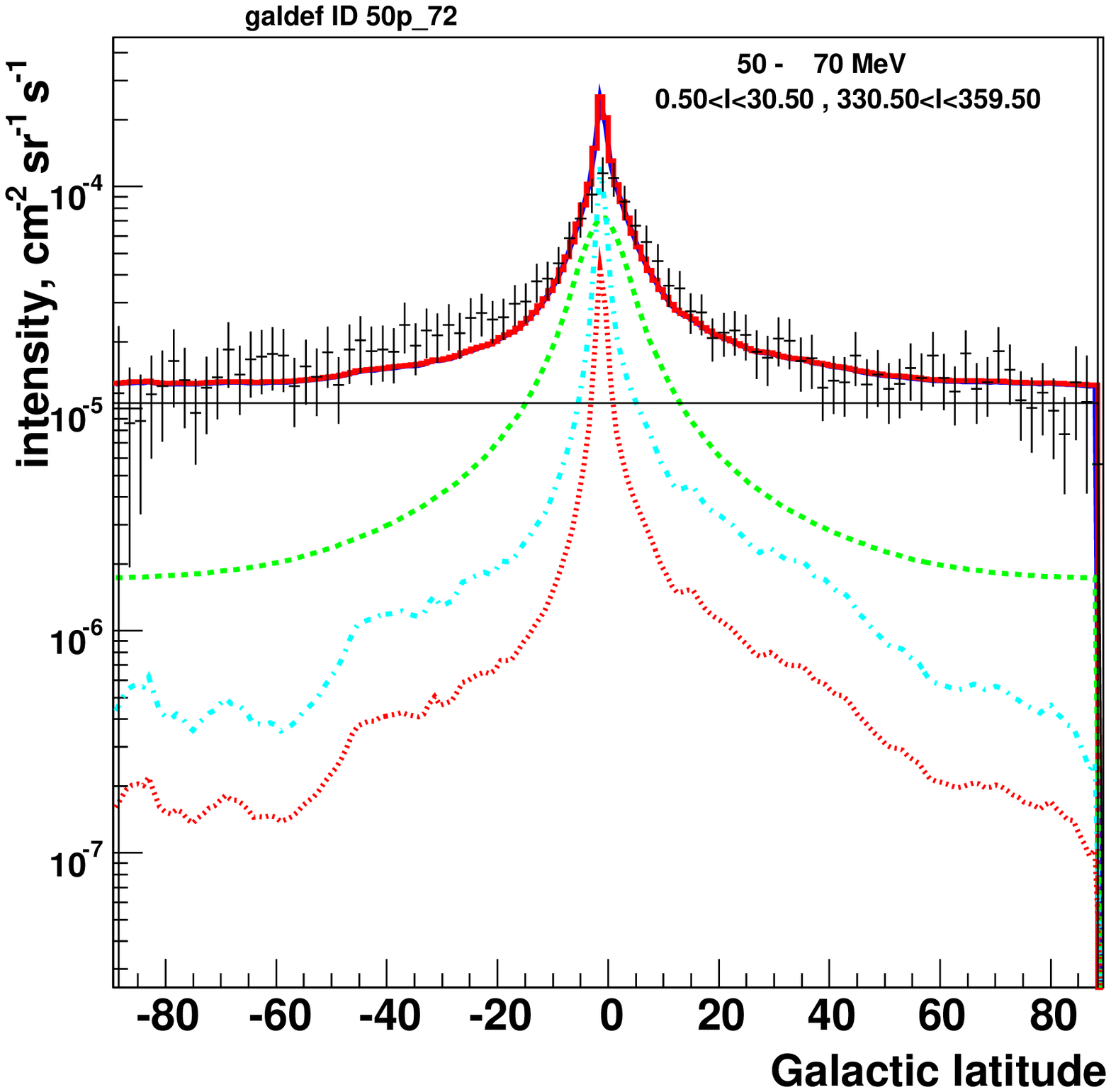}
\includegraphics[scale=0.25]{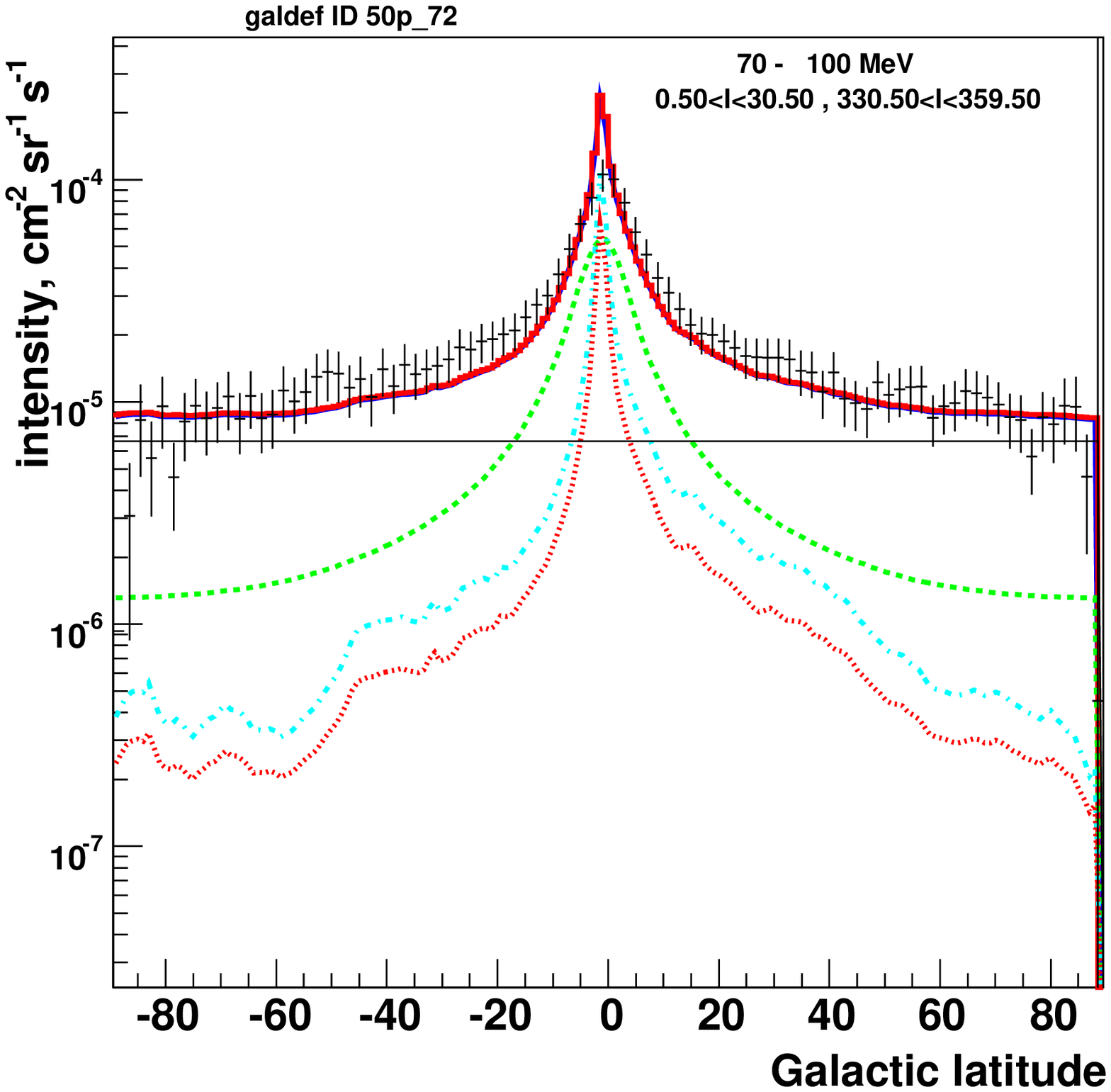}
\includegraphics[scale=0.25]{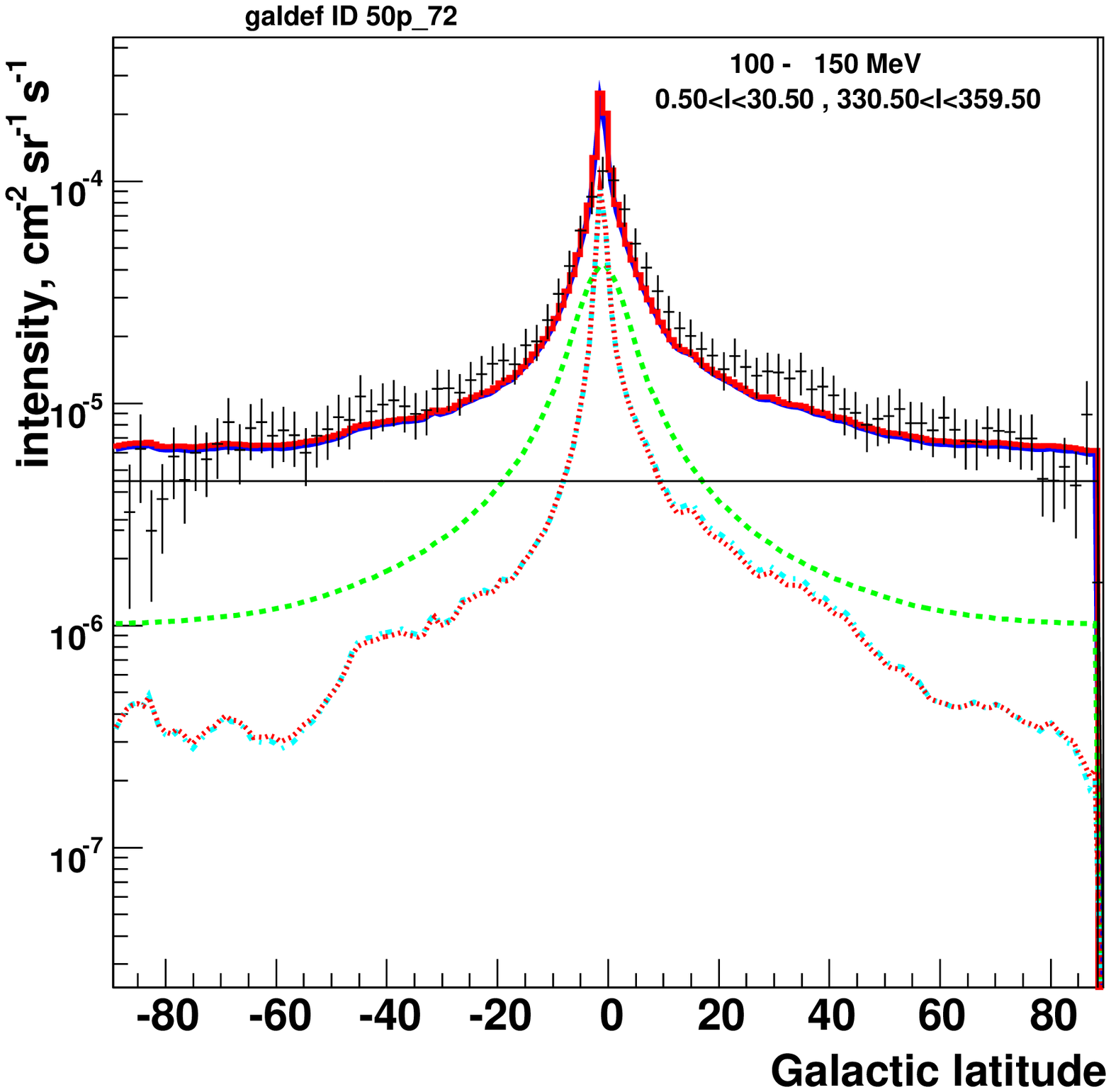}
\includegraphics[scale=0.25]{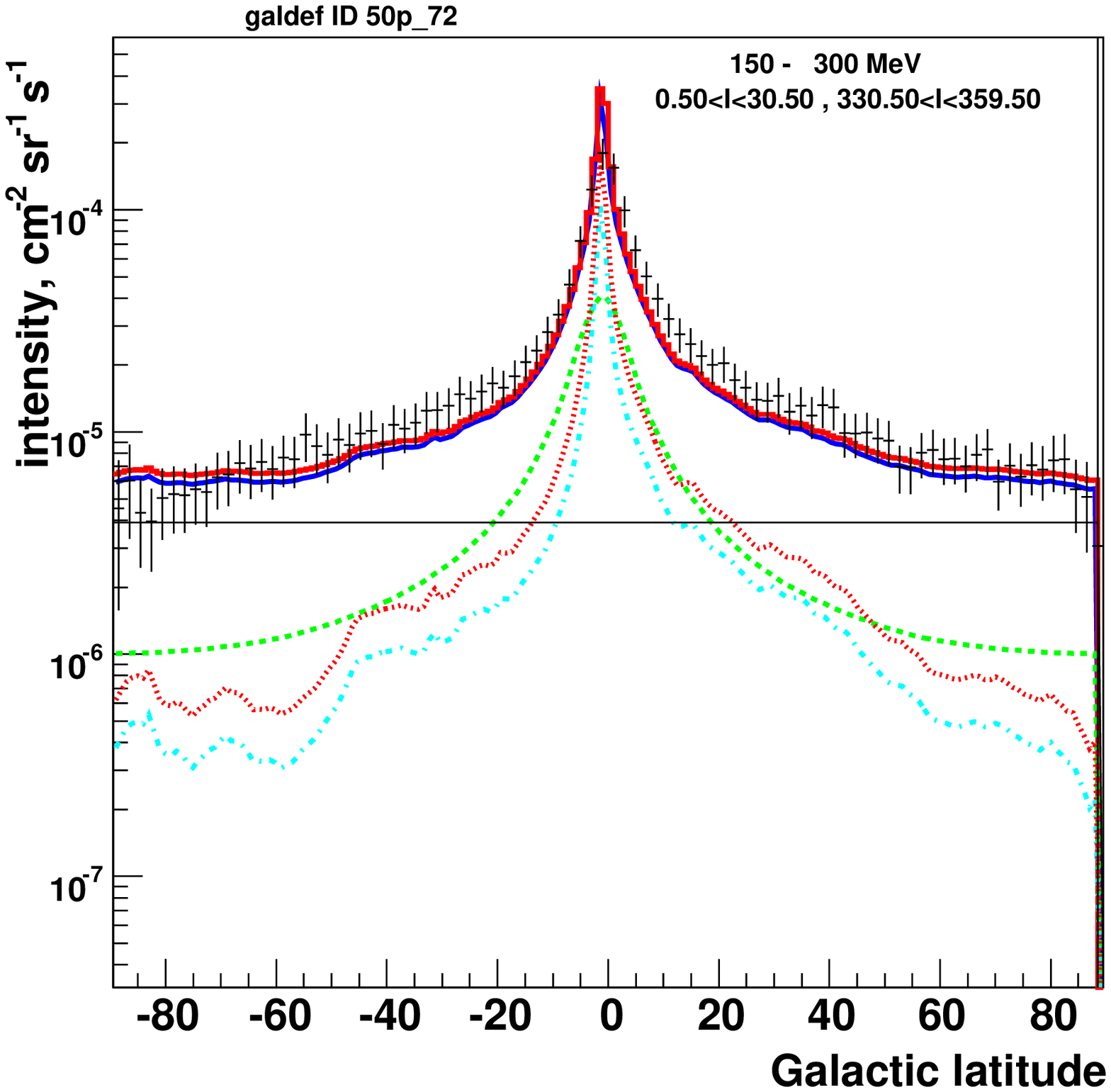}
\includegraphics[scale=0.25]{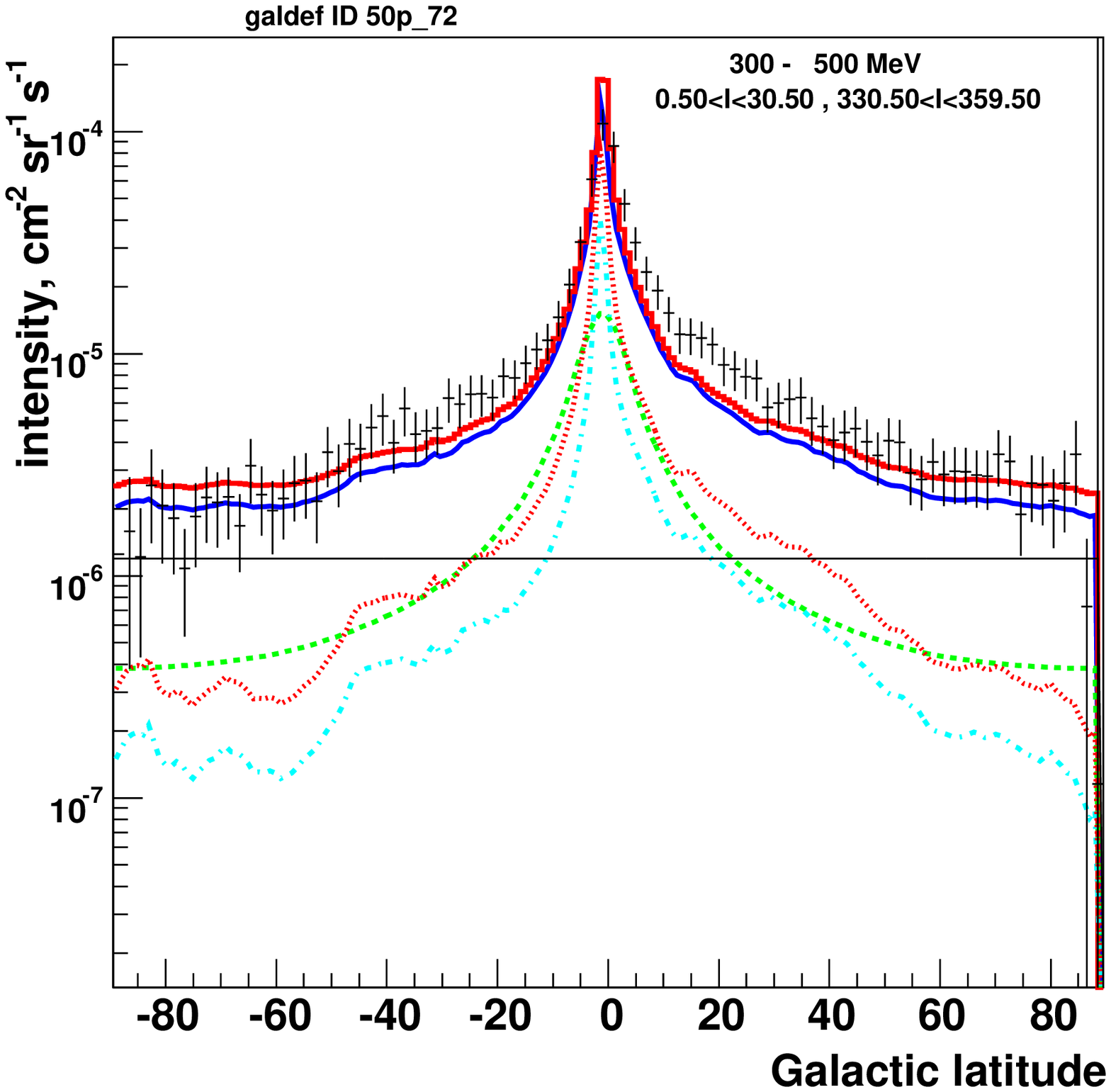}
\includegraphics[scale=0.25]{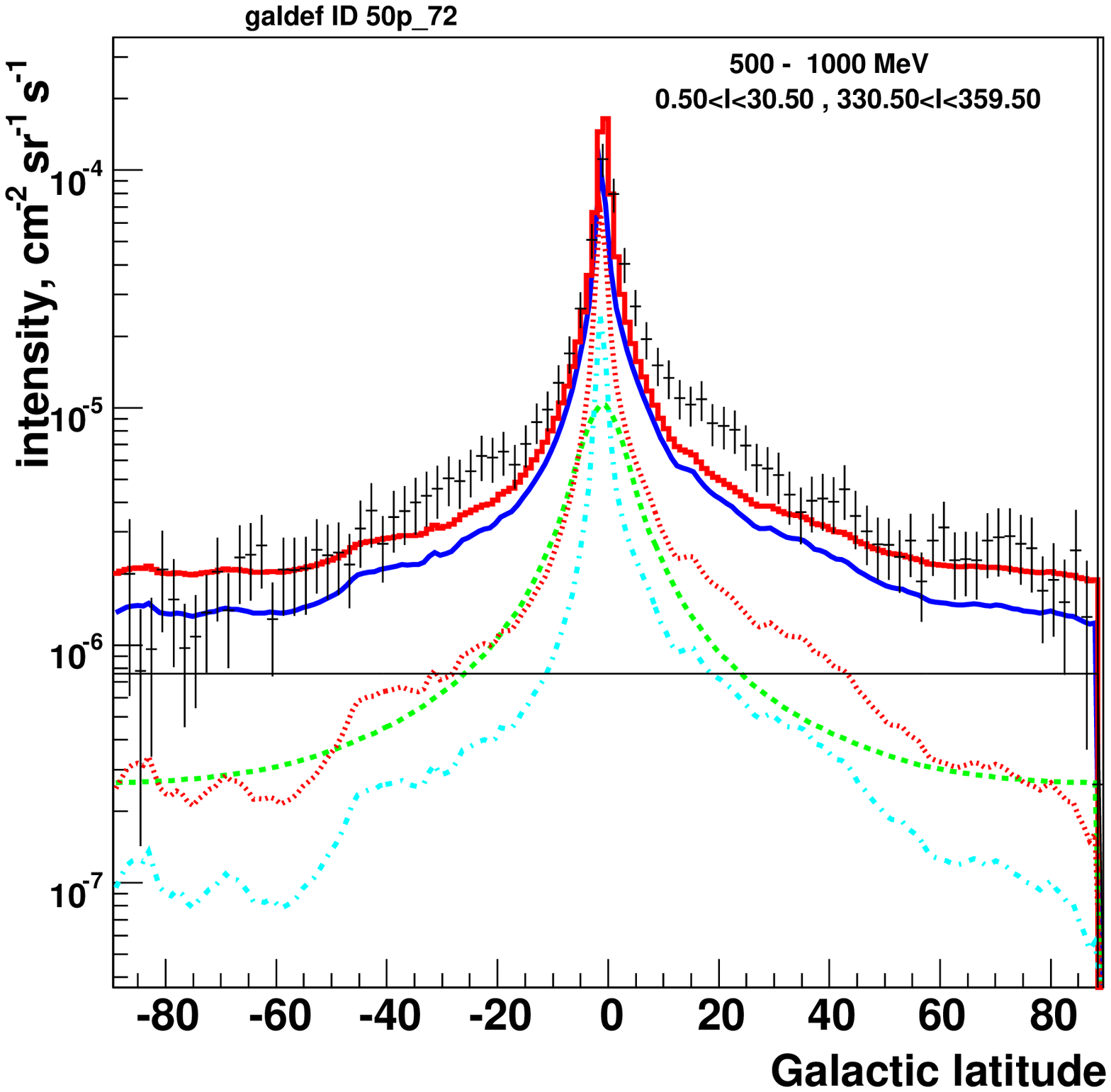}
\includegraphics[scale=0.25]{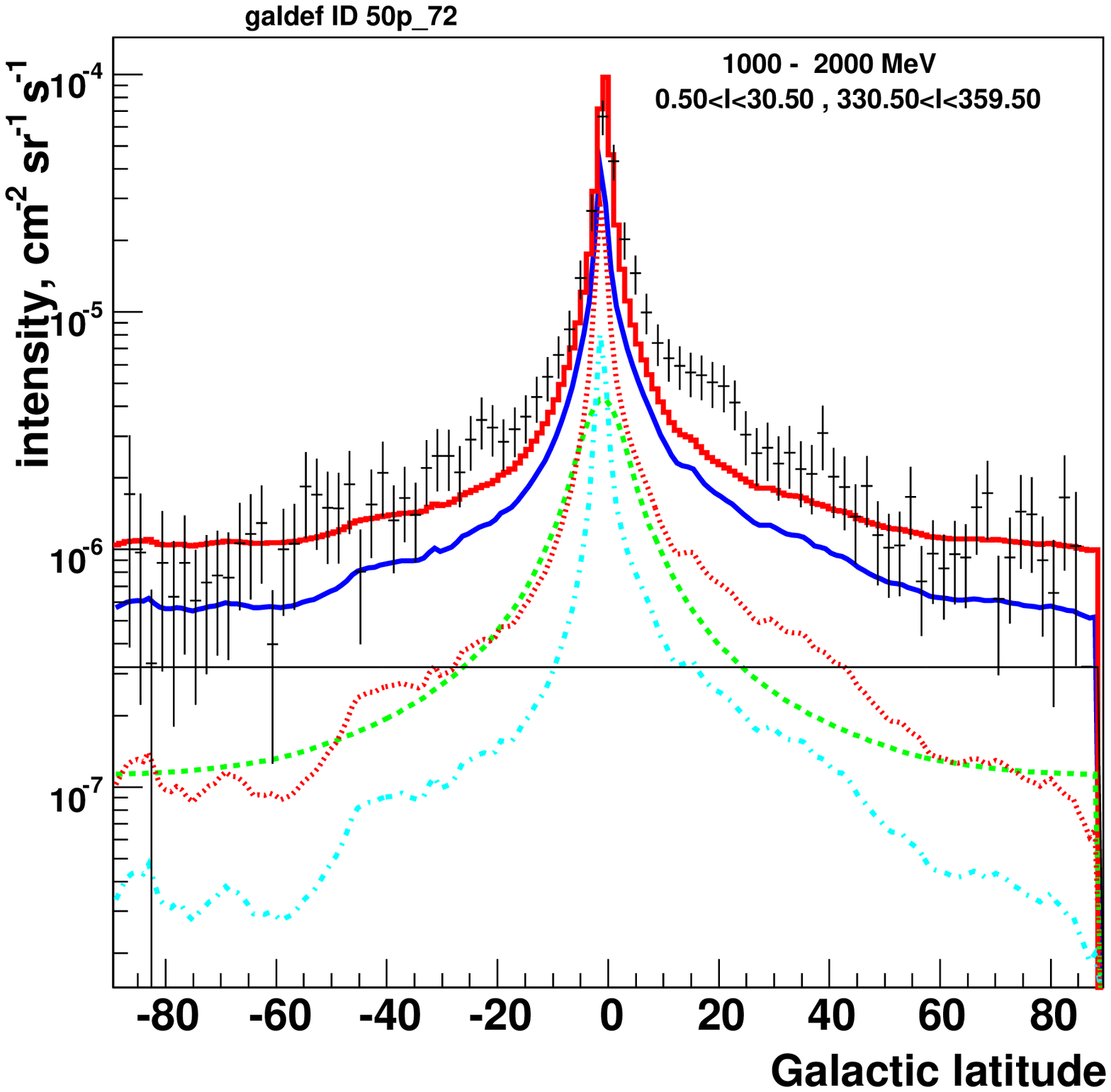}
\includegraphics[scale=0.25]{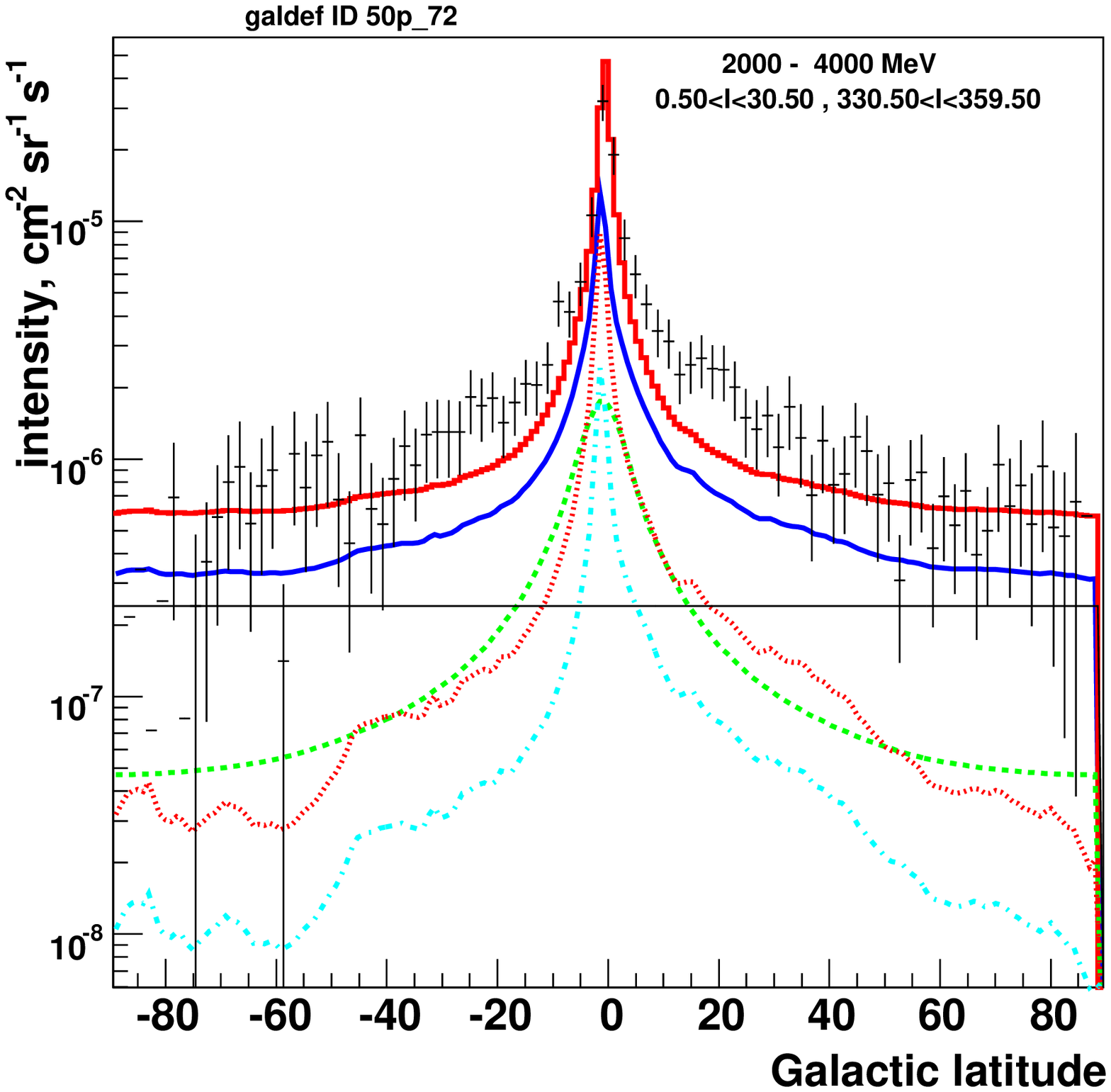}
\includegraphics[scale=0.25]{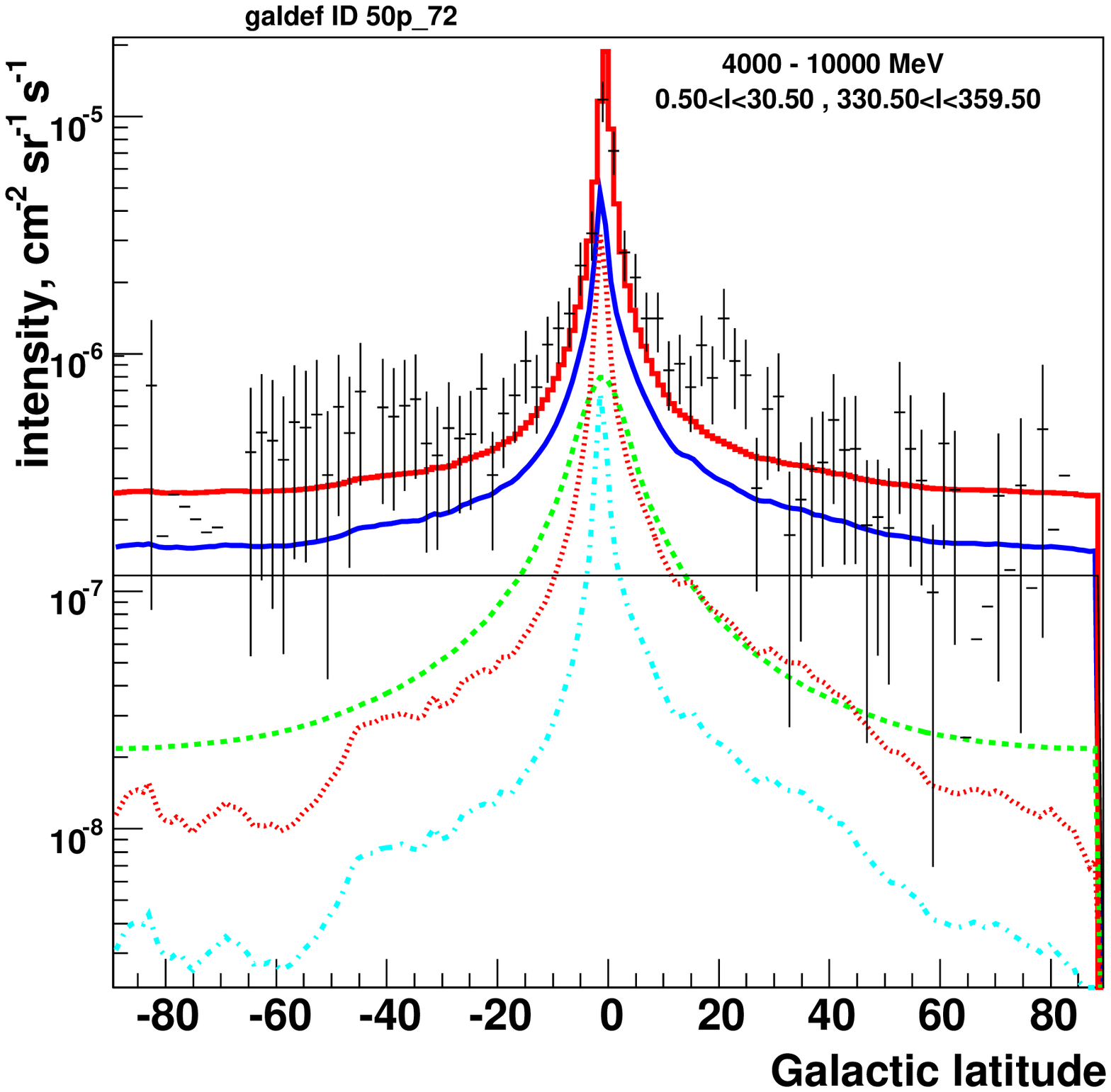}
\caption{\label{in_bpf}
Latitude profiles in the inner Galaxy $|l| < 30^\circ$
in the new propagation model,
compared with EGRET data in 10 energy ranges from 30 MeV to 10GeV.
}
\end{figure}

\begin{figure}
\centering
\includegraphics[scale=0.25]{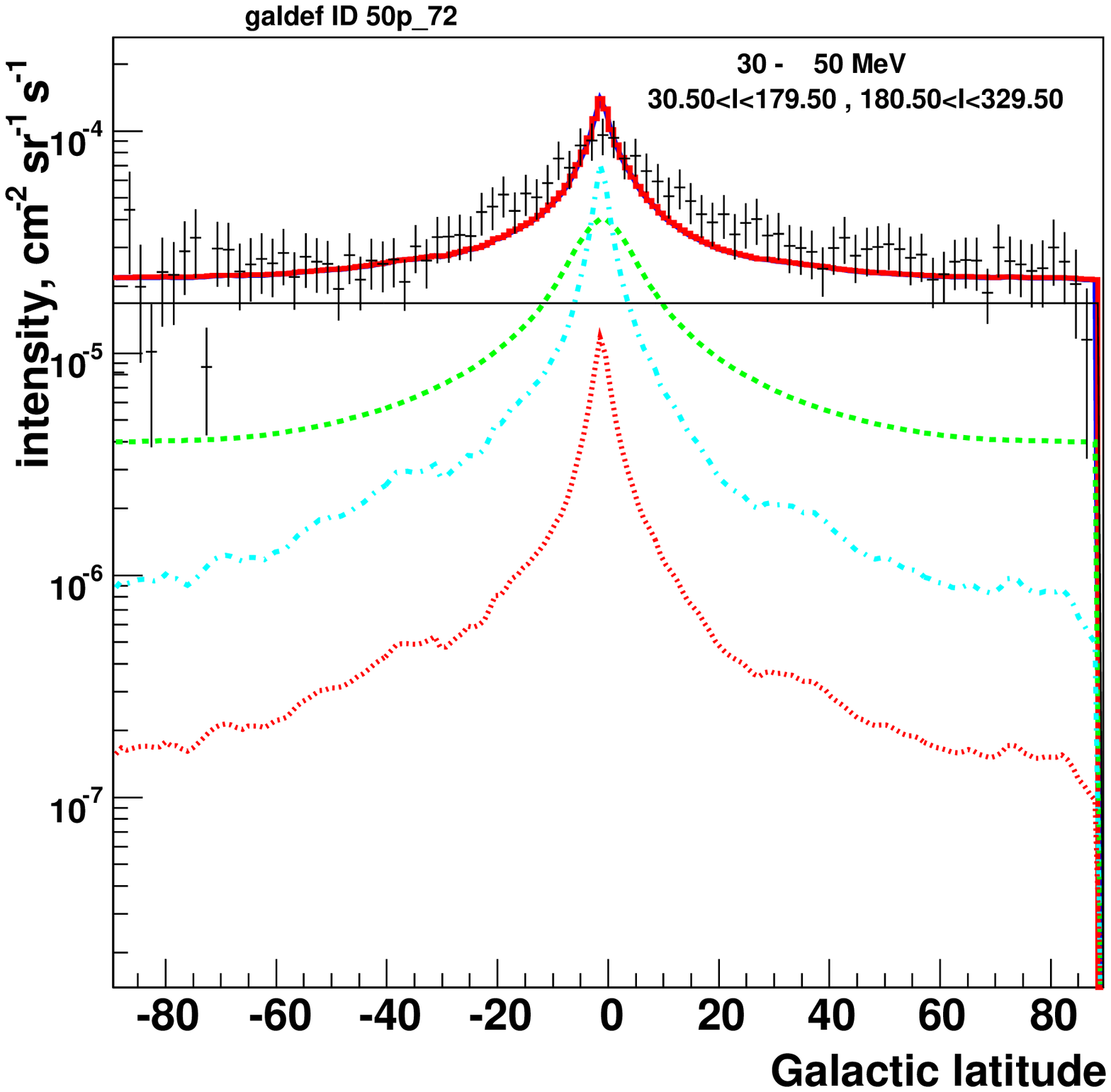}
\includegraphics[scale=0.25]{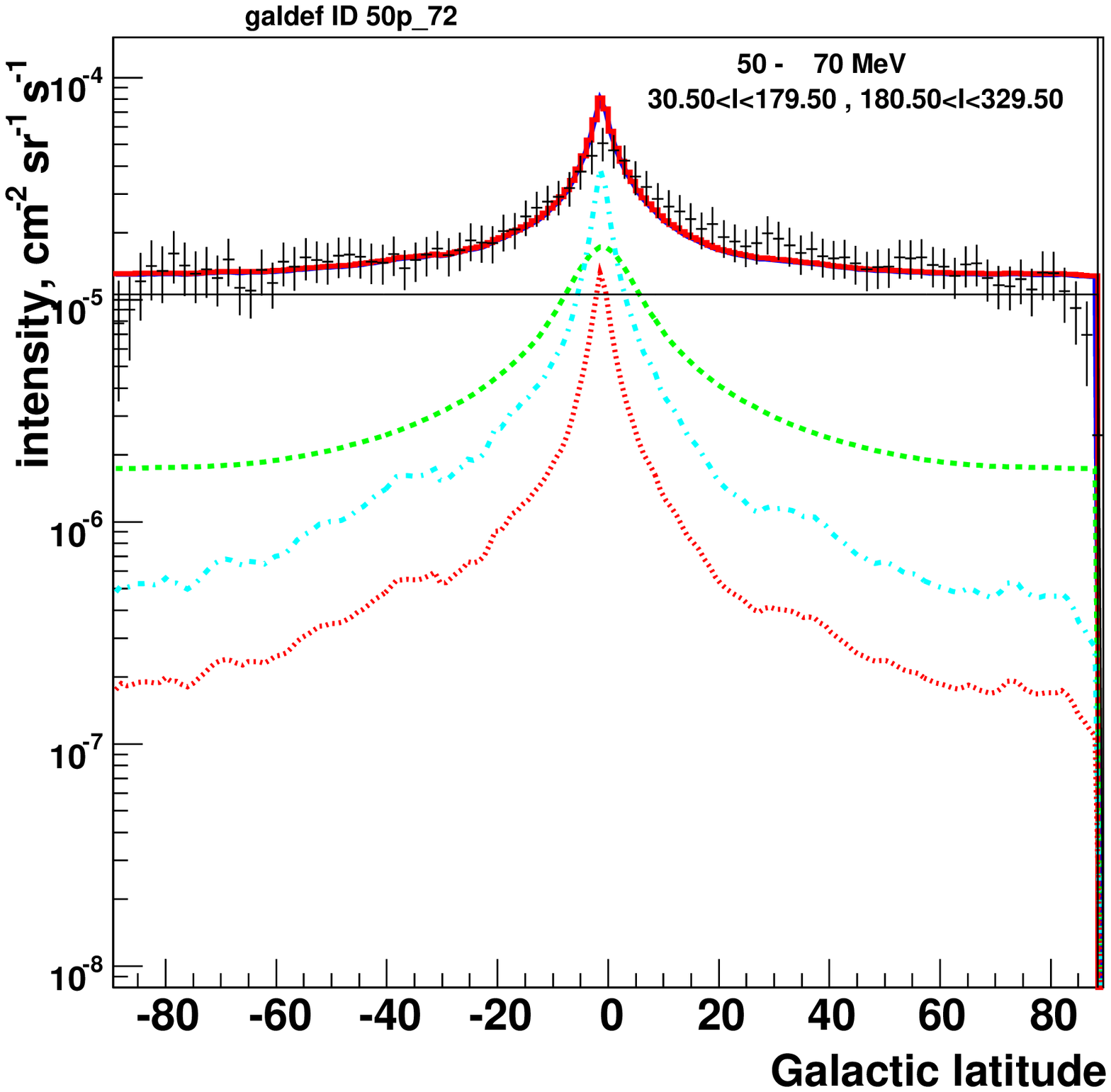}
\includegraphics[scale=0.25]{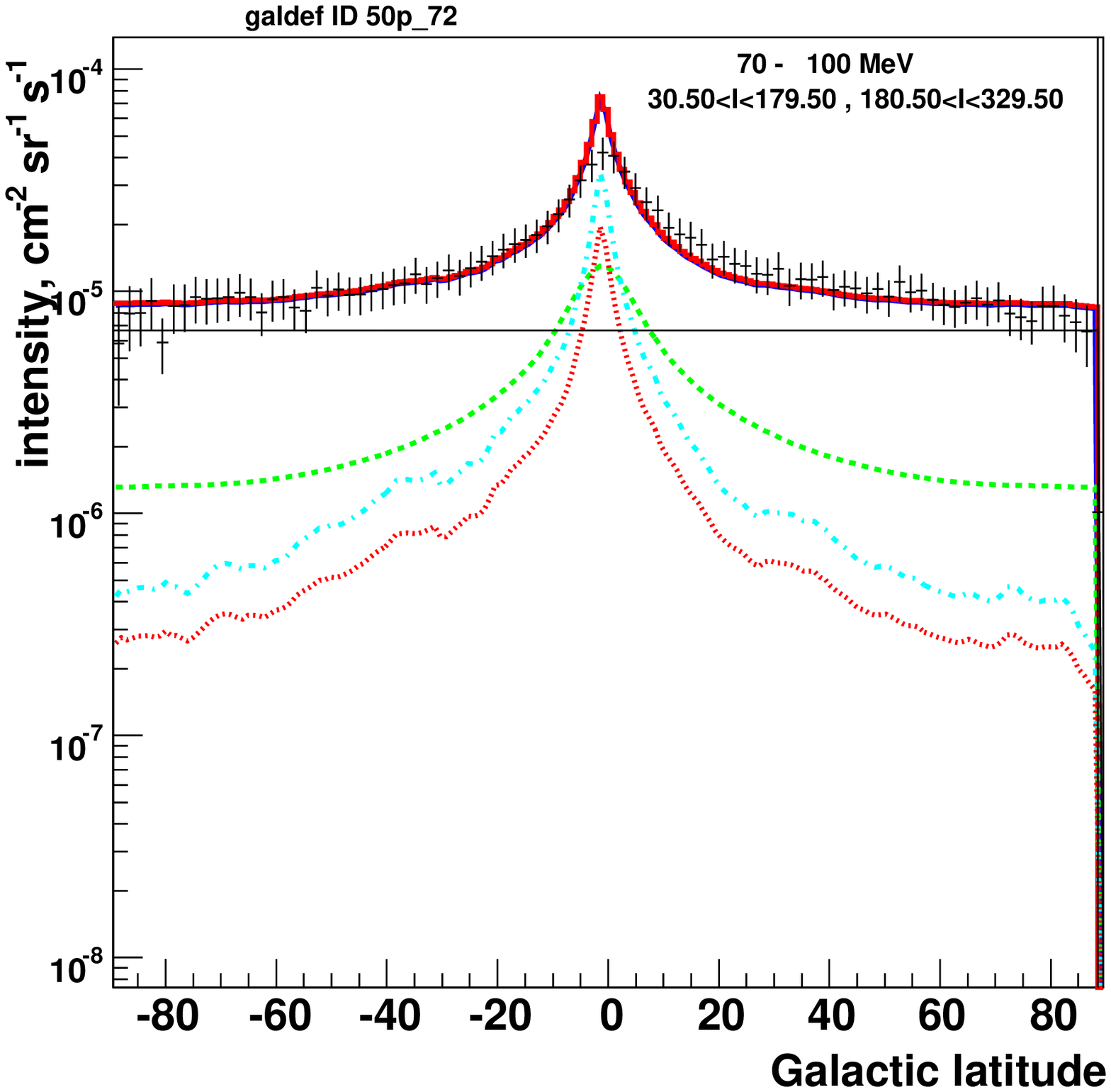}
\includegraphics[scale=0.25]{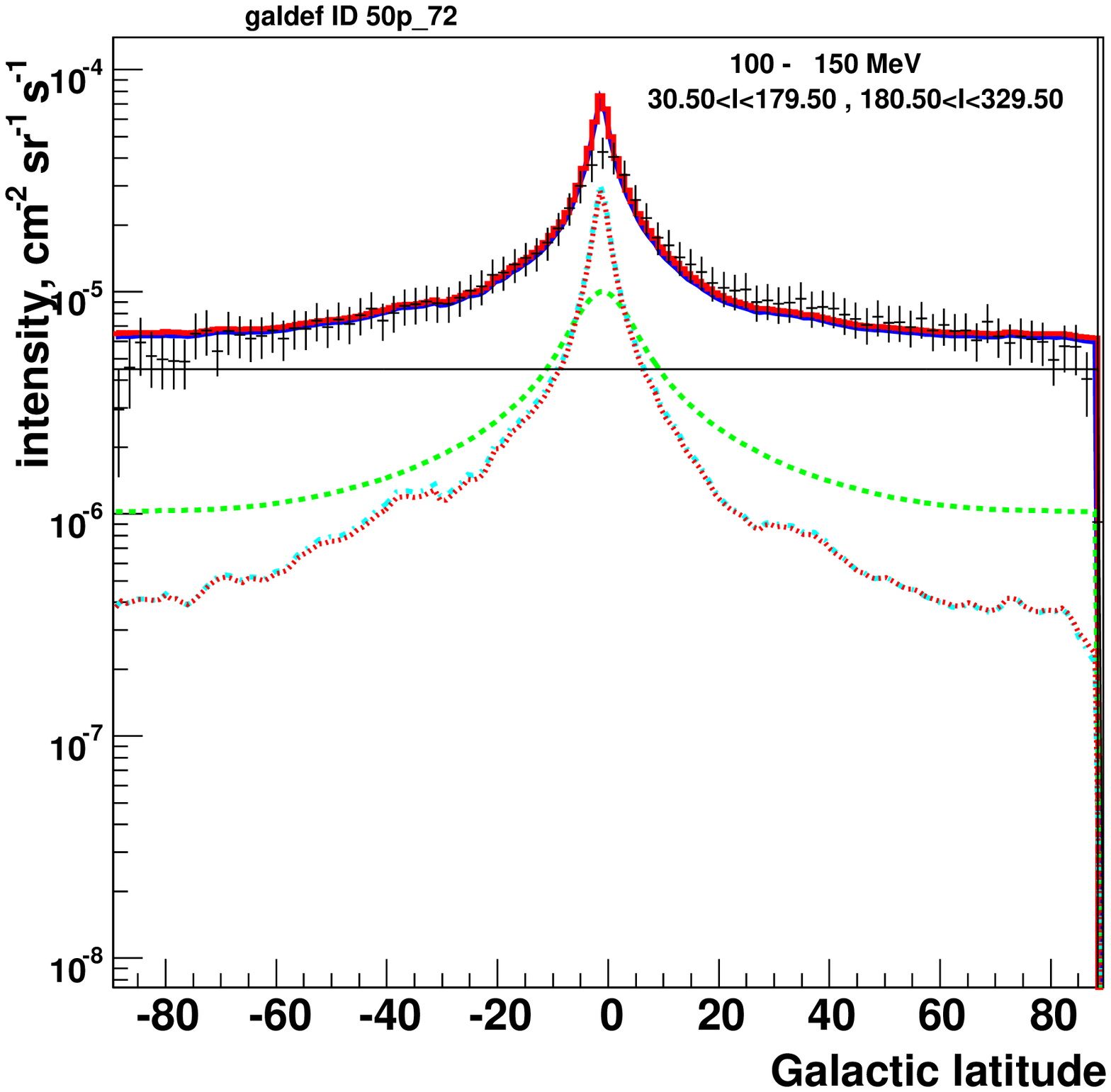}
\includegraphics[scale=0.25]{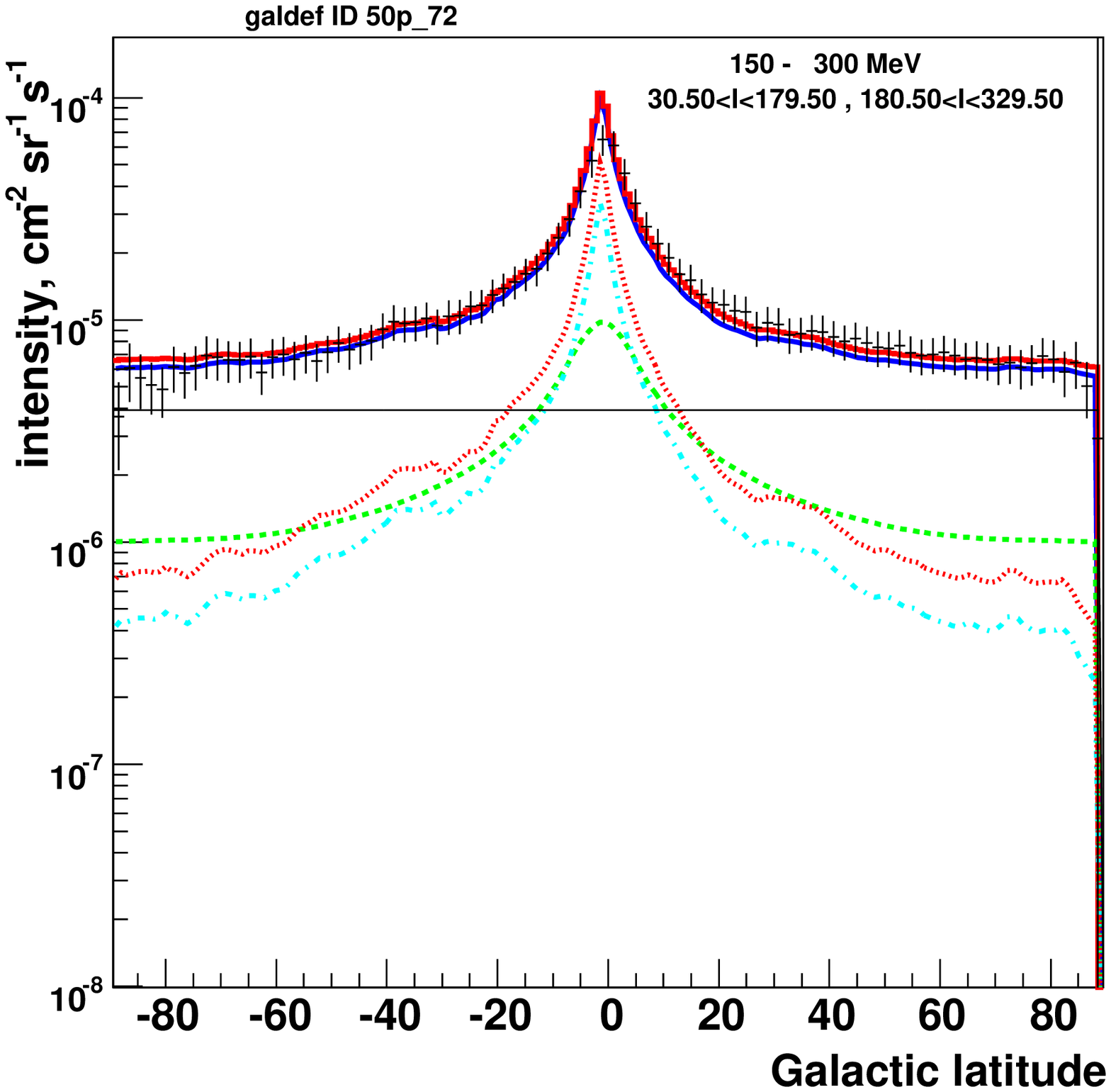}
\includegraphics[scale=0.25]{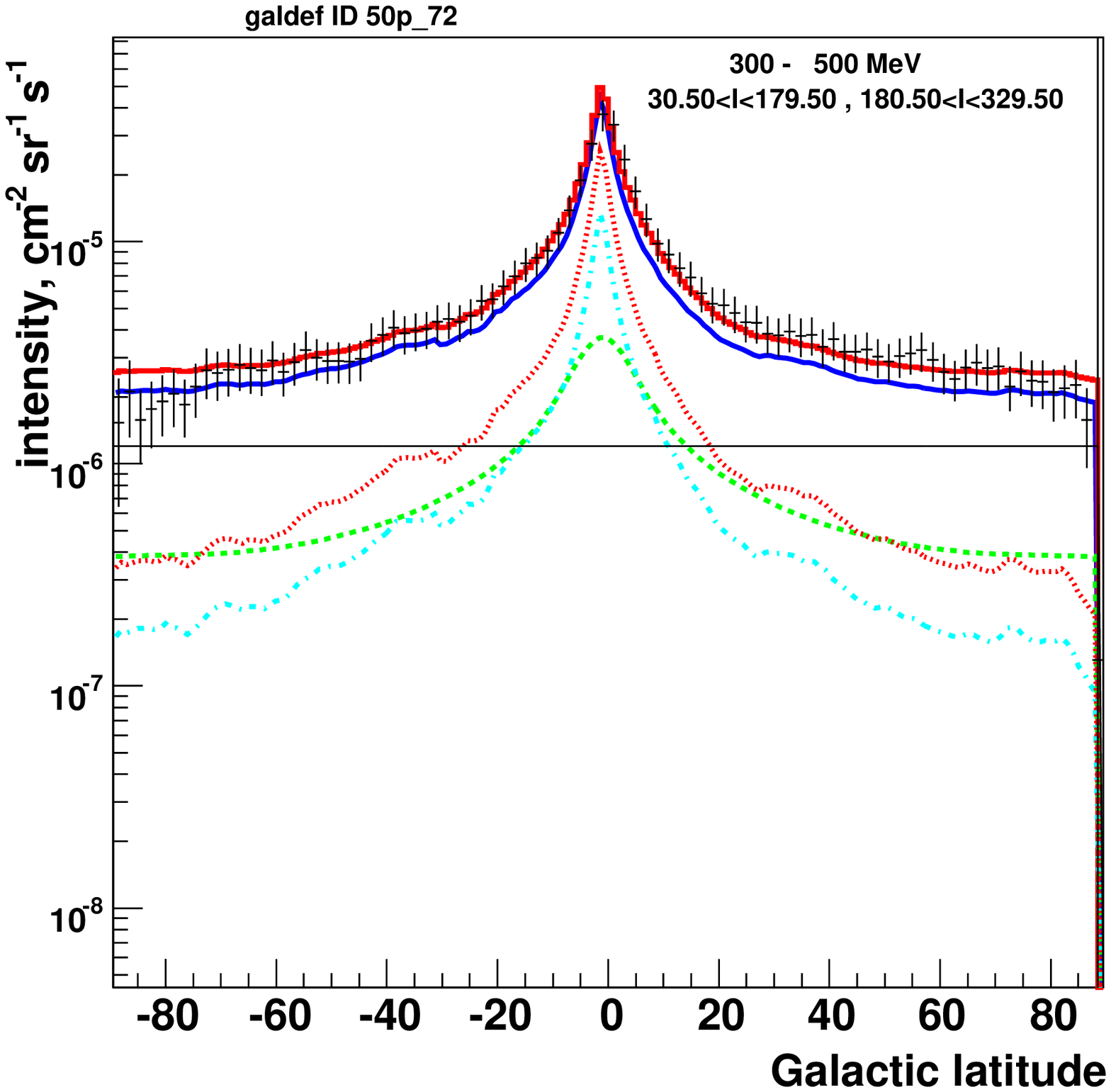}
\includegraphics[scale=0.25]{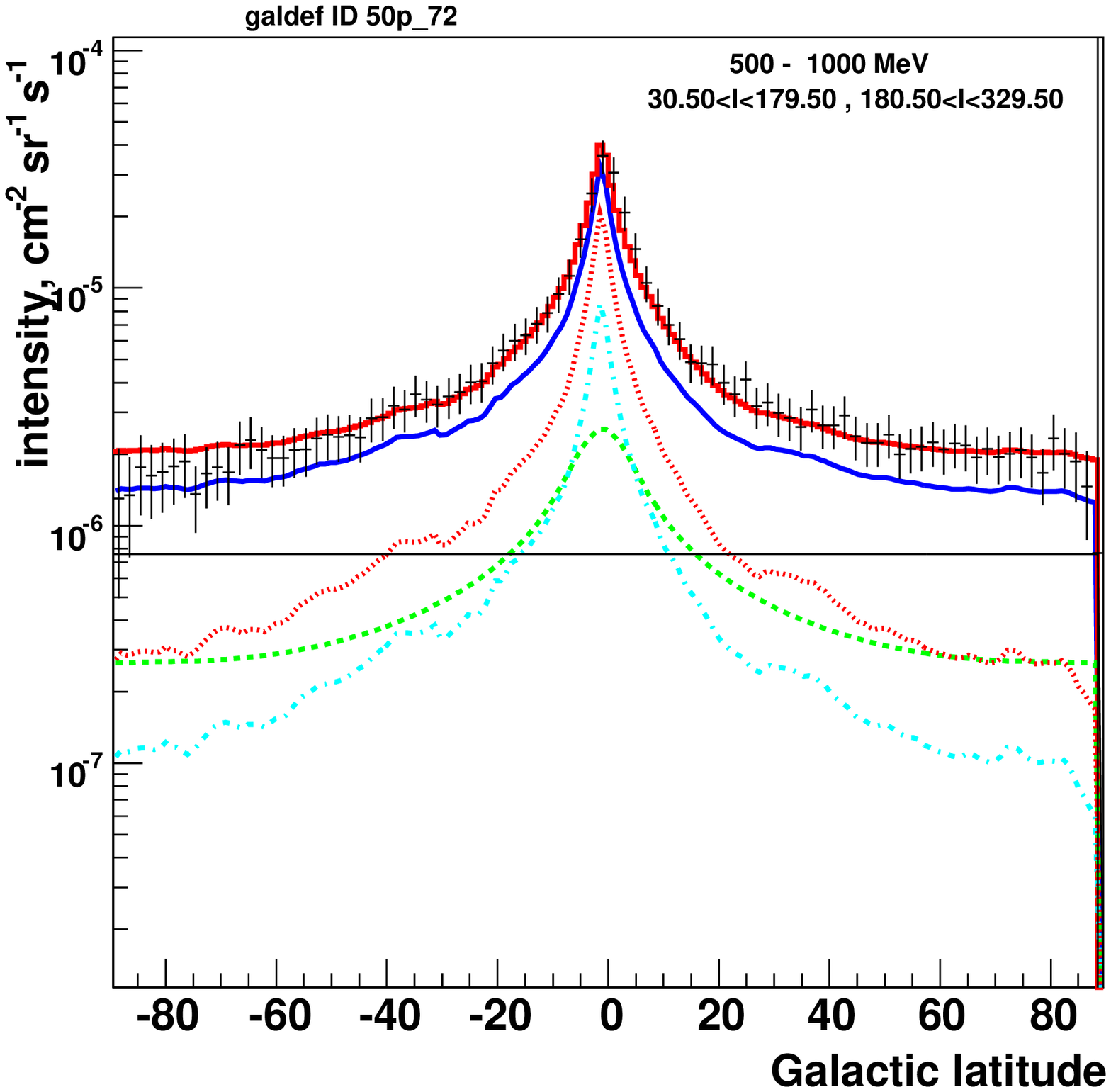}
\includegraphics[scale=0.25]{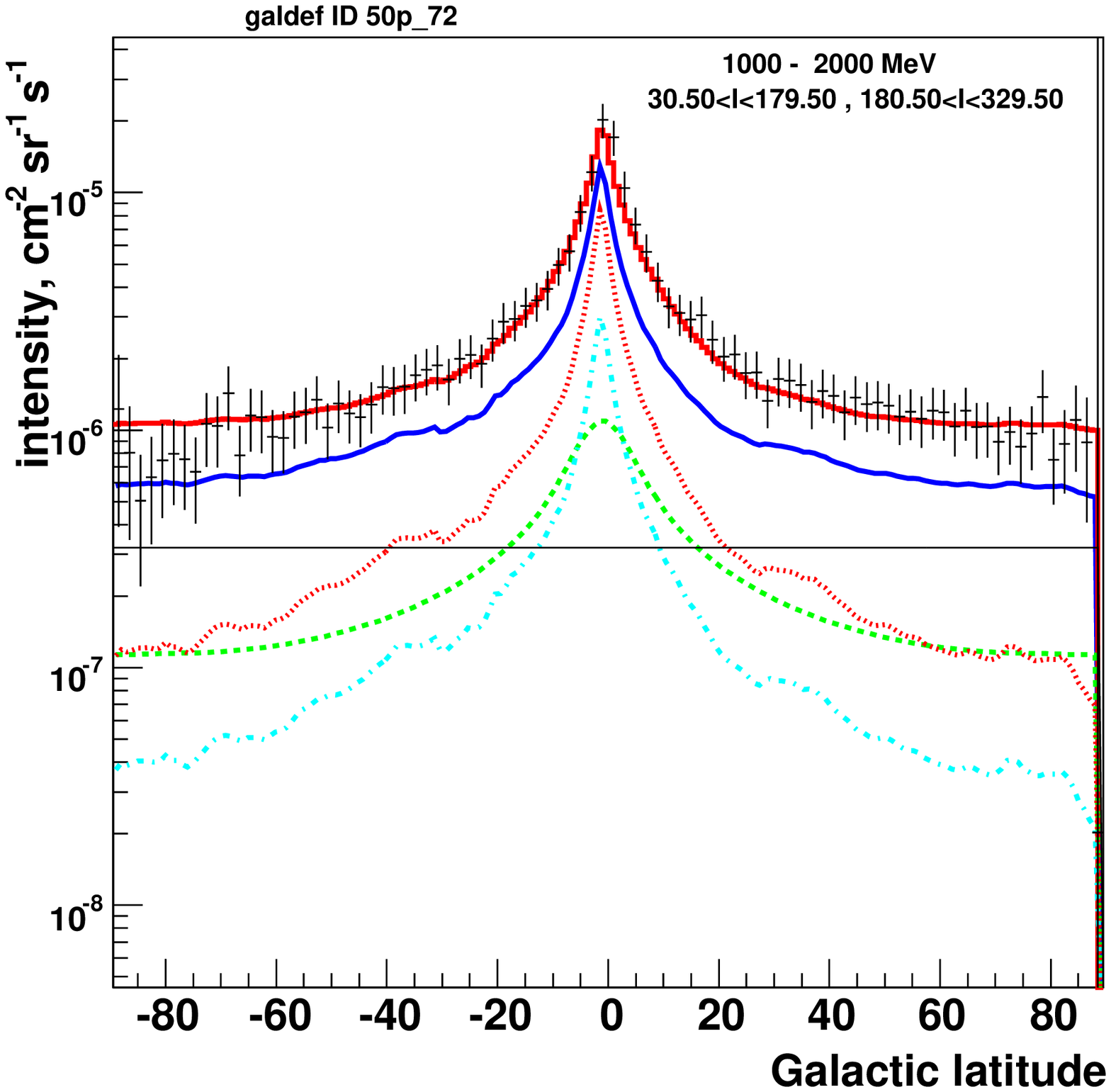}
\includegraphics[scale=0.25]{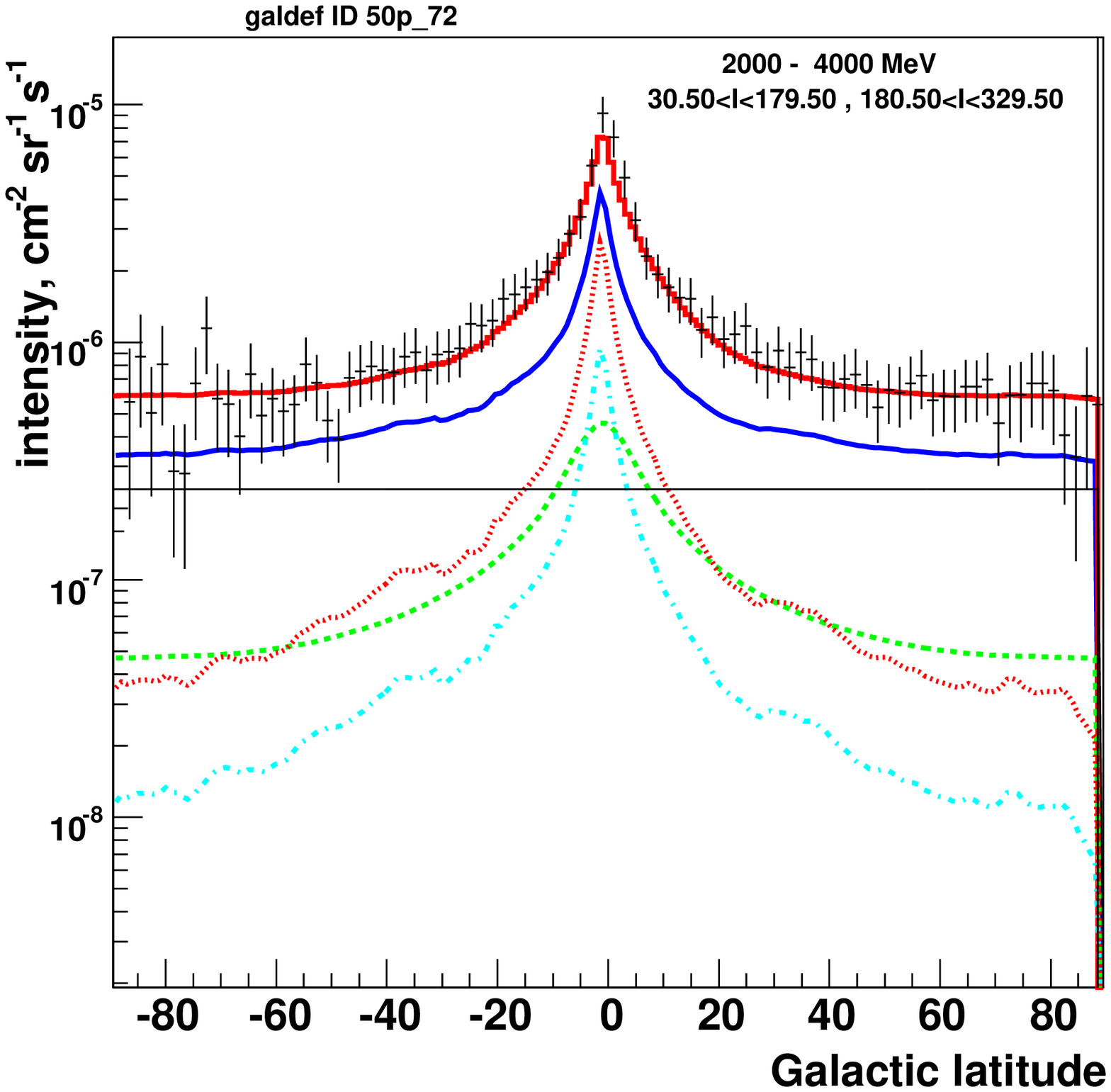}
\includegraphics[scale=0.25]{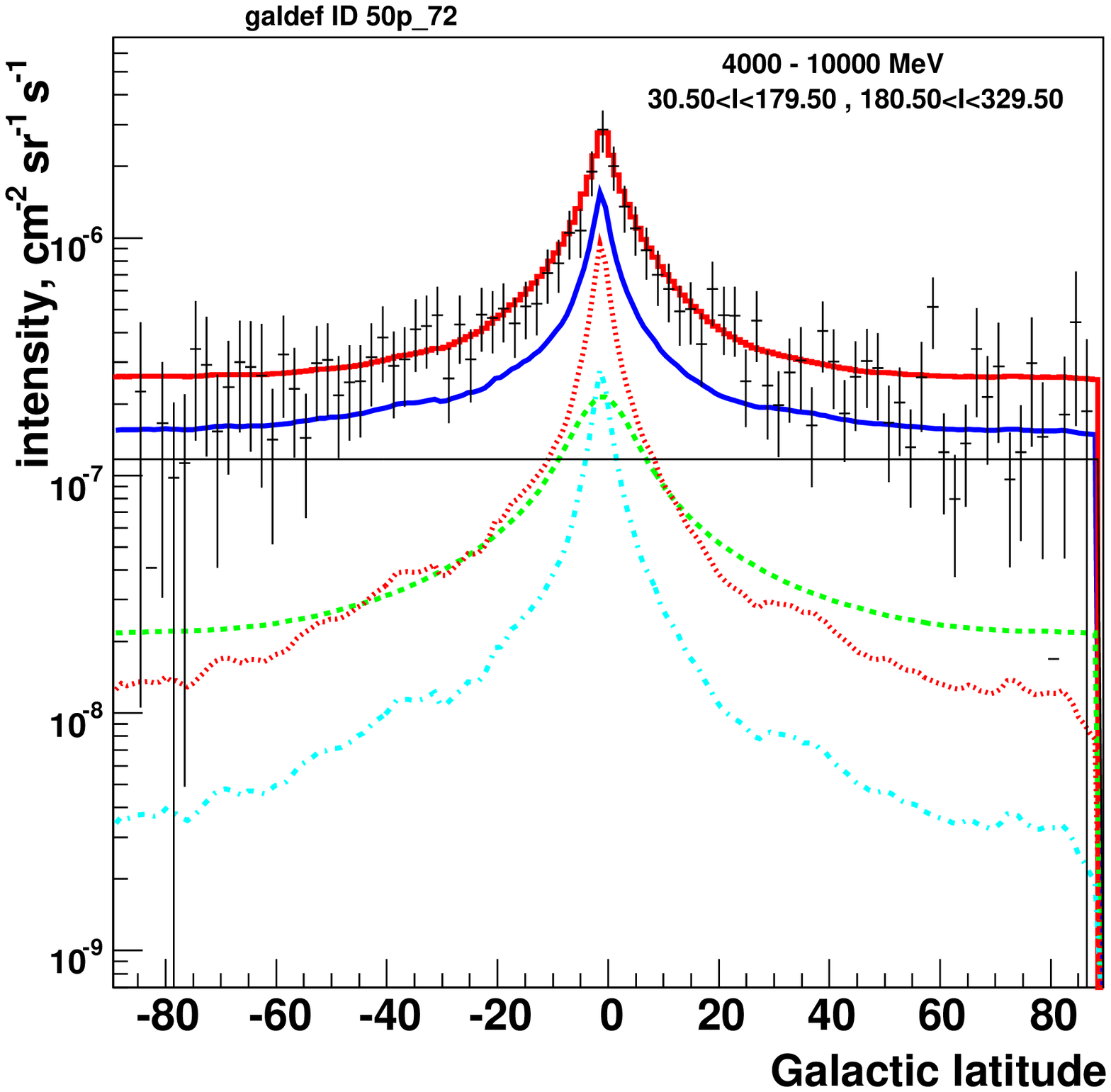}
\caption{\label{out_bpf}
Latitude profiles in the longitude ranges $30^\circ < l < 330^\circ$
in the new propagation model,
compared with EGRET data in 10 energy ranges from 30 MeV to 10GeV.
}
\end{figure}

Figs.~\ref{l_pf},~\ref{in_bpf} and \ref{out_bpf} display the diffuse $\gamma$
longitudinal and latitudinal profiles in our present model.
The line styles in these figures are the same as those given in 
Figs.~\ref{result} and \ref{gamma}, representing contributions from $\pi^0$ 
decay, IC, bremsstrahlung and EGRB respectively. 
The solid red line is the total contribution from 
the background $\gamma$ rays (solid blue) and DMA signals.
From these figures,
it is obvious that the DMA component is essential to explain the
profiles above $500\sim 1000$ MeV.
We notice that the longitudinal profiles at low latitude ($|b|< 5^{\circ}$)
and the latitudinal profiles in the outer Galaxy with 
$30^\circ < l <330^\circ$
are in fairly good agreement with the EGRET data.
However, for the latitudinal profiles in the inner Galaxy 
($330^\circ < l <30^\circ$), we also find that 
the intensities at intermediate latitudes 
$10^\circ\lesssim b \lesssim 30^\circ$
are lower than measurements at energies from 500 MeV to 4000 MeV.
A similar excess is present in the ``optimized model'' \cite{opt},  
in which it is pointed out that
this may be related to an underestimate of the ISRF in the Galactic halo and 
that a factor of $\sim 2$ uncertainty on the
ISRF is quite possible due to the complexity of its calculation \cite{opt}.
Furthermore, the smaller $z_h$ adopted here shallows the 
distribution of electrons, which may also contribute to this discrepancy.

\begin{figure}
\includegraphics[scale=0.6]{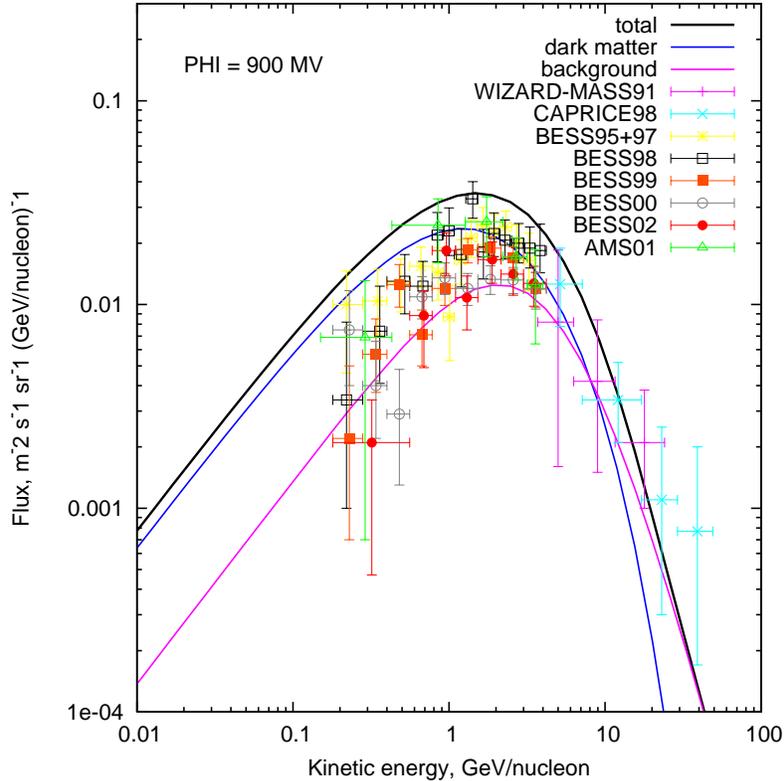}
\caption{\label{pbar2}
Flux of antiprotons after solar modulation in the new propagation model.
Lines and data are the same as in Fig.~\ref{pbar}.
}
\end{figure}

Finally, the antiproton flux is given in Fig.~\ref{pbar2}.
Below $\sim 8$ GeV, the $\bar{p}$ flux from DMA dominates the
CR secondary $\bar{p}$. 
The total $\bar{p}$ flux is a bit higher than the best fit value
of the experimental data \footnote{Here we take the solar modulation 
potential $\Phi=900$ MV. In fact, the modulation
parameter is not a free parameter for different CR species, 
but depends on the solar activities.
The measurements of the $\bar{p}$ flux were taken from the solar 
maximum (BESS00, BESS02) to the solar minimum (BESS95).  }.
However due to large errors of the present experimental data,
the prediction  is still consistent with data within $1\sigma$.
Forthcoming high precision $\bar{p}$ measurements  by 
PAMELA \cite{pamela} and AMS02 \cite{ams} will be helpful
to determine if the present model is viable. 
It should be noted that our model has the potential to further suppress the
$\bar{p}$ flux, by changing the rings position and the diffusion
height $z_h$.

In \cite{boer-prop}, de Boer et al. 
introduced an anisotropic propagation model to greatly suppress the
$\bar{p}$ flux. In their model, only the $\bar{p}$ from DMA within the
gaseous disk is confined by magnetic field and contributes to the $\bar{p}$ flux
at Earth. The $\bar{p}$ from DMA above the gaseous disk is
blown away and has no contribution to the local $\bar{p}$ flux.
In order to reproduce the B/C data, they introduced a 
grammage parameter $c=12$ so that the secondaries and the resident time
are increased by this factor, as a result of the molecular clouds confinement
of charged particles.
However,
we think it may be hard to reproduce the diffuse $\gamma$ ray
data at intermediate and high latitudes in this model, since
the CRs above the gas disk are quickly blown away and produce very low
$\gamma$ ray emissivity.
Our model gives consistent Galactic diffuse $\gamma$ rays and $\bar{p}$
flux with experimental data simultaneously without drastic modifications 
of the GALPROP model. 

\section{Summary and discussion}

In this work we propose to solve the ``GeV excess'' problem
of the Galactic diffuse $\gamma$ rays by developing a new propagation
model which includes contributions from DMA.
We have shown that this propagation model can well reproduce the 
B/C, $^{10}$Be/$^{9}$Be data and spectra of protons and electrons.
The Galactic diffuse $\gamma$ ray spectrum at different sky regions and
its profile as a function of longitude and latitude
are also shown to describe the EGRET data very well, if the
DMA contribution is included.
The $\bar{p}$ flux in this model is consistent with experimental data
within $1\sigma$.
Compared with previous works \cite{opt,boer-fit}, our model does not introduce 
a normalization of the interstellar proton intensity different
from the local one (as it should be universal due to the negligible energy
loss of protons). Furthermore, neither the ``boost factors'' to the DMA 
signals nor the arbitrary renormalization of the Galactic $\gamma$-ray 
background are needed in our model to explain the EGRET data. 



In our model, the $\bar{p}$ flux from DMA is suppressed by the following
changes compared with the de Boer model \cite{boer-fit}: 
1) the smaller $z_h$ helps to suppress the $\bar{p}$ flux 
from the smooth component of DM.
The average gas density CRs crossed may be smaller 
than the average gas density in the disk, as shown in \cite{chandran}.
This fact tends to favor a smaller height of the CR diffusion region. 
2) We do not adopt a universal ``boost factor'' for $\gamma$ rays and
$\bar{p}$. Since the enhancement by subhalos is larger at large radii, 
the boost of $\bar{p}$ at the solar neighborhood is
smaller than the boost of $\gamma$ rays. 
3) The ring parameters are slightly adjusted, which does not change
the $\gamma$ ray profile while greatly suppressing the $\bar{p}$ flux.
This is because the distance dependence of the ${\bar p}$ propagation  
is steeper (exponential decrease) than $r^{-2}$ of $\gamma$-rays \cite{david}.

A potential problem of this model is related to the $\bar{p}$ flux, which
is consistent with data, but cannot best fit the data. Adjusting the propagation
or DM parameters (such as the rings) can further lower the $\bar{p}$ flux.
However, more fine-tuning is required to fit the $\gamma$-ray and rotation curve data.
Note a new explanation of the ``GeV excess'' was given
as an instrumental biases contaminating the EGRET data 
\cite{hunter07}. However, whether this conjecture is correct or not cannot 
be confirmed at the moment. Forthcoming precise observations, such
as GLAST \cite{glast}, PAMELA \cite{pamela} and AMS02 \cite{ams},
will be decisive to validate or disprove this model.
At present, because of the fundamental importance of the DM problem,
we think that any possible implication of DM signals deserves a   
serious treatment.


\begin{acknowledgments}
We thank the anonymous referee for helpful comments on
the manuscript and D. Maurin and J. Lavalle for improvement on English.
This work is supported by the NSF of China under the grant
Nos. 10575111, 10773011 and supported in part by the Chinese Academy of
Sciences under the grant No. KJCX3-SYW-N2.
\end{acknowledgments}

~

\end{document}